%% file: main.tex
\documentclass[aps,prx,10pt,twocolumn,showpacs,amsmath,amssymb,floatfix,superscriptaddress]{revtex4-2}
\usepackage{times}
\usepackage{graphicx}
\usepackage{epsfig}
\usepackage{amsfonts}
\usepackage{amsmath}
\usepackage{amssymb}
\usepackage{color}
\usepackage{multirow}
\usepackage[colorlinks=true,linkcolor=blue,citecolor=blue,urlcolor=blue]{hyperref}
\usepackage[utf8]{inputenc}
\usepackage[T1]{fontenc}
\usepackage{braket}

\newcommand{\tr}[1]{\textcolor{black}{#1}}

\begin{document}
	

\title{Topological Josephson parametric amplifier array: A proposal for directional, broadband, and low-noise amplification}

\author{Tom\'as Ramos}
\email{t.ramos@csic.es}
\affiliation{Institute of Fundamental Physics IFF-CSIC, Calle Serrano 113b, 28006 Madrid, Spain.}

\author{\'Alvaro G\'omez-Le\'on}
\affiliation{Institute of Fundamental Physics IFF-CSIC, Calle Serrano 113b, 28006 Madrid, Spain.}

\author{Juan Jos\'e Garc\'ia-Ripoll}
\affiliation{Institute of Fundamental Physics IFF-CSIC, Calle Serrano 113b, 28006 Madrid, Spain.}

\author{Alejandro Gonz\'alez-Tudela}
\affiliation{Institute of Fundamental Physics IFF-CSIC, Calle Serrano 113b, 28006 Madrid, Spain.}

\author{Diego Porras}
\email{diego.porras@csic.es}
\affiliation{Institute of Fundamental Physics IFF-CSIC, Calle Serrano 113b, 28006 Madrid, Spain.}


\begin{abstract}
Low-noise microwave amplifiers are crucial for detecting weak signals in fields such as quantum technology and radio astronomy. However, designing an ideal amplifier is challenging, as it must cover a wide frequency range, add minimal noise, and operate directionally—amplifying signals only in the observer’s direction while protecting the source from environmental interference. In this work, we demonstrate that an array of non-linearly coupled Josephson parametric amplifiers (JPAs) can collectively function as a directional, broadband quantum amplifier by harnessing topological effects. By applying a collective four-wave-mixing pump with inhomogeneous amplitudes and linearly increasing phase, we break time-reversal symmetry in the JPA array and stabilize a topological amplification regime where signals are exponentially amplified in one direction and exponentially suppressed in the opposite. We show that compact devices with few sites $N\sim 11-17$ can achieve exceptional performance, with gains exceeding 20 dB over a bandwidth ranging from hundreds of MHz to GHz, and reverse isolation suppressing backward noise by more than 30 dB across all frequencies. The device also operates near the quantum noise limit and provides topological protection against up to 15\% fabrication disorder, effectively suppressing gain ripples. The amplifier’s intrinsic directionality eliminates the need for external isolators, paving the way for fully on-chip, near-ideal superconducting pre-amplifiers.
\end{abstract}
    
\maketitle

\section{Introduction}

Low-noise amplification is essential for high-sensitivity applications, spanning from astronomical instrumentation \cite{smith_low_2013} to nano-mechanical sensing \cite{cleland_nanomechanical_2002}. In superconducting quantum technology \cite{blais_circuit_2021, Ripoll2022}, microwave signals carrying quantum information are extremely weak, and thus near-quantum-limited amplification is indispensable to enable reliable single-shot measurements \cite{vijay_observation_2011, walter_rapid_2017, dassonneville_fast_2020, pereira_parallel_2022}. The Josephson parametric amplifier (JPA) \cite{aumentado_superconducting_2020} has emerged as a paradigmatic solution to this challenge. These amplifiers are nonlinear resonators in which Kerr anharmonicity arises from Josephson junctions (JJs) \cite{castellanos-beltran_amplification_2008} or, more recently, from materials with high kinetic inductance \cite{ho_eom_wideband_2012,parker_degenerate_2022,Frasca2024}. By activating their Kerr nonlinearities through three- or four-wave mixing processes \cite{eichler_controlling_2014, roy_introduction_2016,parker_degenerate_2022}, JPAs achieve excellent amplification characteristics, with gains above 20 dB and near-quantum-limited noise performance, though typically within a moderate bandwidth on order $\sim 10$ MHz.

Traveling-wave parametric amplifiers (TWPAs) present a compelling alternative to JPAs by using similar non-linearities but distributed along a transmission line \cite{macklin_nearquantum-limited_2015, white_traveling_2015, winkel_nondegenerate_2020, planat_photonic-crystal_2020, ranzani_kinetic_2018, malnou_three-wave_2021,renberg_nilsson_high-gain_2023,kuznetsov_ultra-broadband_2024}. As non-resonant devices, TWPAs enable high-gain, low-noise amplification in transmission across significantly broader bandwidths, reaching up to the GHz range \cite{esposito_perspective_2021}. However, effective wave propagation along the nonlinear transmission line demands phase-matching, achieved through dispersion engineering \cite{macklin_nearquantum-limited_2015, planat_photonic-crystal_2020} or alternative methods \cite{bell_traveling-wave_2015, kow_self_2024}. This requirement inevitably introduces small impedance mismatches during fabrication, preventing TWPAs from being directional. As a result, parasitic signals and vacuum fluctuations can become back-amplified, potentially contaminating the quantum source. To mitigate these effects, TWPAs are typically used with external isolation components, such as isolators \cite{esposito_perspective_2021}.

Directional amplification without external elements has been achieved using few-mode resonant devices, which are inherently narrowband \cite{abdo_directional_2013, sliwa_reconfigurable_2015, lecocq_nonreciprocal_2017,metelmann_nonreciprocal_2015, fang_generalized_2017, pucher_atomic_2022}. Broadband amplification is crucial for implementing large-scale quantum information processors, where multiple signals need to be multiplexed and simultaneously detected \cite{heinsoo_rapid_2018}. Therefore, unifying the broadband properties of TWPAs with the compactness of a directional amplifier with built-in backward isolation is one of the holy grails in the field \cite{esposito_perspective_2021}.

In this work, we combine concepts of Josephson parametric amplification \cite{eichler_controlling_2014,roy_introduction_2016,esposito_perspective_2021} and topological photonics \cite{peano_topological_2016,porras_topological_2019,wanjura_topological_2020,ramos_topological_2021,ozawa_topological_2019} to demonstrate that an array of non-linearly coupled JPAs can collectively operate as high-performance broadband amplifier in transmission as TWPAs, while being intrinsically directional and resilient to disorder due its non-trivial topology. The key to building such a device is realizing topological amplification \cite{porras_topological_2019,wanjura_topological_2020,ramos_topological_2021} in the superconducting array. This requires non-linear cross-Kerr couplings to realize non-local squeezing interactions between adjacent JPAs $\sim \bar{g}_c\delta a_j^\dag \delta a_{j+1}^\dag$ in addition to standard local squeezing terms $\sim \bar{g}_s\delta a_j^\dag \delta a_j^\dag$. We also require to collectively drive all JPAs with a 4-wave mixing pump with phase gradient $\bar{\varphi}\neq 0$ such that we induce a hopping with complex phase between them $\sim \bar{J} e^{i\bar{\varphi}}\delta a^\dag_{j+1}\delta a_j$ and break time-reversal symmetry. We show that this collective pump can be implemented by $N$ independent microwave drives or by a single one distributed via an auxiliary waveguide. The pump amplitudes must smoothly increase from the boundaries toward the array's center to suppress non-linear boundary effects. Combining this inhomogeneous collective pump, local decay, and the squeezing and complex hopping interactions, the JPA array stabilizes a topologically amplifying steady state. Remarkably, in this configuration, microwave excitations propagating along the array are exponentially amplified in one direction and exponentially suppressed in the opposite. This directional amplification is moreover near-quantum-limited, broadband with bandwidth on the order of the effective hopping and decay rate, and is topologically protected against disorder in all system parameters, particularly imperfections inherent in the fabrication of JJs \cite{kreikebaum_improving_2020}.

Our design fundamentally differs from the previous topological amplification proposals \cite{peano_topological_2016} where the topology comes fully from the coherent dynamics and thus requires a 2D configuration whose boundary admits directionally propagating edge states. Our proposed setup is intrinsically 1D and the topology is associated with a driven-dissipative steady-state with directional current.

Beyond proposing the design of this novel topological JPA array, we also characterize its performance at various operation points and demonstrate the feasibility of implementing it with current superconducting circuit technology \cite{eichler_quantum-limited_2014,mutus_strong_2014,kounalakis_tuneable_2018}. We predict that compact devices with $N\sim 11-17$ sites can provide near quantum-limited amplification with more than 20 dB of gain and $30$ dB of reverse isolation over a bandwidth of $\sim 300$ MHz. This is more than an order of magnitude larger in bandwidth and directionality than previous resonant non-reciprocal amplifiers \cite{abdo_directional_2013,sliwa_reconfigurable_2015,metelmann_nonreciprocal_2015,lecocq_nonreciprocal_2017,fang_generalized_2017,pucher_atomic_2022} and is on the same order as TWPAs with isolation, which use frequency conversion and external diplexers to protect the source \cite{ranadive_traveling_2024,malnou_traveling-wave_2024}. Increasing the bandwidth further requires using non-linearities with higher dynamic range such as arrays of kinetic inductance parametric amplifiers \cite{ho_eom_wideband_2012,parker_degenerate_2022,Frasca2024,jouanny_band_2024}. Another route is implementing Quarton non-linearities instead of conventional JJs, for which we estimate that the amplification bandwidth can be increased up to $\sim 1 {\rm GHz}$, keeping all the high-performance properties mentioned above. 

Other remarkable properties of our proposed device are that the reverse isolation acts on all frequencies, not only on the amplification band, and the intrinsic topological protection to disorder and inhomogeneities allows for effective suppression of gain ripples. It also eliminates the need for dispersion engineering for phase-matching \cite{macklin_nearquantum-limited_2015,white_traveling_2015,planat_photonic-crystal_2020}, Floquet engineering  \cite{roushan_chiral_2017,Peng2022,carrasco_effect_2022} or external magnetic fields. Our work thus provides a promising route towards the scalable integration of intrinsically directional and broadband amplifiers and quantum processors on the same on-chip, avoiding external isolators.

The paper is structured as follows. In Sec.~\ref{sec:CircuitModel}, we provide the superconducting circuit model to realize the topological JPA array, detailing the configuration of Josephson junctions (JJs), capacitive couplings, transmission lines, and microwave pumps. In Sec.~\ref{Sec:4WM_tot}, we describe how four-wave mixing processes are implemented in the JPA array using a strong collective pump, including tapering techniques to suppress boundary effects and stabilize the operation of the topological amplification. In Sec.~\ref{auxiliaryarray}, we propose a scalable way of implementing the global pump with phase gradient by distributing a single pump through an auxiliary waveguide. In Sec.~\ref{sec:directional_amp_topology}, we characterize the emergence of topological amplification in the JPA array and discuss its main properties. In Sec.~\ref{Peformance_ExpParameters}, we quantify the broadband amplification performance that can be achieved with the JPA array for four state-of-the-art experimental parameter sets. In Sec.~\ref{topProtectionP1}, we demonstrate the topological protection against disorder in the device. Finally, in Sec.~\ref{conclusions}, we end the work with conclusions and outlook.

\section{Modelization of the topological Josephson parametric amplifier array}\label{sec:CircuitModel}

In this section, we describe the setup for realizing a topological JPA array. We first introduce the Lagrangian model for the required Josephson junction array (JJA) (see~Sec.~\ref{sec:setupCircuitMain}), which we quantize and obtain a Hamiltonian describing a non-linearly coupled Kerr-resonator array (see~Sec.~\ref{quantumH_setup}). We then include the coupling to transmission lines and coherent pumps on the JJA (see~Sec.~\ref{Sec:dissipation}), and finally, we comment on a method to control saturation in the JJA (see~Sec.~\ref{sec:saturation_main}).  

\begin{figure*}
    \centering
    \includegraphics[width=0.9\textwidth]{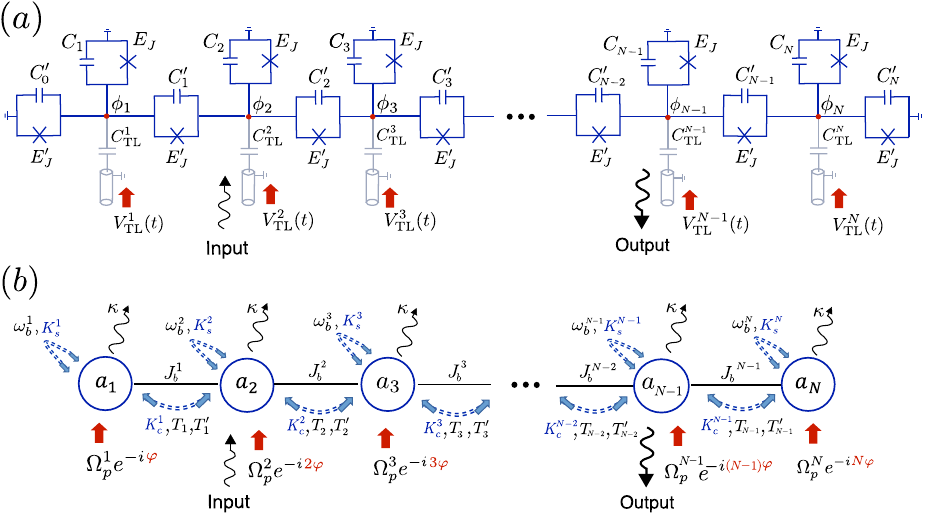}
    \caption{Realization of a topological Josephson parametric amplifier array (TJPAA). (a) Superconducting circuit scheme consisting of a Josephson junction array (JJA), where at each node $j=1,\dots,N$, a local capacitance $C_j$ and a local Josephson junction $E_J$ defines a non-linear oscillator mode. These local modes interact via linear capacitive couplings $C'_j$ and nonlinear inductive couplings $E'_J$. Finally, external transmission lines provide local dissipation and voltage pump $V_{\rm TL}^j(t)$ at each site $j$. (b) Quantum optics model implemented by the superconducting circuit in (a). The JJA behaves as an array of weakly non-linear modes $a_j$ of frequency $\omega_b^j$ and self-Kerr anharmonicity $K_s^j$, coupled linearly via hopping $J_b^j$ and non-linearly via cross-Kerr interactions $K_c^j$, $T_j,$ and $T_j'$. The transmission lines provide local dissipation $\kappa$ to each site as well as inhomogeneous coherent pumps with amplitude $\Omega_p^j$ and spatially increasing phase $\sim j\varphi$. These pumps are used to perform four-wave-mixing, breaking time-reversal symmetry and inducing a topological parametric amplifying phase in the array (see text).}
    \label{Fig:Setup1}
\end{figure*}
 
\subsection{Superconducting circuit design}\label{sec:setupCircuitMain}

The topological Josephson parametric amplifier array we propose can be realized with the superconducting circuit shown in Fig.~\ref{Fig:Setup1}(a). At each node $j=1,\dots, N$, we define independent flux variables $\phi_j(t)$, which locally behave as weakly anharmonic transmons or non-linear microwave oscillators \cite{blais_circuit_2021} consisting of an on-site JJ with Josephson energy $E_{J}$ coupled in parallel to an on-site capacitance $C_j$, and connected to ground. In addition, each of these non-linear oscillators couples inductively to its neighbors via an inter-site JJ with Josephson energy $E'_J$ and capacitively via an inter-site capacitance $C'_j$. The Lagrangian describing the dynamics of the resulting Josephson junction array (JJA) reads,
\begin{align}
     {\cal L}_{\rm JJA}= {}&\sum_{j=1}^N \frac{C_j}{2}(\dot{\phi}_j)^2 +\sum_{j=0}^{N}\frac{C'_j}{2}(\dot{\phi}_{j+1}-\dot{\phi}_j)^2\label{LagJJ}\\
    +{}&\sum_{j=1}^N E_J\cos\left(\frac{\phi_j}{\Phi_0}\right)+\sum_{j=0}^{N}E'_J\cos\left(\frac{\phi_{j+1}-\phi_j}{\Phi_0}\right)\nonumber,
\end{align}
where $\Phi_0=\hbar/(2e)$ is the flux quantum with $\hbar$ the reduced Planck's constant and $e$ the electron's charge (constants of nature). Note that we include inter-site Josephson energies $E'_J$ and capacitances $C'_0$ and $C'_N$ at both ends of the JJA, coupling to $\phi_1$ and $\phi_N$. This does not add extra dynamical variables as these ends are also coupled to the ground, thus having $\phi_0 = \phi_{N+1}=0$ as boundary conditions. 

To provide local dissipation and externally pump each non-linear oscillator, we couple each node $j=1,\dots, N$ to an independent transmission line via a capacitance $C_{\rm TL}^j$ [See Fig.~\ref{Fig:Setup1}(a)]. On each of these lines, the oscillators interact with an external voltage $V_{\rm TL}^j(t)$, so that the dynamics is described by the Lagrangian ${\cal L}_{\rm TL}$,
\begin{align}
     {\cal L}_{\rm TL}= {}&\sum_{j=1}^N \frac{C_{\rm TL}^j}{2}(\dot{\phi}_j-V_{\rm TL}^j(t))^2.\label{Lagp}
\end{align}
For simplicity, we assume here that the $N$ voltages $V_p^j(t)$ are generated by independent microwave sources at each line [see Fig.~\ref{Fig:Setup1}]. However, in Sec.~\ref{auxiliaryarray}, we also provide a more sophisticated design where $V_{\rm TL}^j(t)$ are generated by distributing a single voltage source via an auxiliary array [see Fig.~\ref{Fig:setup_waveguide}]. Both alternatives are feasible for sizes $N\lesssim 17$ that we require in this work, but the auxiliary array approach is more scalable. 

Finally, we also use two of the pump transmission lines to send and retrieve amplified signals at the fixed sites denoted by $j=I$ and $j=O$, respectively. As shown in Fig.~\ref{Fig:Setup1}(a), we typically send the signal at site $I=2$ and retrieve it at site $O=N-1$. The description of the input signal is given in Sec.~\ref{Sec:dissipation} within the input-output formalism. 

From the total Lagrangian ${\cal L} = {\cal L}_{\rm JJA} + {\cal L}_{\rm TL}$, we derive the total Hamiltonian of the circuit $H=H_{\rm JJA}+H_{\rm TL}$, whose parts for JJA and transmission lines read (see Appendix~\ref{TotalH}):
\begin{align}
    H_{\rm JJA}={}&\sum_{j=1}^N\frac{1}{2C^j_{\rm eq}}q_j^2+\sum_{j=1}^N\frac{1}{2L_{\rm eq}}\phi_j^2\label{Hexpand}\\
    +{}&\sum_{j=1}^{N-1}\frac{C'_j}{C^j_{\rm eq}C^{j+1}_{\rm eq}}q_jq_{j+1}-\sum_{j=1}^{N-1}\frac{1}{L'_{J}}\phi_j\phi_{j+1}\nonumber\\
    -{}&\sum_{j=1}^N \frac{E_J}{24(\Phi_0)^4}\phi_j^4-\sum_{j=0}^{N}\frac{E'_{J}}{24(\Phi_0)^4}(\phi_{j+1}-\phi_j)^4+{\cal O}^6\nonumber,\\
    H_{\rm TL}={}&\sum_{j=1}^N\frac{C^j_{\rm TL}}{C_{\rm eq}^j}q_j V_{\rm TL}^j(t).\label{TRHam}
\end{align}
At each site $j=1,\dots, N$, we define the charge variable $q_j=\partial {\cal L}/\partial \dot{\phi}_j$, as well as the equivalent capacitance $C_{\rm eq}^j$ and inductance $L_{\rm eq}$ as,
\begin{align}
    C_{\rm eq}^j={}&C_j+C^j_{\rm TL}+C'_{j-1}+C'_j,\label{Ceq}\\
    L_{\rm eq} ={}&(L_J L'_J)/(2L_J+L'_J),\label{Leq}
\end{align}
with $L_J=(\Phi_0)^2/E_J$ and $L'_J=(\Phi_0)^2/E'_J$ the on-site and inter-site Josephson inductances, respectively. Notice also that $H_{\rm JJA}$ in Eq.~(\ref{Hexpand}) neglects capacitive couplings beyond nearest neighbors provided $C'_j,C_{\rm TL}^j\ll C_{\rm eq}^j$. We also assume the low phase drop condition, $|\phi_j|/\Phi_0, |\phi_{j+1}-\phi_j|/\Phi_0 \ll 1$ to expand the cosine potentials of the JJs and obtain quartic Kerr-like terms. For more details, see Appendix~\ref{TotalH}.

\subsection{Quantum description of the Josephson junction array}\label{quantumH_setup}

We now quantize the circuit by considering canonical commutation relations between flux and charge variables, $[\phi_j,q_l]=i\hbar \delta_{jl}$, for $j,l=1,\dots, N$, and expressing them in terms of creation $a_j^\dag$ and annihilation $a_j$ operators as
\begin{align}
    \phi_j={}&\sqrt{\frac{\hbar}{2\omega_b^jC_{\rm eq}^j}}(a_j^\dag+a_j),\quad q_j = i\sqrt{\frac{\hbar \omega_b^jC_{\rm eq}^j}{2}}(a_j^\dag-a_j).\label{phia}
\end{align}
Following the standard procedure, each site $j$ of the JJA can be interpreted as a local microwave oscillator of bare frequency,   
\begin{align}
\omega_b^j = 1/\sqrt{L_{\rm eq} C_{\rm eq}^j},\label{omegabFull}
\end{align}
which can be populated by a quantized number of microwave excitations \cite{blais_circuit_2021}. Replacing Eqs.~(\ref{phia}) into (\ref{Hexpand}), we find the quantum Hamiltonian of the JJA. This behaves as an array of $N$ non-linear Kerr resonator modes $a_j$ with anharmonicity $K_s^j$ that are further coupled to the nearest-neighbor sites $a_{j-1}$ and $a_{j+1}$ via linear hopping $J_b^j$ and non-linear cross-Kerr couplings $K_c^j$, $T_j$, and $T_j'$ [see Fig.~\ref{Fig:Setup1}(b)]. In the rotating wave approximation (RWA), provided couplings are much smaller than bare frequencies $\omega_b^j$, the JJA Hamiltonian takes the form (see Appendix \ref{TotalH}):
\begin{align}
    \frac{H_{\rm JJA}}{\hbar} ={}& \sum_{j=1}^N \omega_b^j a_j^\dag a_j -\sum_{j=1}^N\frac{K_s^j}{2}a_j^\dag a_j^\dag a_j a_j\label{H2}\\
    +{}&\sum_{j=1}^{N-1} J_b^j(a_{j+1}^\dag a_{j}+{\rm h.c.})-\sum_{j=1}^{N-1}K_c^j a_{j+1}^\dag a_{j+1} a_{j}^\dag a_{j}\nonumber\\
    -{}&\sum_{j=1}^{N-1}\frac{K^j_c}{4}(a_{j+1}^\dag a_{j+1}^\dag a_j a_j +{\rm h.c.})\nonumber\\
    +{}&\sum_{j=1}^{N-1}\left[\frac{T_j}{2} a_{j+1}^\dag a_{j+1}^\dag a_j a_{j+1} +\frac{T_j'}{2}a_{j}^\dag a_{j+1}^\dag a_{j} a_{j} +{\rm h.c.}\right]\nonumber.
\end{align}
Self-Kerr non-linearities are originated from intra- $E_{J}$ and inter-site $E'_{J}$ Josephson energies and read,
\begin{align}
    K_s^j = {}&\hbar[(E_{J}+2E'_J)/8(\Phi_0)^4](Z_j)^2,\label{Ks_full}
\end{align}
where $Z_j = \sqrt{L_{\rm eq}/C_{\rm eq}^j}$ is the impedance of the JJA at site $j$. Linear couplings between adjacent sites are given by the bare hopping rates,
\begin{align}
J_b^j ={}& \frac{1}{2}\sqrt{\omega_b^j\omega_b^{j+1}}\left[C'_{j}/\sqrt{C_{\rm eq}^j C_{\rm eq}^{j+1}}-L_{\rm eq}/L'_J\right],\label{bareHopping}
\end{align}
which contain a capacitive (positive) part originated by the inter-site capacitance $C'_j$, as well as an inductive (negative) part coming from the linear part of Josephson inductance $L_J'$. In addition, the quartic contribution of the inter-site Josephson energies $E_J'$ gives rise to cross-Kerr non-linear couplings between nearest neighbor sites with rate,
\begin{align}
    K_c^j = {}&\hbar( E'_{J}/4(\Phi_0)^4)Z_jZ_{j+1},\label{fullKc}
\end{align}
as well as non-linear photon number-dependent hopping with rates closely related to the cross-Kerr coupling as $T_j = K_c^j \sqrt{Z_{j+1}/Z_j}$ and $T_j' = K_c^j \sqrt{Z_{j}/Z_{j+1}}$. There are also small renormalizations of bare frequencies and hopping due to the non-linear couplings, given by $\omega^j_b \rightarrow \omega^j_b - K^j_s - (K^j_c + K^{j-1}_c)/2$ and $J_b^j\rightarrow J_b^j+ (T_j+T_j')/2$. See Appendix \ref{TotalH} for more details.

\subsection{Inhomogeneous pumps and local dissipation}\label{Sec:dissipation}

To drive and control each local mode $a_j$ of the JJA, we send a specific sinusoidal voltage pulse $V_{\rm TL}^j(t)$ of frequency $\omega_p$ on each transmission line [see Fig.~\ref{Fig:Setup1}(a)-(b)]. As explained below, the key to stabilizing a topological amplifying phase is to allow these pumps to have inhomogeneous amplitudes $A_p^j$ and phases $\theta_p^j$. Therefore, we consider external voltage operators of the form, 
\begin{align}
V_{\rm TL}^j(t)=A^j_p\cos(\omega_p t + \theta_p^j)+\delta V_{\rm TL}^j, \quad j=1,\dots,N,\label{Vpj}
\end{align}
where $\delta V_{\rm TL}^j$ is the operator describing quantum fluctuations of voltage at each line. The voltage amplitudes $A_p^j$ are related to the average source power $P_p^j$ as $A_p^j=\sqrt{2Z_{\rm TL}P_p^j}$, where $Z_{\rm TL}$ is the impedance of the transmission lines. Unless specified, we typically consider $Z_{\rm TL}=50\Omega$.

On the one hand, the classical part of the voltage (\ref{Vpj}) leads to a Hamiltonian $H_p$ describing the coherent pump on each site $j$ of the JJA. In a frame rotating with the pump frequency $\omega_p$ and considering the RWA, this Hamiltonian reads (see Appendix \ref{FulldissipationModel})
\begin{align}
    \frac{H_p}{\hbar} =  i\sum_{j=1}^N\Omega_p^j (e^{-i\theta^j_p}a_j^\dag-e^{i\theta_p^j}a_j)-\omega_p\sum_{j=1}^N a_j^\dag a_j,\label{HpRWA}
\end{align}
where the strengths $\Omega_p^j$ of the coherent drives are related to the coupling capacitance $C_{\rm TL}^j$ and the local power $P_p^j$ as
\begin{align}
\Omega_p^j={}&(C_{\rm TL}^j/2C_{\rm eq}^j)\sqrt{(Z_{\rm TL}/Z_j)(P_p^j/\hbar)}.\label{Omegap_Circuit}
\end{align}

On the other hand, the quantum fluctuations of the voltage $\delta V_{\rm TL}^j$ in Eq.~(\ref{Vpj}) lead to local dissipation $\kappa$ at each site of the JJA, which we describe in a fully quantum description via the input-output formalism \cite{yurke_quantum_1984,QuantumNoise,ramos_topological_2021}, see Appendix~\ref{FulldissipationModel} for details. In this framework, the rate $\kappa$ of microwave excitations decaying to each transmission line is given by,
\begin{align}
    \kappa={}&(C_{\rm TL}^j/2C_{\rm eq}^j)^2(Z_{\rm TL}/L_{\rm eq}).\label{kappap}
\end{align}
Note that dissipation scales with the coupling capacitance as $(C_{\rm TL}^j/C_{\rm eq}^j)^2$ but, by design, we choose this ratio constant so that all local decay rates $\kappa$ are homogenous along the JJA. Taking into account dissipation and coherent pumps, the dynamics of the JJA are governed by quantum Langevin equations for the Heisenberg operators $a_j(t)$: 
\begin{align}
    \dot{a}_j={}&\frac{i}{\hbar}[H_{\rm JJA}+H_p,a_j]-\frac{\kappa}{2}a_j+\sqrt{\kappa}a_{\rm in}^j(t).\label{QLE1disp}
\end{align}
Here, the coherent dynamics is described by the commutator with the Hamiltonian $H_{\rm JJA}+H_p$, whereas dissipation is described by input noise operators, $a_{\rm in}^j(t)$ and $[a_{\rm in}^j(t)]^\dag$, which destroy and create photons at transmission line $j$ at the above rate $\kappa$. 

The input operators $a_{\rm in}^j(t)$ also describe the input field of microwave excitations entering the JJA via the transmission line $j$. In particular, for a monochromatic coherent signal of amplitude $\alpha_{\rm sig}$ and frequency $\omega_s$ entering at site $j=I$, the input operator at a frame rotating with $\omega_p$ takes the form,
\begin{align}
    a^j_{\rm in}(t) = \delta_{jI} \alpha_{\rm sig} e^{-i(\omega_s-\omega_p) t} + \delta a_{\rm in}^j(t).\label{inputsignal}
\end{align}
Here, $|\alpha_{\rm sig}|^2=P_s/(\hbar \omega_s)$ corresponds to the photon flux generated by the voltage source of power $P_s$ and $\delta a_{\rm in}^j(t)$ describes quantum vacuum noise at the input of each site $j$ (see also Appendix~\ref{FulldissipationModel}). 

Finally, the output fields $a_{\rm out}^j(t)$ and $[a_{\rm out}^j(t)]^\dag$ describe the microwave excitations leaving the JJA via the transmission line at site $j$. These can be accessed by solving the driven-dissipative dynamics in Eqs.~(\ref{QLE1disp}) with initial condition (\ref{inputsignal}) and then using the input-output relations,
\begin{align}
    a^j_{\rm out}(t) ={}& a^j_{\rm in}(t) - \sqrt{\kappa} a_j(t),\quad j=1,\dots, N.\label{dispInOut1}
\end{align}
Below, solve these equations to characterize the topological amplification performance of the device.

\subsection{Control of saturation of the Josephson junction array}\label{sec:saturation_main}

Modeling the JJA as non-linearly coupled Kerr resonators [see Fig.~\ref{Fig:Setup1}(b)] relies on the low phase drop approximation, $\langle \phi_j^2\rangle/\Phi_0^2\ll 1$ and $\langle (\phi_{j+1}-\phi_{j})^2\rangle/\Phi_0^2\ll 1$, which allows us to approximate the cosine potentials of JJs by quartic terms, leading to Hamiltonians (\ref{TRHam}) and (\ref{H2}). As shown in Ref.~\cite{eichler_controlling_2014}, this approximation demands that the mean number of excitations on each site is bounded as $\langle a_j^\dag a_j\rangle\ll (\Phi_0/2\phi_{\rm zpf}^j)^2$, with $\phi_{\rm zpf}^j=[\hbar/(2C_{\rm eq}^j\omega_b^j)]^{1/2}$ the zero-point fluctuation of the JJA. When the array is populated close to or beyond the above limit, higher-order non-linear effects take place in the device, leading to saturation and limiting the dynamic range of the amplifier. 

Nevertheless, it is possible to reduce the saturation by employing standard techniques already experimentally demonstrated for JPAs \cite{eichler_controlling_2014,eichler_quantum-limited_2014,planat_understanding_2019}. In particular, if we replace each JJ in the JJA by a sub-array of $M$ JJs in series with $M$ larger Josephson energies, i.e.~$E_J,E_J'\rightarrow ME_J,ME_J'$, all linear properties of the circuit remain the same, but the phase drop on each physical JJ reduces by a factor $M$. As a result, Kerr non-linearities effectively reduce as $K_c^j,K_s^j\rightarrow K_c^j/M^2,K_s^j/M^2$, allowing the device to sustain $M^2$ times more photons before saturating. As shown in Appendix~\ref{ControlSaturation}, the upper bound on photon occupation then reads: 
\begin{align}
\langle a_j^\dag a_j\rangle\ll M^2(\Phi_0/2\phi_{\rm zpf}^j)^2=M^2(\hbar\omega_b^jC_{\rm eq}^j)/(8e^2). \label{upperbound} 
\end{align}

We remark that using sub-arrays of JJs as non-linearities is optional in the design we propose, and below we consider cases with and without it. However, for reaching the best amplifier's performance, we show that it is desirable to implement sub-arrays with $M\sim 15-30$ JJs, as already demonstrated for two coupled JPAs \cite{eichler_controlling_2014}. 

\section{Parametric amplification and artificial gauge fields via four-wave mixing}\label{Sec:4WM_tot}

In this section, we generalize the theory of four-wave mixing processes to arrays of non-linearly coupled Kerr resonators subject to a strong pump on all sites. In Sec.~\ref{Sec:4wm}, we first derive the effective linearized dynamics of the array, showing that it behaves as a parametric amplifier lattice with complex hopping terms controlled by the pump's phases. Subsequently, in Sec.~\ref{Sec:Stabilization_inhomogeneities}, we show how to stabilize a topological amplifying phase by engineering inhomogeneous parameters on the boundaries such as the pump amplitudes and resonator frequencies.

\subsection{Linearized dynamics for quantum fluctuations}\label{Sec:4wm}

Four-wave mixing processes are well understood in non-linear quantum optics \cite{WallsMilburnBook}. By applying a strong pump on a non-linear medium, two coherent photons coming from the pump at frequency $\omega_p$ are converted into signal and idler photons of frequencies $\omega_s$ and $2\omega_p-\omega_s$, respectively. In the context of JPAs \cite{aumentado_superconducting_2020} or TWPAs \cite{esposito_perspective_2021}, these processes generate squeezing and parametric amplification of both idler and signal fields. In the following, we show that the topological amplifier array also employs squeezing terms to generate parametric amplification, but it also requires breaking time-reversal symmetry via the pump's phases to render this amplification directional and topologically protected.

To perform four-wave mixing on each mode $a_j$ of the JJA, we consider a strong coherent pump on all sites, namely $\Omega_p^j\gg \kappa,|\Delta_b^j|$, with $\Delta_b^j=\omega_p-\omega_b^j$, the detuning between the pump and the bare resonator frequency at site $j$. In this limit, each mode is well described via a displacement transformation,
\begin{align}
    a_j= \alpha_j + \delta a_j,\label{displacement}
\end{align}
where the coherent component $\alpha_j$ is highly populated compared to the small quantum fluctuation around it, i.e. $\langle \delta a_j^\dag \delta a_j\rangle \ll |\alpha_j|^2$. Using the decomposition (\ref{displacement}) in Eq.~(\ref{QLE1disp}), we can derive a system of classical non-linear equations for $\alpha_j(t)$, as well as quantum Langevin equations for the quantum fluctuations $\delta a_j(t)$, describing an effective linear dynamics. 

\begin{figure*}
    \centering
    \includegraphics[width=0.8\textwidth]{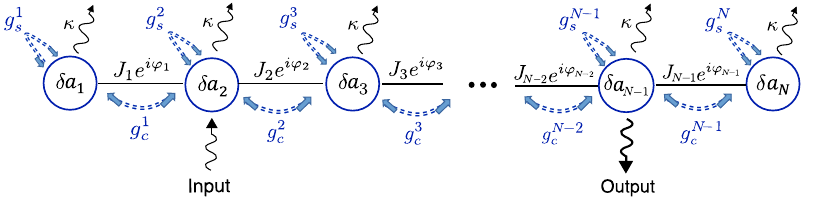}
    \caption{Effective parametric amplifier lattice model describing the linearized dynamics of the quantum fluctuations $\delta a_j$ of the JJA. The combination of complex hopping $J_je^{i\varphi_j}$, local decay $\kappa$, and parametric local and non-local pumping terms $g_s^j$ and $g_c^j$, leads the system to a topological steady-state phase, where external signals entering at site $j=I=2$ are directionally amplified and leave at the right channel $j=O=N-1$.}
    \label{Fig_Setup_4WM}
\end{figure*}

The classical coupled equations for the mean displacements $\alpha_j$ take the form,
\begin{align}
    \dot{\alpha}_j ={}& -\left(\frac{\kappa}{2}-i\Delta_b^j\right)\alpha_j +iK^j_{s}|\alpha_j|^2\alpha_j +\Omega_p^je^{-i\theta^j_p},\label{classicalTime}\\
    {}&-i\left(J_b^j\alpha_{j+1}+J_b^{j-1}\alpha_{j-1}\right)+ F_{c}^j(\alpha_j,\alpha_{j\pm 1})\nonumber,
\end{align}
with boundary conditions $\alpha_0(t)=\alpha_{N+1}(t)=0$. The first line in Eqs.~(\ref{classicalTime}) describes Duffing oscillator equations for each displacement $\alpha_j$, with local decay, detuning, self-Kerr non-linearity, and pump. On top of this, the second line includes linear and non-linear couplings between the modes, coming from bare hoppings $J_b^j$ and cross-Kerr non-linearities $F_{c}^j(\alpha_j,\alpha_{j\pm 1})$, which explicitly read: 
\begin{align}
    F_{\rm c}^j(\alpha_j,\alpha_{j\pm 1}){}&=iK^j_{c}|\alpha_{j+1}|^2\alpha_j+iK^{j-1}_{c}|\alpha_{j-1}|^2\alpha_j\nonumber\\
    {}&-i(T_j'|\alpha_{j}|^2+\frac{T_j}{2}|\alpha_{j+1}|^2)\alpha_{j+1}\nonumber\\
    {}&-i(T_{j-1}|\alpha_{j}|^2+\frac{T_{j-1}'}{2}|\alpha_{j-1}|^2)\alpha_{j-1}\nonumber\\
    {}&+i\frac{K^j_{c}}{2}\alpha_j^\ast\alpha_{j+1}^2+i\frac{K^{j-1}_{c}}{2}\alpha_j^\ast\alpha_{j-1}^2\nonumber\\
    {}&-i\frac{T_{j-1}}{2}\alpha_{j-1}^\ast\alpha_j^2-i\frac{T_j'}{2}\alpha_{j+1}^\ast\alpha_j^2.\label{classicalTimef}
\end{align}
The system of equations (\ref{classicalTime}) can be numerically integrated in time via standard numerical methods such as Runge-Kutta. The steady-state configuration corresponds to a solution $\alpha_{\rm ss}^j=\alpha_j(t\rightarrow\infty)$ at which the mean displacement converges at long times. For convenience, we decompose the steady-state displacements in modulus and phase, namely 
\begin{align}
\alpha_{\rm ss}^j=|\alpha_{\rm ss}^j|e^{-i\theta_{\rm ss}^j}.\label{SSansatz} 
\end{align}
Importantly, not all steady-state solutions are compatible with topological amplification, and in the next subsection, we provide a method to induce suitable profiles of steady-state amplitudes $|\alpha^j_{\rm ss}|$ and phases $\theta_{\rm ss}^j$ to ensure proper stabilization.

Around the classical mean-field solution, the quantum fluctuations $\delta a_j(t)$ undergo driven-dissipative dynamics given by the quantum Langevin equations (\ref{QLE1disp}). Provided $\langle \delta a_j^\dag \delta a_j\rangle/|\alpha^j_{\rm ss}|^2\ll 1$, we linearize these equations, obtaining
\begin{align}
    \delta\dot{a}_j={}&\frac{i}{\hbar}\left[H_{\rm pa},\delta a_j\right]-\frac{\kappa}{2}\delta a_j+\sqrt{\kappa} a_{\rm in}^j(t).\label{QLE2}
\end{align}
The coherent part of the effective dynamics is governed by a quadratic bosonic lattice Hamiltonian $H_{\rm pa}$ that reads
\begin{align}
    \frac{H_{\rm pa}}{\hbar} = {}&-\sum_{j=1}^N\Delta_j \delta a_j^\dag \delta a_j -\sum_{j=1}^{N} \left(\frac{g^j_{s}}{2}\delta a_{j}^\dag \delta a_j^\dag + {\rm h.c.}\right) \label{effH}\\ 
    {}&+  \sum_{j=1}^{N-1}J_je^{-i\varphi_j}(\delta a_{j+1}^\dag \delta a_{j} + {\rm h.c.} )\nonumber\\
    {}&-\sum_{j=1}^{N-1} \left(g^j_{c}\delta a_{j+1}^\dag \delta a_j^\dag + {\rm h.c.}\right) + {\cal O}(|\alpha_j|\delta a_j^3).\nonumber
\end{align}
Eqs.~(\ref{QLE2})-(\ref{effH}) are expressed in terms of the re-defined quantum fluctuation operators, $\delta a_j\rightarrow e^{i\theta_{\rm ss}^j}\delta a_j$, and quantum noise operators $a_{\rm in}^j(t)\rightarrow e^{i\theta_{\rm ss}^j}a_{\rm in}^j(t)$, so that only the phase differences between neighboring sites,
\begin{align}
\varphi_j ={}& \theta^{j+1}_{\rm ss}-\theta^{j}_{\rm ss}, \qquad j\in [1,N-1],
\end{align}
appear explicitly in all equations. These phase differences $\varphi_j\neq 0$ play the role of an artificial gauge field in the lattice \cite{Koch2010,roushan_chiral_2017} and determine the phase of the complex hopping $J_je^{-i\varphi_j}$ in Eq.~(\ref{effH}). This effective Hamiltonian $H_{\rm pa}$ also contains local and non-local parametric squeezing terms $g^j_{s}$ and $g^j_{c}$, as well as effective detunings $\Delta_j$. All these coefficients of the effective linearized model depend on the JJA parameters as well as on the steady-state amplitudes $|\alpha_{\rm ss}^j|$ and phase differences $\varphi_j$. Explicitly, they are given by
\begin{align}
\Delta_j ={}& \Delta^j_b + 2 K^j_{s}|\alpha^j_{\rm ss}|^2 + K^j_{c}|\alpha_{\rm ss}^{j+1}|^2 + K^{j-1}_{c}|\alpha_{\rm ss}^{j-1}|^2 \nonumber\\
-{}&2|\alpha_{\rm ss}^j|\left[T_{j-1}|\alpha_{\rm ss}^{j-1}|\cos(\varphi_{j-1})+T_{j}'|\alpha_{\rm ss}^{j+1}|\cos(\varphi_{j})\right],\label{EffectiveDet}\\
J_j = {}& J^j_b - 2 K^j_c |\alpha_{\rm ss}^j| |\alpha_{\rm ss}^{j+1}| \cos(\varphi_{j})+T_j|\alpha_{\rm ss}^{j+1}|^2+T_j'|\alpha_{\rm ss}^{j}|^2,\\
g^j_{c} = {}& K^j_{c}|\alpha_{\rm ss}^{j+1}||\alpha_{\rm ss}^{j}|-\frac{T_j}{2}|\alpha_{\rm ss}^{j+1}|^2 e^{-i\varphi_{j}}-\frac{T_j'}{2}|\alpha_{\rm ss}^{j}|^2 e^{i\varphi_{j}},\label{effectivegc}\\
g^j_{s} = {}& K^j_{s}|\alpha_{\rm ss}^{j}|^2+\frac{K^j_{c}}{2}|\alpha_{\rm ss}^{j+1}|^2e^{-2i\varphi_{j}}+\frac{K^{j-1}_{c}}{2}|\alpha_{\rm ss}^{j-1}|^2e^{2i\varphi_{j-1}}\nonumber\\
-{}&T_{j-1}|\alpha_{\rm ss}^j||\alpha_{\rm ss}^{j-1}|e^{i\varphi_{j-1}}-T_{j}'|\alpha_{\rm ss}^j||\alpha_{\rm ss}^{j+1}|e^{-i\varphi_{j}},\label{effectivegs}
\end{align}
with $|\alpha_{\rm ss}^0|=|\alpha^{N+1}_{\rm ss}|=0$ as boundary conditions.

Figure~\ref{Fig_Setup_4WM} summarizes the physical processes involved in the effective parametric amplifier array model for $\delta a_j(t)$. Here, the squeezing interactions $\sim g_s^j$ and $\sim g_c^j$ are responsible for generating amplification of microwave signals propagating along the JJA, which in combination with complex hopping $\sim J_j e^{-i\varphi_j}$ and local dissipation $\sim \kappa $ leads to directional amplification when the topological parameter regime is met \cite{ramos_topological_2021,GomezLeon2022}. Importantly, this requires having a non-zero phase $\varphi_j\neq 0$ to break time-reversal symmetry and that all effective quantities are homogeneous or quasi-homogeneous along the array. Below, we identify the parameter regimes and the experimental conditions under which the JJA can manifest topological amplification of microwave signals.

\subsection{Stabilization of quasi-homogeneous steady-state via inhomogeneous parameters on boundaries}\label{Sec:Stabilization_inhomogeneities} 

In this subsection, we provide a method to stabilize a steady state configuration of mean-field displacements $\alpha_{\rm ss}^j=\alpha_j(t\rightarrow\infty)$ that leads to quasi-homogeneous profiles of all effective parameters in Eqs.~(\ref{EffectiveDet})-(\ref{effectivegs}) and is compatible with topological amplification.

First, we note that in a JJA with open boundaries such as in Figs.~\ref{Fig:Setup1}-\ref{Fig_Setup_4WM}, the sites $j=1$ and $j=N$ at the boundaries have only one neighbor while all the rest have two. Inspecting the non-linear equations (\ref{classicalTime}), this implies that even for a fully homogeneous array, the mean-field dynamics will unavoidably have asymmetric non-linear shifts $\sim (K_c^j |\alpha_{j+1}|^2 + K_c^{j-1} |\alpha_{j-1}|^2)\alpha_j$ at the boundaries. One may naively expect that this effect is not important for a large array. However, four-wave mixing requires large displacements $|\alpha_{\rm ss}^j|^2\gg 1$, such that the JJA will be deep in the non-linear regime, where non-linear shifts can not be neglected. Consequently, if boundary effects are not properly suppressed, the resulting steady-state configuration can be highly inhomogeneous or there may be no steady-state, forbidding topological amplification in either case.

\begin{figure}
\centering
\includegraphics[width=1\columnwidth]{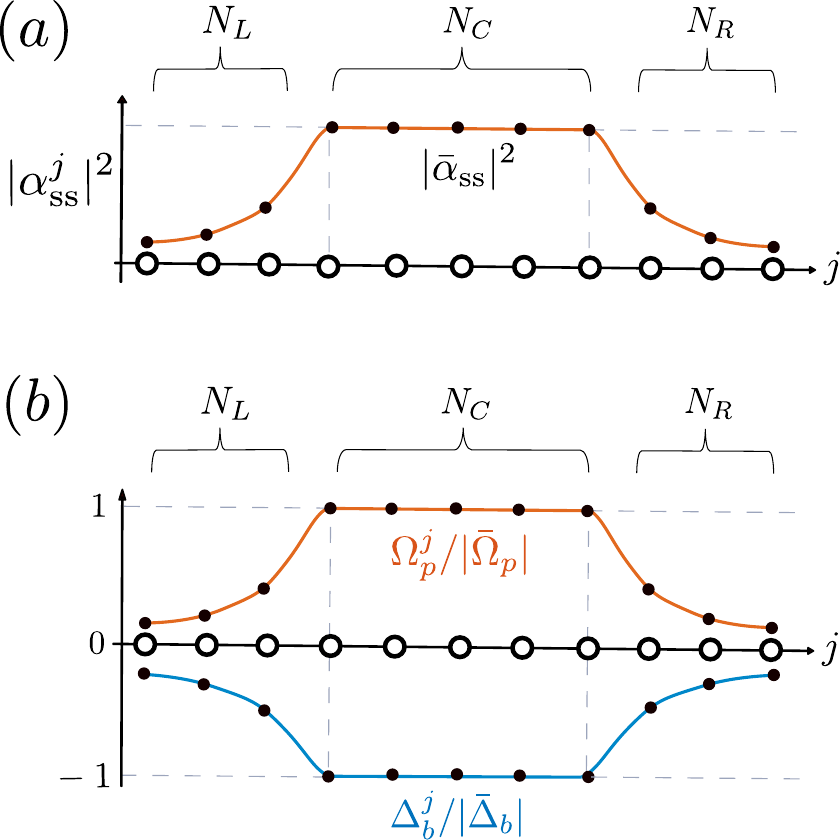}
\caption{Engineering boundary conditions to stabilize a steady-state configuration of mean-field displacements $|\alpha_{\rm ss}^j|^2$ compatible with topological amplification. (a) Separation of JJA in three parts: The left buffer region of $N_L$ sites, the central region of $N_C$ sites, and the right buffer region of $N_R$ sites. At the buffer regions, $|\alpha_{\rm ss}^j|^2$ slowly increase from boundaries towards the center to suppress non-linear boundary effects. At the central region, $|\alpha_{\rm ss}^j|^2$ are quasi-homogenous (up to imperfections). (b) Inhomogeneous profiles of pump strengths $\Omega_p^j$ (red) and bare detunings $\Delta_b^j$ (blue) engineered to stabilize the steady state profile in (a).}
\label{Fig_boundary_conditions}
\end{figure}

Our aim here is to engineer inhomogeneous parameters of the JJA to stabilize a quasi-homogeneous steady-state configuration as depicted in Fig.~\ref{Fig_boundary_conditions}(a), where mean-field displacements are strongly suppressed at the boundaries and they grow slowly towards a homogeneous region at the center. To do so, we divide the $N$ sites of the array into three regions: (i) the left buffer region with $N_L$ sites (ii) the central region with $N_C$ sites, and (iii) the right buffer region of $N_R$ sites, such that $N=N_L+N_C+N_R$ (see Fig.~\ref{Fig_boundary_conditions}). 

At the buffer regions, the key is to smoothly increase the pump amplitudes $\Omega_p^j$ from the boundaries towards a constant value $\bar{\Omega}_p$ at the central region.  We model this tapering as
\begin{align}
    \Omega_p^j = f_j \bar{\Omega}_p,
\end{align}
where $f_j$ is a smooth function that increases from the boundaries to the center up to value $1$. In particular, we consider
\begin{align}
f_j={}&
        \begin{cases}
        \sin^2( \frac{\pi}{2}\frac{j}{(N_L+1)}) & \text{if } 1\leq j\leq N_L\\
        1 & \text{if } N_L+1\leq j \leq N-N_R\\
        \sin^2( \frac{\pi}{2}\frac{(N+1-j)}{(N_R+1)}) & \text{if } N-N_R+1\leq j\leq N\label{tapering}
    \end{cases}
\end{align}
so that the pump strength looks like the red curve in Fig.~\ref{Fig_boundary_conditions}(b).

\begin{figure*}[t]
\centering
\includegraphics[width=\textwidth]{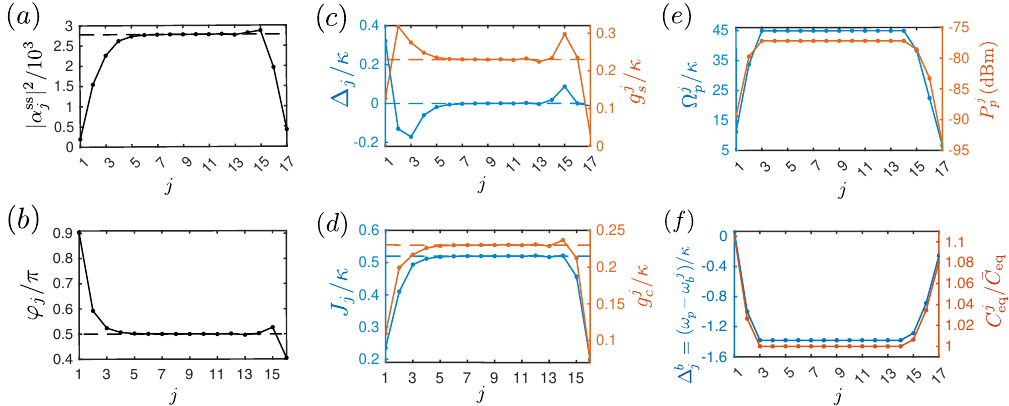}
\caption{Stable quasi-homogeneous configuration of mean-field displacements $|\alpha_{\rm ss}^j|^2$ leading to topological amplification in the JJA. (a) Inhomogenous profile of mean-field displacements $|\alpha_{\rm ss}^j|^2$. The dashed line corresponds to the homogeneous estimation at the center, $|\alpha_{\rm ss}^j|^2=|\bar{\alpha}_{\rm ss}|^2\gg 1$. (b) Profile of phase differences $\varphi_j$ between neighboring sites as a function of site index $j$. The dashed line corresponds to $\bar{\varphi}=\pi/2$. (c) Variation of effective detuning $\Delta_j/\kappa$ over the array (left, blue) and variation of local squeezing parameters $g_s^j/\kappa$ (right, red). (d) Variation of effective hopping $J_j/\kappa$ over the array (left, blue) and variation of non-local squeezing parameters $g_c^j/\kappa$ (right, red). (e) Inhomogeneous pump amplitudes $\Omega_p^j$ as a function of site index $j$ (left, blue), realized by an inhomogeneous power $P_p^j$ (right, red). (f) Inhomogeneous bare detunings $\Delta_b^j=\omega_p-\omega_b^j$ vs site index $j$ (left, blue), obtained by an inhomogeneous capacitance profile $C_{\rm eq}^j/\bar{C}_{\rm eq}=h_j$ (right, red). For all panels, parameters are $\bar{J} = 0.52\kappa$ and $\bar{g}_c = 0.23\kappa$, and $\bar{g}_s/\bar{g}_c=\lambda=1$, $\bar{\Delta}=0$, $\varphi = \pi/2$, and $N=17$ (with $N_L=2$, $N_R=3$, $N_C=12$ and $M=25$).}
\label{Fig_inhomogeneities}
\end{figure*}

At the central region, all parameters of the JJA are constant, leading to homogeneous bare detunings $\Delta_b^j=\bar{\Delta}_b$, bare hoppings $J_b^j=\bar{J}_b$, self-Kerr couplings $K_s^j=\bar{K}_s$, cross-Kerr couplings $K_c^j=T_j=T_j'=\bar{K}_c$, and also driving strengths $\Omega_p^j=\bar{\Omega}_p$. Note that here and throughout this work, the bar symbol $\bar{x}$ denotes a constant profile of the corresponding quantity $x_j$. Within this central region, boundary effects are suppressed and it is a good approximation to assume periodic boundary conditions and look for homogeneous steady-state solutions. In particular, we use the ansatz (\ref{SSansatz}) with constant amplitudes $|\alpha_{\rm ss}^j|=|\bar{\alpha}_{\rm ss}|$ and linearly increasing phase $\theta_{\rm ss}^j=\bar{\varphi} j$, such that the phase differences are also constant $\bar{\varphi}=\theta_{\rm ss}^{j+1}-\theta_{\rm ss}^{j}$. In this case, the set of non-linear equations (\ref{classicalTime}) reduces to a single non-linear Duffing equation \cite{eichler_controlling_2014,planat_understanding_2019} for the target steady-state amplitude $|\bar{\alpha}_{\rm ss}|$, which reads:
\begin{align}
    \left\lbrace\kappa/2-i(\bar{X}+\bar{Y}|\bar{\alpha}_{\rm ss}|^2)\right\rbrace |\bar{\alpha}_{\rm ss}|e^{-i\bar{\varphi} j} 
    = \bar{\Omega}_p e^{-i\theta^j_p}.\label{SSperiodic}
\end{align}  
Inspecting Eq.~(\ref{SSperiodic}), we see that the required steady state phase profile $\theta_{\rm ss}^j=\bar{\varphi} j$ can be conveniently imprinted by the pump's phase by setting: 
\begin{align}
\theta_{p}^j=\bar{\varphi} j.\label{theta_pump}   
\end{align}
Note that the pump phases (\ref{theta_pump}) are taken over the whole array, $j=1,\dots, N$, and not only over the central region. The coefficients $\bar{X}$ and $\bar{Y}$ entering the Duffing equation are strongly influenced by the externally imprinted phase difference $\bar{\varphi}$. Explicitly, the coefficient $\bar{X} = \bar{\Delta}_b - 2\bar{J}_b\cos(\bar{\varphi})$ depends on linear properties of the JJA at the central region such as bare detuning and hopping, whereas $\bar{Y} = \bar{K}_s +\bar{K}_c[2+\cos(2\bar{\varphi})-4\cos(\bar{\varphi})]$ contains the non-linear Kerr contributions. As shown in Appendix~\ref{central_parameters}, Eq.~(\ref{SSperiodic}) can be numerically and even analytically solved. In terms of this homogeneous solution for $|\bar{\alpha}_{\rm ss}|$, the effective quantities in Eqs.~(\ref{EffectiveDet})-(\ref{effectivegs}) take a simple form in the central region:
\begin{align}
    \bar{\Delta} ={}& \bar{\Delta}_b + 2(\bar{K}_s+\bar{K}_c[1-2\cos(\bar{\varphi})])|\bar{\alpha}_{\rm ss}|^2,\label{DeltaHom}\\
    \bar{J} = {}& \bar{J}_b + 2\bar{K}_c|\bar{\alpha}_{\rm ss}|^2[1-\cos(\bar{\varphi})],\label{JeffHom}\\
    \bar{g}_c = {}& \bar{K}_c|\bar{\alpha}_{\rm ss}|^2[1-\cos(\bar{\varphi})],\label{gchom}\\
    \bar{g}_s = {}& (\bar{K}_s + \bar{K}_c[\cos(2\bar{\varphi})-2\cos(\bar{\varphi})])|\bar{\alpha}_{\rm ss}|^2.\label{gsHom}
\end{align}

In practice, to find suitable circuit parameters that stabilize topological amplification, we first target the effective parameters at the central region, namely $\bar{\Delta}$, $\bar{J}$,  $\bar{g}_c$, $\bar{g}_s$, $\kappa$, $\bar{\varphi}$, and $|\bar{\alpha}_{\rm ss}|\gg 1$. Given these values, we use relations (\ref{DeltaHom})-(\ref{gsHom}), and the Duffing equation (\ref{SSperiodic}) to determine the actual circuit parameters that originate them such as $\bar{\Omega}_p$, $\bar{\Delta}_b$, $\bar{J}_b$, $\bar{K}_c$, $\bar{K}_s$, etc. The exact procedure is detailed in Appendix~\ref{sec:inverse_procedure}.

We perform a numerical test of the steady-state stabilization method explained above. For a JJA with $N=17$ sites, we evolve numerically the non-linear equations (\ref{classicalTime}) up to a steady-state configuration with target central values: $|\bar{\alpha}_{\rm ss}|= 2780$, $\bar{\varphi}=\pi/
2$, $\bar{\Delta}=0$, $\bar{J} = 0.52\kappa$, and $\bar{g}_c=\bar{g}_s=0.23\kappa$. In Figs.~\ref{Fig_inhomogeneities}(a)-(d), we show the resulting inhomogeneous profiles of mean-field displacements $|\alpha_{\rm ss}^j|^2$, phase differences $\varphi_j$, as well as effective detunings $\Delta_j$ (blue), hoppings $J_j$ (blue), and squeezing terms $g_s^j$, $g_c^j$ (red), demonstrating that all of them stabilize well to the target values in the central region (compare solid with dashed lines). Regarding the buffer regions, we considered $N_L=2$ sites at the left boundary and $N_R=3$ at the right boundary, as we checked heuristically that this already provides a smooth enough tapering function $f_j$ in Eq.~(\ref{tapering}) to reach the target quasi-steady state effectively. At the interface between central and buffer regions, we can notice small spikes in the exact numerical solutions which are due to residual asymmetric non-linear shifts. However, as we show below, these small inhomogeneities do not alter the directional amplifying properties of the device since it is topologically protected against disorder and local imperfections.

For the above example, Fig.~\ref{Fig_inhomogeneities}(e), blue, indicates the required inhomogeneous pump profile $\Omega_p^j$ to reach the smooth quasi-homogenous configuration with $\bar{\Omega}_p=44.9\kappa$. This corresponds to varying the pump powers in the range $P_p^j\in -95$ to -75 dBm, for typical superconducting circuit parameters, as indicated in Fig.~\ref{Fig_inhomogeneities}(e), red. All circuit parameters considered in this example are detailed below in Sec.~\ref{Peformance_ExpParameters} and Table~\ref{tab:circuit_param}.

An important technical requirement of the stabilization method used above, and that we have not commented on yet, is that to achieve a quasi-homogenous profile of effective detunings $\Delta_j$ it is also necessary to engineer inhomogenous bare detunings $\Delta_b^j$ at the buffer regions. This is because when having small $|\alpha_{\rm ss}^j|^2$ at the boundaries, the non-linear frequency shifts in Eq.~(\ref{EffectiveDet}) due to self- and cross-Kerr non-linearities are also smaller. Therefore, we need to compensate for this by having a smaller absolute value $|\Delta_b^j|$ at the boundaries and increase it towards the center until it reaches the maximum negative value $\bar{\Delta}_b=-|\bar{\Delta}_b|$ targeted at the central region [see blue curve in Fig.~\ref{Fig_boundary_conditions}(b)].

To engineer inhomogeneous bare detunings $\Delta_b^j=\omega_p-\omega_b^j$ in the JJA, we require inhomogeneous local frequencies $\omega_b^j$ at its boundaries. To do so, we can build either inhomogeneous capacitances or inductances. In this work, we leave all the JJs homogeneous as already assumed in the setup section \ref{sec:setupCircuitMain}, but we consider a tapering of the equivalent capacitances as:
\begin{align}
C_{\rm eq}^j= h_j \bar{C}_{\rm eq}.    
\end{align}
Here, $\bar{C}_{\rm eq}$ is the central value of the equivalent capacitance and $h_j$ is a dimensionless function to be determined that smoothly decreases from the boundaries towards the central region, where it becomes $1$. In our setup, inducing the above profile for $C_{\rm eq}^j=C_j+C_{\rm TL}^j+C'_j+C'_{j-1}$ requires varying the individual capacitances $C_j$, $C_j'$, and $C_{\rm TL}^j$ in the form:
\begin{align}
    C_j ={}& h_j \bar{C},\qquad C_{\rm TL}^j = h_j \bar{C}_{\rm TL}, \qquad
    C_j' = q_j \bar{C}',\label{tapering_individual}\\
    {}&{\rm with}\qquad h_j = (q_j+q_{j-1})/2.\label{hjviaqj}
\end{align}
Here, $q_j$ is a dimensionless and smoothly decreasing function from the boundaries to the central region, which determines $h_j$ via Eq.~(\ref{hjviaqj}). In addition, $\bar{C}$, $\bar{C}_{\rm TL}$, and $\bar{C}'$ are the homogenous central values of the individual capacitances, satisfying $\bar{C}_{\rm eq}=\bar{C}+\bar{C}_{\rm TL}+2\bar{C}'$. Note that the tapering (\ref{tapering_individual}) ensures that the ratio $C_{\rm TL}^j/C_{\rm eq}^j$ is constant so that $\kappa$ in Eq.~(\ref{kappap}) is homogenous over the whole JJA. Additionally, it ensures that the hopping rates $J_b^j$ as well as the Kerr non-linearities $K_s^j$, $K_c^j$, $T_j$ and $T_j'$ change weakly in the buffer regions, while the bare on-site frequencies vary strongly as $\omega_b^j=h_j^{-1/2}\bar{\omega}_b$.

We obtain $q_j$ by numerically solving Eqs.~(\ref{classicalTime}) and finding the optimal set of $q_j$ that minimizes the deviation of the effective detuning $\Delta_j$ along the whole array with respect to the homogeneous target at the center $\delta \Delta_j = |\Delta_j-\bar{\Delta}|$. Once the optimal $q_j$ is obtained, we determine the profile of $h_j$ via Eq.~(\ref{hjviaqj}). An example of the profile $h_j=C_{\rm eq}^j/\bar{C}_{\rm eq}$ is shown by the red curve in Fig.~\ref{Fig_inhomogeneities}(f), as well as the resulting profile for bare detunings $\Delta_b^j/\kappa$ (blue). The profile for inhomogeneous pump strengths $P_p^j$ can be directly obtained by the profile $P_p^j = (f_j)^2 h_j^{-1/2} \bar{P}_p$, with $\bar{P}_p$ its central value [Fig.~\ref{Fig_inhomogeneities}(e), red]. For more details, in Appendix~\ref{central_parameters} we indicate all the specific profiles for $f_j$, $q_j$, and $h_j$ used in this work (see Table~\ref{tab:tapering_f}). 

Finally, we note that instead of tapering the pump strengths $\Omega_p^j$, it is also possible to taper the decay rates as $\kappa_j=\kappa /f_j$ with the same function $f_j$ in Eq.~(\ref{theta_pump}), and leave $\Omega_p^j = \bar{\Omega}_p$ constant over the whole array. Here, we do not show this alternative explicitly, but it leads to qualitatively similar results.

\begin{figure*}
    \centering
    \includegraphics[width=0.9\textwidth]{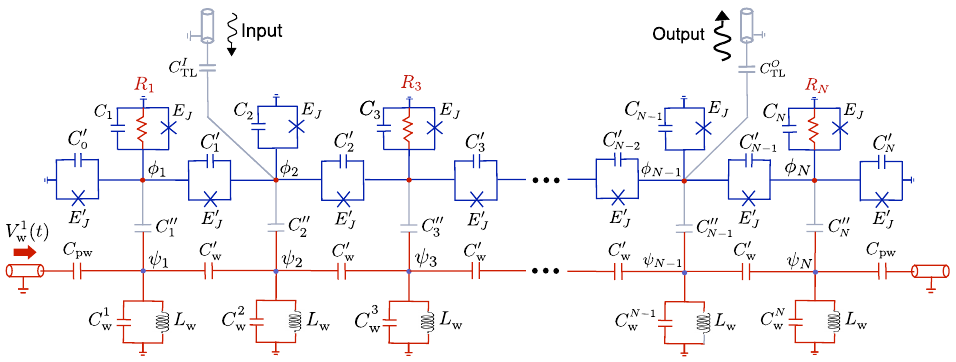}
    \caption{Alternative design of the amplifier array that distributes a coherent pump via an auxiliary waveguide. The pump is sent through a linear resonator waveguide. Inhomogeneous pumps are realized by properly engineering the values of the coupling capacitors $C''_j$. Local dissipation on the JJA is induced by local engineered resistors $R_j$.}
    \label{Fig:setup_waveguide}
\end{figure*}

\section{Distributed pump via auxiliary waveguide}\label{auxiliaryarray}

To send a pump on all sites of the JJA, the setup in Fig.~\ref{Fig:Setup1} requires $N$ independent voltage sources and transmission lines (one per site). This provides direct control of individual pump amplitudes and phases but makes the device too bulky for $N\gtrsim 20$. When scaling to larger $N$ we may require other methods. In this section, we propose to use an auxiliary array of linear superconducting resonators to distribute a single strong pump into $N$ independent sites as shown in Fig.~(\ref{Fig:setup_waveguide}). In Sec.~\ref{Aux_waveguide_circuit}, we introduce the circuit setup, in Sec.~\ref{waveguide_dynamics} we describe its driven-dissipative dynamics, and in Sec.~\ref{4WM_waveguide} we explain the conditions to perform 4-wave-mixing via the auxiliary array, effectively obtaining the same amplifier model discussed in the previous section. 

\subsection{Circuit design for auxiliary waveguide}\label{Aux_waveguide_circuit}

The alternative design of the topological amplifier array with an auxiliary waveguide is shown in Fig.~\ref{Fig:setup_waveguide}. The JJA (shown in blue) is the same as introduced in Sec.~\ref{sec:setupCircuitMain}. The auxiliary waveguide (shown in red) consists of a linear array of superconducting resonators of inductance $L_{\rm w}$ and capacitance $C_{\rm w}^j$, which are coupled via a capacitance $C'_{\rm w}$. Each auxiliary resonator is further coupled to site $j$ of the JJA via an inter-array capacitance $C''_j$. The resulting waveguide Lagrangian ${\cal L}_{\rm w}$ reads:
\begin{align}
    {\cal L}_{\rm w} = {}&\sum_{j=1}^N \frac{C_{\rm w}^j}{2}(\dot{\psi}_j)^2 -\sum_{j=1}^N \frac{(\psi_j)^2}{2L_{\rm w}}\label{LagWaveguide}\\
     +{}&\sum_{j=1}^{N-1}\frac{{C}'_{\rm w}}{2}(\dot{\psi}_{j+1}-\dot{\psi}_j)^2 + \sum_{j=1}^N \frac{C''_j}{2}(\dot{\psi}_j-\dot{\phi}_j)^2.\nonumber
\end{align}
Here, $\psi_j(t)$ are the flux operators at site $j$ of the auxiliary waveguide, and $\phi_j(t)$ are the flux operators of the JJA, which are further subjected to Lagrangian ${\cal L}_{\rm JJA}$ in Eq.~(\ref{LagJJ}). 

In addition, the auxiliary array requires two transmission lines (see Fig.~\ref{Fig:setup_waveguide}): one at the left border ($j=1$) to send a coherent pump, and one at the right border ($j=N$) to engineer impedance matching conditions and dissipate the pump field. The Lagrangian ${\cal L}_{\rm pw}$ describing the capacitive coupling to these transmission lines reads
\begin{align}
    {\cal L}_{\rm pw} = {}&\sum_{j=1,N}\frac{C_{\rm pw}}{2}(\dot{\psi}_j-V_{\rm w}^j(t))^2,
\end{align}
with $C_{\rm pw}$ the coupling capacitance. The voltage on the left border, $V_{\rm w}^1(t)$, corresponds to a coherent pump of amplitude $A_{\rm pw}$ and frequency $\omega_p$ generated with a single microwave source:
\begin{align}
V_{\rm w}^1(t)=A_{\rm pw}\cos(\omega_p t)+\delta V_{\rm w}^1(t), \label{pump_voltage}
\end{align}
where $\delta V_{\rm w}^1(t)$ are the vacuum fluctuations. The voltage on the right border, $V^N_{\rm w}(t)=\delta V_{\rm w}^N(t)$, consists only of vacuum fluctuations which we will use to controllably dissipate the propagating pump when reaching the boundary, as discussed below.

Finally, to induce local losses $\kappa$ on all sites of the JJA, we propose two options: (i) either connect independent transmission lines to all JJA sites as in the original setup (see Fig.~\ref{Fig:Setup1}) or (ii) connect transmission lines only to sites $j=I,O$, where we send and retrieve signals, and in all other sites $j\neq I,O$ use on-chip microwave attenuators $R_j$ \cite{yeh_microwave_2017,yeh_hot_2019} (see Fig.~\ref{Fig:setup_waveguide}). In the first case (i), the coupling to JJA transmission lines is modeled via Lagrangian ${\cal L}_{\rm TL}$ as in Sec.~\ref{sec:setupCircuitMain}, which leads to local dissipation on all sites with rate $\kappa$ in Eq.~(\ref{kappap}). In the second case (ii), we induce the same local dissipation rate $\kappa$ by engineering the resistance $R_j$ of the attenuators as \cite{yurke_quantum_1984},
\begin{align}
    \kappa= 1/(C_{\rm eq}^jR_j),\qquad j\neq I,O.\label{dissipationResistor}
\end{align}
Only at sites $j=I,O$ we use transmission lines to induce decay $\kappa$ as we also require real microwave ports to send and retrieve the signals. The corresponding Lagrangian reads
\begin{align}
    {\cal L}_{\rm TL}' = {}&\sum_{j=I,O}\frac{C_{\rm TL}^j}{2}(\dot{\phi}_j-V_{\rm TL}^j(t))^2,
\end{align}
which we use instead of ${\cal L}_{\rm TL}$ in Eq.~(\ref{Lagp}). Regardless of the method to induce homogeneous decay on all JJA sites, having an auxiliary waveguide allows us to reduce the number of independent pump generators from $N$ to $1$. Option (i) requires a total of $N+2$ transmission lines, while (ii) reduces this to $4$, independent of array size.

\subsection{Auxiliary waveguide Hamiltonian and driven-dissipative dynamics of coupled system}\label{waveguide_dynamics}

The total Hamiltonian $H'=H_{\rm JJA}+H_{\rm w}+H_{\rm pw}$ of the composite system composed of JJA coupled to the auxiliary waveguide is derived in Appendix~\ref{auxWaveguide}. In particular, the auxiliary waveguide can be effectively modeled as a linearly coupled resonator array with Hamiltonian,
\begin{align}
    \frac{H_{\rm w}}{\hbar} ={}& \sum_{j=1}^N \omega_{\rm w} b_j^\dag b_j + \sum_{j=1}^{N-1} J_{\rm w} (b_{j+1}^\dag b_{j} + b_{j}^\dag b_{j+1})\nonumber\\
{}&+\sum_{j=1}^N J'_{\rm w}{}^j(a_{j}^\dag b_{j}+b_{j}^\dag a_{j}).\label{Hb}
\end{align}
Here, $b_j$ and $b_j^\dag$ destroy and create microwave excitations of frequency $\omega_{\rm w}=(C_{\rm eq}^{\rm w}L_{\rm w})^{-1/2}$ at each site $j$ of the auxiliary array. These auxiliary linear resonators have intra-array nearest neighbor couplings $J_{\rm w}$ as well as inter-array couplings $J'_{\rm w}{}^j$ to site $j$ of the JJA. In terms of the circuit parameters, these hopping rates read
\begin{align}
J_{\rm w} ={}& \frac{C'_{\rm w}}{2(C_{\rm eq}^{\rm w})^2Z_{\rm w}}, \qquad
J_{\rm w}'{}^j =\frac{C_j''(Z_jZ_{\rm w})^{-1/2}}{2C_{\rm eq}^jC_{\rm eq}^{\rm w}}.\label{Jwcouplings_hom}
\end{align}
The equivalent capacitance at site $j$ of the auxiliary waveguide is given by
$C_{\rm eq}^{\rm w}=C_{\rm w}^j+C_j''+2C_{\rm w}'+(C_{\rm pw}-C_{\rm w}')(\delta_{j1}+\delta_{jN})$, and it will be constrained to be homogemeous by engineering $C_{\rm w}^j$ and $C_{\rm w}'{}^j$.

In addition, the classical part of the pump voltage (\ref{pump_voltage}) leads to a coherent drive Hamiltonian $H_{\rm pw}$ acting only at the first auxiliary site $j=1$. In the rotating frame with the pump frequency $\omega_p$, it takes the form,
\begin{align}
\frac{H_{\rm pw}}{\hbar} = {}&i\Omega_{\rm pw} (b_1^\dag-b_1)-\omega_p\sum_{j=1}^N (a_j^\dag a_j+b_j^\dag b_j).\label{HpRWA2_main}
\end{align}
with $\Omega_{\rm pw}= (C_{\rm pw}/2C_{\rm eq}^{\rm w})\sqrt{(Z_{\rm pw}/Z_{\rm w})(P_{\rm pw}/\hbar)}$ the strength of the drive, determined by the coupling capacitance $C_{\rm pw}$, the impedance of the line $Z_{\rm pw}$, the impedance of the auxiliary resonator $Z_{\rm w}=\sqrt{L_{\rm w}/C_{\rm eq}^{\rm w}}$, and the pump power $P_{\rm pw}=A_{\rm pw}^2/(2Z_{\rm pw})$. For simplicity, both Hamiltonians $H_{\rm w}$ and $H_{\rm pw}$ assume the RWA provided the couplings are much smaller than free frequencies $\omega_b$, $\omega_{\rm w}$, and $\omega_p$. However, this approximation is not strictly necessary and in Appendix~\ref{auxWaveguide} we show that qualitatively same results are obtained without it when slightly adapting the required pump strength on the auxiliary array $\Omega_{\rm pw}$.

The dissipation due to transmission lines at the boundaries of the auxiliary array can be also included in the input-output formalism (see Appendix~\ref{auxWaveguide}). In the Markov approximation, the driven-dissipative dynamics of the coupled system of JJA and waveguide reads
\begin{align}
    \dot{a}_j={}&\frac{i}{\hbar}[H_{\rm JJA}+H_{\rm w},a_j]-\frac{\kappa}{2}a_j+\sqrt{\kappa}a_{\rm in}^{j}(t),\label{QLEw1}\\
    \dot{b}_j={}&\frac{i}{\hbar}[H_{\rm w}+H_{\rm pw},b_j]-\frac{\kappa^j_{\rm w}}{2}b_j+\sqrt{\kappa_{\rm w}^j} b_{\rm in}^j(t).\label{QLEw2}
\end{align}
Note that Eq.~(\ref{QLEw1}) is the same as (\ref{QLE1disp}), except for the coupling to the auxiliary waveguide in $H_{\rm w}$. The auxiliary waveguide has its own linear dynamics and, in particular, it is subjected to input operators $b_{\rm in}^j(t)$ describing the decay of photons into the transmission lines on its boundaries $j=1,N$. The associated decay decay rate $\kappa_{\rm w}^j$ is given by, 
\begin{align}
    \kappa_{\rm w}^{j}=(C_{\rm pw}/2C_{\rm eq}^{\rm w})^2(Z_{\rm pw}/Z_{\rm w})\omega_{\rm w}(\delta_{j1}+\delta_{jN}).\label{kappa_waveguide}
\end{align}
Notice that in the case of inducing local dissipation via local resistors $R_j$, the input operators $a_{\rm in}^j(t)$ at sites $j\neq I,O$ do not correspond to physical channels, but for they must be formally included for consistency of the Langevin Eq.~(\ref{QLEw1}).

\subsection{Four-wave-mixing via auxiliary waveguide}\label{4WM_waveguide}

This subsection generalizes the four-wave mixing analysis in Sec.~\ref{Sec:4WM_tot} by including the auxiliary waveguide in the device. Instead of directly driving all JJA sites, here we apply a strong pump $\Omega_{\rm pw}$ only on the left input port of the waveguide $b_1$ as given by $H_{\rm pw}$ in Eq.~(\ref{HpRWA2_main}). Since JJA modes $a_j$ and auxiliary modes $b_j$ are coupled, the strong local pump displaces all of them as 
\begin{align}
    a_j= \alpha_j + \delta a_j,\qquad b_j=\beta_j + \delta b_j,\label{displacements_ab}
\end{align}
with $\delta a_j$ and $\delta b_j$ the quantum fluctuations around the large mean-field displacements $\alpha_j$ and $\beta_j$. Using the above decompositions in Eqs.~(\ref{QLEw1})-(\ref{QLEw2}), we find coupled classical equations for $\alpha_j$ and $\beta_j$, as well as coupled quantum equations for $\delta a_j$ and $\delta b_j$, which generalize Eqs.~(\ref{classicalTime}) and (\ref{QLE2}), respectively (See Appendix~\ref{auxWaveguide}). For the validity of the mean-field theory and linearization, the solutions must satisfy $\langle \delta a_j^\dag \delta a_j\rangle \ll |\alpha_j|^2$ and $\langle \delta b_j^\dag \delta b_j\rangle \ll |\beta_j|^2$.

The non-linear equations for the mean-field displacements including the coupling to the auxiliary waveguide read:
\begin{align}
    \dot{\alpha}_j ={}& -\left(\frac{\kappa}{2}-i\Delta_b^j\right)\alpha_j +iK^j_{s}|\alpha_j|^2\alpha_j -iJ_{\rm w}'{}^j\beta_j,\label{classicalTime_coupled}\\
    {}&-i\left(J_b^j\alpha_{j+1}+J_b^{j-1}\alpha_{j-1}\right)+ F_{c}^j(\alpha_j,\alpha_{j\pm 1}),\nonumber\\   
    \dot{\beta}_j={}&-(\frac{\kappa_{\rm w}^j}{2}-i\Delta_{\rm w})\beta_j-iJ_{\rm w}(\beta_{j-1}+\beta_{j+1})\label{betaEq}\\
    {}&+\Omega_{\rm pw}\delta_{j1}-i J'_{\rm w}{}^j\alpha_j.\nonumber
\end{align}
Here, $\Delta_{b}^j=\omega_p-\omega^j_b$ and $\Delta_{\rm w}=\omega_p-\omega_{\rm w}$ are the detunings between pump frequency and bare frequencies of JJA and waveguide, respectively. As discussed in Sec.~\ref{Sec:Stabilization_inhomogeneities}, the topological amplifying regime requires a steady-state $\alpha_{\rm ss}^j=\alpha_j(t\rightarrow\infty)$ of the form (\ref{SSansatz}) with linearly increasing phase $\theta_{\rm ss}^j \approx \bar{\varphi} j$. Remarkably, here we show that Eqs.~(\ref{classicalTime_coupled})-(\ref{betaEq}) allows us to stabilize this solution by inducing a compatible steady-state $\beta_{\rm ss}^j=\beta_j(t\rightarrow\infty)=|\beta_{\rm ss}^j|e^{-i\theta_{\rm w, ss}^j}$ in the auxiliary waveguide with the same phase dependence $\theta_{\rm w,ss}^j\approx \bar{\varphi} j$.

\begin{figure*}[t]
\centering
\includegraphics[width=\textwidth]{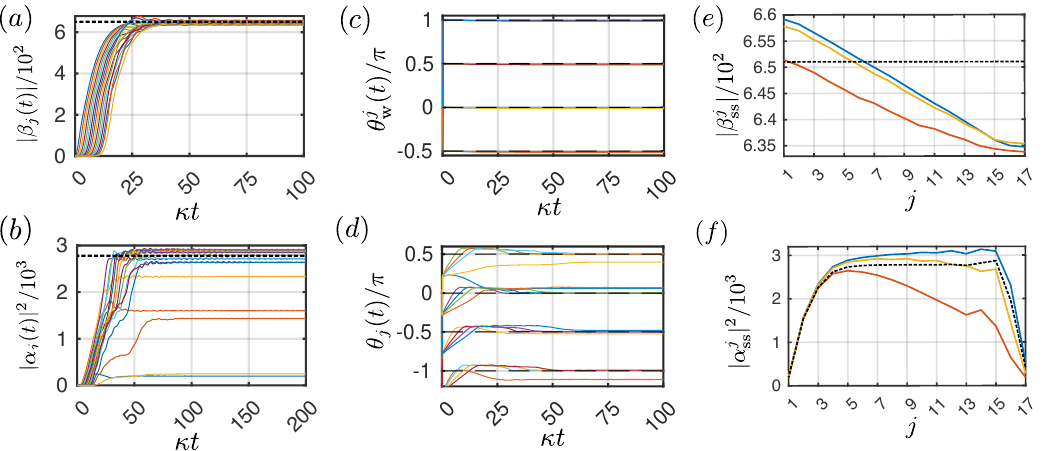}
\caption{Distributing a non-local inhomogeneous pump via an auxiliary waveguide. (a)-(d) Dynamics to reach the steady state of the coupled JJA-waveguide system, showing for all sites $j=1,\dots, N$ the time-dependence of (a) waveguide mean displacements $|\beta_j(t)|$, (b) JJA mean occupations $|\alpha_j(t)|^2$, (c) waveguide phases $\theta_{\rm w}^j(t)$, and (d) JJA phases $\theta_j(t)$, for $\Omega_{\rm pw}=1.01 \Omega_{\rm pw}^{\rm theo}$ with $\Omega_{\rm pw}^{\rm theo}=(2J_{\rm w}/\bar{J}'_{\rm w})\bar{\Omega}_p$. The dashed black lines indicate the steady-state estimations in the central region. (e) Steady-state profile of waveguide displacements $|\beta_j^{\rm ss}|$ vs site index $j$, for three values of pump strengths $\Omega_{\rm pw}=\Omega_{\rm pw}^{\rm theo}$ (red), $\Omega_{\rm pw}=1.01\Omega_{\rm pw}^{\rm theo}$ (yellow), and $\Omega_{\rm pw}=1.012\Omega_{\rm pw}^{\rm theo}$ (blue). The dashed black line indicates the homogeneous analytical estimation $|\beta_{\rm ss}^{\rm theo}|=\Omega_{\rm pw}^{\rm theo}/(2J_{\rm w})$. (f) Inhomogenous profile of JJA mean displacements $|\alpha_{\rm ss}^j|^2$ induced by the coupling to the waveguide. The color code corresponds to the same cases as in panel (e). As a reference, the dashed black curve indicates the steady-state profile obtained by directly pumping the JJA in Fig.~\ref{Fig_inhomogeneities}(a). All JJA parameters are the same as in that figure, and waveguide parameters read $\bar{J}_{\rm w}'=0.1J_{\rm w}=0.05|\bar{\Delta}_b|=0.069\kappa$, $\Delta_{\rm w}=0$, $\kappa_{\rm w}^{1,N}/J_{\rm w}=2$, and $Z_{\rm pw}/Z_{\rm w}=2$.}
\label{Fig_inhomogeneities_auxiliary}
\end{figure*}

To achieve this, the auxiliary waveguide must fulfill certain conditions. First, it must be in the weak coupling and strong driving regime, $J_{\rm w}'{}^j\ll J_{\rm w}, |\bar{\Delta}_b|\ll \Omega_{\rm pw}$ so that its dynamics approximately decouples from the JJA. In addition, the pump frequency must be on resonance with the waveguide, $\Delta_{\rm w}=0$, and the dissipation on the boundaries must be engineered as $\kappa_{\rm w}^{1}=\kappa_{\rm w}^{N}=2J_{\rm w}$. Under these conditions, we can solve Eq.~(\ref{betaEq}) analytically, and find that the waveguide steady-state develops a plane-wave profile with phase difference $\bar{\varphi}=\pi/2$: 
\begin{align}
    \beta_{\rm ss}^j\approx i(\Omega_{\rm pw}/2J_{\rm w})e^{-i\bar{\varphi} j}.\label{betaSSapprox}
\end{align}
Physically, this steady-state builts because the resonant pump inputs a large stream of photons on the left boundary of the waveguide ($j=1$), which propagates with phase $\bar{\varphi}=\pi/2$, and leaks out without reflections at the right boundary ($j=N$) due to perfect impedance matching conditions. See Appendix~\ref{auxWaveguide} for more details. 

Replacing Eq.~(\ref{betaSSapprox}) in the coupling term $-iJ_{\rm w}'{}^j\beta_j$ of Eq.~(\ref{classicalTime_coupled}), we see that in steady state the waveguide induces a constant coherent drive on all sites of the JJA with the phase difference $\bar{\varphi}=\pi/2$ and whose strength reads
\begin{align}
   \Omega_{p}^j \approx (J_{\rm w}'{}^j/2J_{\rm w})\Omega_{\rm pw}.\label{EffectivePump}
\end{align}
This effective pump further induces the desired steady state profile on the JJA mean-field displacements $\alpha_{\rm ss}^j=|\alpha_{\rm ss}^j|e^{-i\theta_{\rm ss}^j}$ with phase $\theta_{\rm ss}^j=\bar{\varphi}j$ imprinted by the waveguide. The quasi-homogeneous configuration in $|\alpha_{\rm ss}^j|$ can be controlled by an inhomogeneous profile in the inter-array couplings, $J'_{\rm w}{}^j = f_j \bar{J}'_{\rm w}$, with $\bar{J}'_{\rm w}$ the homogenous value at the center and $f_j$ the same smooth function defined in Eq.~(\ref{tapering}). In practice, this profile is realized by engineering the inter-array capacitances as
$C_j'' = f_j h_j^{3/4} \bar{C}''$, and choosing the local waveguide capacitances $C_{\rm w}^j$ such that $C_{\rm eq}^{\rm w}$ is constant. To induce the required effective pump $\bar{\Omega}_p$ at the central region, we estimate a waveguide pump on order $\Omega_{\rm pw}^{\rm theo}=(2J_{\rm w}/\bar{J}'_{\rm w})\bar{\Omega}_p$. Finally, the required inhomogeneous bare detunings $\Delta_b^j$ are obtained by engineering the JJA capacitances using the same procedure as in Sec.~\ref{Sec:Stabilization_inhomogeneities}. 

An unwanted effect of using the auxiliary waveguide is that it leads to super-radiant dissipation on the JJA with the rate (see Appendix~\ref{auxWaveguide}),
\begin{align}
    \gamma_{j}=(J_{\rm w}'{}^j)^2/(2J_{\rm w}).
\end{align}
However, this effect is small in the weak off-resonant coupling regime $J_{\rm w}'{}^j\ll J_{\rm w}, |\bar{\Delta}_b|$, and can be compensated by increasing the pump strength $\Omega_{\rm pw}$ slightly above the estimation $\Omega_{\rm pw}^{\rm theo}$, as shown below.

With all these considerations, we can perform 4-wave mixing via the waveguide and stabilize the required quasi-homogenous configurations compatible with topological amplification. We confirm the feasibility of this approach by numerically solving the coupled non-linear Eqs.~(\ref{classicalTime_coupled})-(\ref{betaEq}) for the same JJA of $N=17$ sites as in Fig.~\ref{Fig_inhomogeneities}. The auxiliary waveguide is weakly coupled with $\bar{J}_{\rm w}'=0.1 J_{\rm w}=0.05|\bar{\Delta}_b|$, and is driven with strength $\Omega_{\rm pw}=1.01\Omega_{\rm pw}^{\rm theo}$. Figs.~\ref{Fig_inhomogeneities_auxiliary}(a)-(d) display the dynamics of (a) waveguide mean displacements $|\beta_j(t)|$, (b) JJA mean occupations $|\alpha_j(t)|^2$, (c) waveguide phases $\theta_{\rm w}^j(t)$, and (d) JJA phases $\theta_j(t)$ towards reaching the steady state. Each curve corresponds to a site $j=1,\dots, N$, and the horizontal dashed lines indicate the homogenous estimation for amplitudes and phases, which agree well for all waveguide sites and the central sites of the JJA. The steady state is reached appreciably at times $t>100/\kappa$, but to ensure a precise convergence, we evolve up to $t_{\rm ss}=300/\kappa$, switching on the waveguide pump over a finite ramp-up time of $T_{\rm ramp}=10/\kappa$.

To understand the dispersion around the estimated steady-state values in Figs.~\ref{Fig_inhomogeneities_auxiliary}(a)-(b), we analyze the spatial dependence of the waveguide profiles $|\beta_{\rm ss}^j|$ and JJA profiles $|\alpha_{\rm ss}^j|^2$ for different values of the waveguide pump $\Omega_{\rm pw}$ [see Fig.~\ref{Fig_inhomogeneities_auxiliary}(e)-(f)]. For $\Omega_{\rm pw}=\Omega_{\rm pw}^{\rm theo}$ (red), $|\beta_{\rm ss}^j|$ agrees with the homogeneous estimation $|\beta_{\rm ss}^{\rm theo}|=\Omega_{\rm pw}^{\rm theo}/(2J_{\rm w})$ (dashed line) only at the left border and then reduces towards the right border due to residual decay induced by the JJA. This reduction is of only $\sim 3\%$, and might be thought negligible. However, we see in Fig.~\ref{Fig_inhomogeneities_auxiliary}(f, red) that it causes a reduction of order $\sim 50\%$ in the JJA displacements $|\alpha_{\rm ss}^j|^2$ with respect to the case without the auxiliary waveguide (dashed lines). To compensate for this effect, we can pump the waveguide slightly above the estimation. For $\Omega_{\rm pw}=1.01\Omega_{\rm pw}^{\rm theo}$ [see Figs.~\ref{Fig_inhomogeneities_auxiliary}(e-f), yellow], we can induce the same target profile in $|\alpha_{\rm ss}^j|^2$ as without waveguide up to small inhomogeneities. For $\Omega_{\rm pw}=1.012\Omega_{\rm pw}^{\rm theo}$ [see Figs.~\ref{Fig_inhomogeneities_auxiliary}(e-f), blue], the JJA displacements can even surpass the target profile preserving its shape. In addition, we show in Appendix~\ref{SimNonRWA} that slightly modifying $\Omega_{\rm pw}$ can be also used to compensate for non-RWA effects or additional dissipation in the auxiliary waveguide. Therefore, the exact value of $\Omega_{\rm `w}\sim \Omega_{\rm `w}^{\rm theo}$ should be ultimately calibrated in the actual device to achieve the best overall amplification performance.

Regarding the dynamics of the quantum fluctuations, the same conditions discussed above, $J_{\rm w}'{}^j\ll |\Delta_b|, J_{\rm w}=\kappa_{\rm w}^{1,N}/2$, allows us to adiabatically eliminate the waveguide fluctuations $\delta b_j$, which evolve slowly and independently compared to the JJA fluctuations $\delta a_j$. As shown in Appendix~\ref{auxWaveguide}, we can thus obtain the same effective linearized equations for the quantum fluctuations $\delta a_j$ in Eqs.~(\ref{QLE2})-(\ref{effH}) provided $\gamma_j\ll\kappa$. Therefore, the auxiliary waveguide can merely behave as a multiplexer \cite{li_scalable_2024} to distribute the desired collective pump on the JJA, alleviating the need of one transmission line and signal generator per site. In the next subsection \ref{sec:directional_amp_topology}, we solve for the dynamics of quantum fluctuations and quantum amplification properties of the device.

Finally, we comment on the possibility of a hybrid design, which uses a fully homogeneous auxiliary waveguide of size $N_C$ coupled only to the central region of the JJA, while on the left and right boundary regions, we use standard independent transmission lines to send the required inhomogeneous pump drives $\Omega_p^j$. This has the advantage of avoiding the engineering of intra-array coupling capacitances $C_j''$, and yet it is scalable as it requires $N_L+N_R+1\sim 6$ microwave generators for any $N$.

\section{Directional amplification via topology}\label{sec:directional_amp_topology}

So far, we have discussed the conditions to induce a quasi-homogeneous steady state of the mean-field displacements $|\alpha_{\rm ss}^j|^2$, which is a necessary but not sufficient condition to have topological amplification via four-wave mixing.

In this section, we solve for the steady state of the quantum fluctuations $\delta a_j(t)$ and demonstrate the emergence of directional amplification in the topological JPA array. In Sec.~\ref{Sec:TopAmp}, we shortly introduce methods to quantify the gain, noise, and bandwidth of the device, and in Sec.~\ref{Characterizing_TopAmp} we describe the main topological amplification properties.

\subsection{Quantification of amplification properties: Gain, reverse gain, noise, and bandwidth}\label{Sec:TopAmp}

Under the conditions discussed in Secs.~\ref{Sec:4WM_tot}-\ref{auxiliaryarray}, the dynamics of quantum fluctuations $\delta a_j(t)$ can be well approximated by the effective parametric amplifier array model in Eqs.~(\ref{QLE2})-(\ref{effH}), including or not the auxiliary waveguide. To solve these linear quantum Langevin equations, it is convenient to write them in matrix form as,  
\begin{align}
     \delta\dot{\vec{a}}(t)= -iH_{\rm nh}\delta\vec{a}(t) \tr{+} \sqrt{\kappa}  \vec{a}_{\rm in}(t),\label{FullQLangevin}
\end{align}
where $\delta\vec{a}(t)=[\delta a_j(t),\delta a_j^\dag(t)]^T$ and $\vec{a}_{\rm in}(t)=[a_j^{\rm in}(t),a_j^{\rm in}{}^\dag(t)]^T$ \tr{are the Nambu} vectors of quantum fluctuations and input fields, respectively [cf.~Fig.~\ref{Fig_Setup_4WM}]. The non-Hermitian matrix $H_{\rm nh}$ it is of size $2N\times 2N$, and describes coherent interactions in $H_{\rm pa}$ and the dissipative processes $\kappa$ in a unified way. In our setup it takes the form:
\begin{align}H_{\rm nh}=\begin{pmatrix}
    {\cal M}_{jl}-i\frac{\kappa}{2}\delta_{jl} & -{\cal K}_{jl} \\
    {\cal K}_{jl} & -{\cal M}^\ast_{jl}-i\frac{\kappa}{2}\delta_{jl}\label{Hnh}
    \end{pmatrix}.
\end{align}
Here, the photon-conserving interactions, ${\cal M}_{jl}=-\tr{\Delta_j}\delta_{jl}+J_j(e^{-i\varphi_j}\delta_{j+1,l}+e^{i\varphi_j}\delta_{j-1,l})$, contain the detuning $\Delta_j$ and the complex hopping terms $J_j$ with phase \tr{differences $\varphi_j\neq 0$}. The local and non-local squeezing terms $g_s^j$ and $g_c^j$ appear in the off-diagonal components, ${\cal K}_{jl}=g_s^j\delta_{jl}+g_c^j(\delta_{j+1,l}+\delta_{j-1,l})$, while local decay $-i\kappa/2$ appears in the diagonal.

Importantly, the Green's function matrix $G(\omega)$ can be obtained from $H_{\rm nh}$ as
\begin{align}
G(\omega)=\frac{1}{\omega-\omega_p-H_{\rm nh}},\label{eq:GreensFunctionMatrix}
\end{align}
which determines all the dynamical and spectral properties of the amplifier (see Appendix~\ref{app:amplifier_properties}), as well as the stability and topology of the steady-state (see next subsection). For instance, if we Fourier transform Eq.~(\ref{FullQLangevin}) and use the input-output relation (\ref{dispInOut1}), we can solve for the frequency-resolved output operator $a_{\rm out}^j(\omega)$ at site $j$ as (see Appendix~\ref{app:amplifier_properties})
\begin{align}
    a_{\rm out}^j(\omega) =\!{}&\sum_l [\delta_{jl}\!-i\kappa G_{jl}(\omega)]a_{\rm in}^l(\omega)\!-\!\sqrt{2\pi\kappa}|\alpha_{\rm ss}^j|\delta(\omega-\omega_p)\nonumber\\
    {}&-i\kappa\sum_l G_{j,N+l}(\omega)a_{\rm in}^l{}^\dag(2\omega_p-\omega)\label{eq:aout},
\end{align}
with $a^j_{\rm in}(\omega) = (2\pi)^{-1/2}\int dt e^{i\omega t}a_{\rm in}^j(t)$ the Fourier transform of the input operator. We operate the device as a phase-insensitive linear amplifier, so the first term $\sim a^l_{\rm in}(\omega)$ describes the amplification of signals at frequency $\omega$. The second term, proportional to $|\alpha_{\rm ss}^j|$, describes the scattered pump field at frequency $\omega_p$, and the last term $\sim a^l_{\rm in}{}^\dag$ describes the idler field generated at frequency $2\omega_p-\omega$, which appears due to energy conservation in the four-wave mixing process \cite{WallsMilburnBook}.

From Eq.~(\ref{eq:aout}) we can extract all figures of merit of the amplifier \cite{ramos_topological_2021,caves_quantum_2012}. In particular, the gain of the amplifier for the coherent signal (\ref{inputsignal}) of frequency $\omega_s$, entering at site $j=I$ and leaving at any site $j>I$, is given by the pre-factor squared of the first term in Eq.~(\ref{eq:aout}), namely
\begin{align}
    {\cal G}_j(\omega_s)= \kappa^2|G_{jI}(\omega_s)|^2,\qquad j\neq I.\label{gainjI}
\end{align}
The reverse gain corresponds to the case where the same signal propagates in the opposite direction through the array, from site $j>I$ to $j=I$, obtaining 
\begin{align}
    {\cal G}^{(R)}_j(\omega_s) = \kappa^2|G_{Ij}(\omega_s)|^2, \qquad j\neq I.\label{reversegainjI}
\end{align}
A directional amplifier is characterized by having asymmetric direct and reverse gains, ${\cal G}_j\neq {\cal G}_j^{(R)}$, and below we show this occurs in the topological amplifying regime. 

The idler field is also amplified when propagating through the array with gain and reverse gain similarly given by the pre-factor of the last term in Eq.~(\ref{eq:aout}) (see Appendix~\ref{app:amplifier_properties}). When operating as a phase-insensitive amplifier, the idler field will unavoidably add noise to the output signal \cite{caves_quantum_2012}. The number of photons per unit frequency $\omega$ generated by the amplifier at output channel $j$ can be computed as (see Appendix~\ref{app:amplifier_properties})
\begin{align}
    n_{\rm amp}^j(\omega) = \kappa^2 \sum_{l=1}^N |G_{j,N+l}(\omega)|^2.\label{amp_photons}
\end{align}
To quantify how close to the quantum limit the amplification is, it is convenient to compute the added noise $n_{\rm add}^j(\omega)$ by normalizing the total added photons in Eq.~(\ref{amp_photons}) by the gain:
\begin{align}
    n_{\rm add}^j(\omega) = \frac{n_{\rm amp}^j(\omega)}{{\cal G}_j(\omega)}.\label{Added_Noise_Main}
\end{align}
In Sec.~\ref{Peformance_ExpParameters}, we show that the topological JPA array can be near-quantum-limited, reaching an added noise close to the fundamental lower bound $n_{\rm add}^j(\omega)\rightarrow 1$ \footnote{We use the convention of normally ordered noise moments leading to the fundamental lower bound $n_{\rm add}^j(\omega)\geq 1$ \cite{QuantumNoise}. Alternatively, using the symmetrized convention, the same limit corresponds to $n_{\rm add}^j(\omega)\geq 1/2$ \cite{caves_quantum_1982}}.

\begin{figure*}
    \centering
    \includegraphics[width=0.95\textwidth]{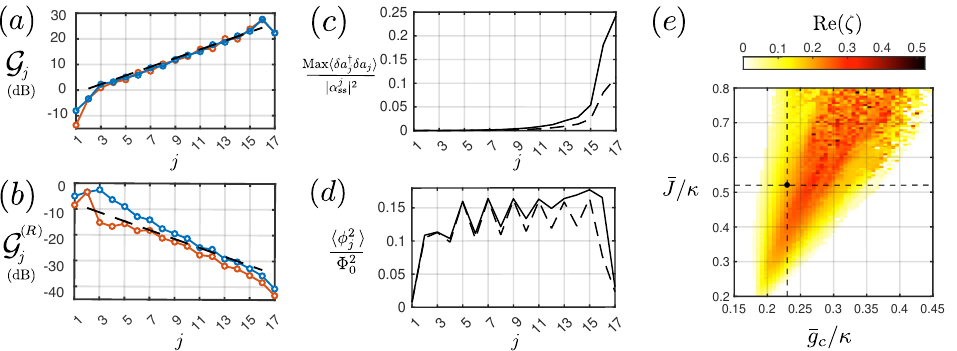}
    \caption{{Directional amplification in the topological \tr{parameteric amplifier array}.} (a) Exponential growth of the gain ${\cal G}_j$ (in dB) for a signal of frequency $\omega_s=\tr{\omega_p+0.3\kappa}$ entering at input site \tr{$I=2$} and leaving at site $j$. \tr{Blue and red circles indicate the gain at the signal and idler frequencies, respectively. The black dashed line shows the prediction using the exponential approximation (\ref{exponentialdep}).} (b) Exponential suppression of reverse gain ${\cal G}_j^{(R)}$ (in dB) for the same signal propagating backward from the site $j$ to the \tr{input} site \tr{$I=2$}. \tr{The color code of the three curves is the same as in panel (a).} (c) Maximum occupation of fluctuation ${\rm Max}|\langle \delta a_j^\dag(t) \delta a_j(t)\rangle|^2$ normalized by the occupation of the mean-field displacement vs site index $j$. (d) Flux mean square $\langle \phi_j^2 \rangle$ normalized by the flux quantum squared $\Phi_0^2$ as a function of site index $j$. The Dashed lines in (c)-(d) show only the contribution from the vacuum noise generated by the amplifier and the solid line includes the coherent contribution from a signal of amplitude \tr{$|\alpha_{\rm sig}|^2 =0.1\kappa$}. (e) Phase diagram as a function of \tr{$\bar{J}/\kappa$} and \tr{$\bar{g}_c/\kappa$} indicating the region for stable topological steady-state phases. The color bar shows the inverse localization length ${\rm Re}[\zeta]>0$, which determines the strength of the topological amplification. The black dot indicates the operation point \tr{at panels (a)-(d)}. This corresponds to the mean-field configuration in Fig.~\ref{Fig_inhomogeneities_auxiliary} with the same effective parameters: \tr{$\bar{J} = 0.52\kappa$ and $\bar{g}_c = \bar{g}_s = 0.23\kappa$, $\bar{\Delta}=0$, $\bar{\varphi} = \pi/2$, $\omega_s=\omega_p+0.3\kappa$, and $N=17$, $M=25$ (with $N_L=2$, $N_R=3$, $N_C=12$)}.}
    \label{Fig_directional_amplification}
\end{figure*}

To estimate the effect of saturation in the amplifier array and ensure the validity of the linear regime, we evaluate $ \langle \delta a_j^\dag(t) \delta a_j(t)\rangle \ll |\alpha_{\rm ss}^j|^2$, where the total occupation of the fluctuations at any site $j$ is given by (see Appendix~\ref{app:amplifier_properties})
\begin{align}
    \langle \delta a_j^\dag(t) \delta a_j(t)\rangle {}&=\frac{1}{2\pi\kappa}\int d\omega n_{\rm amp}^j(\omega)+n_{\rm sig}^j(\omega_s).\label{eq:occupation_amp}
\end{align}
Here, $n_{\rm amp}^j(\omega)$ are the photons generated by the amplifier in Eq.~(\ref{amp_photons}), while $n_{\rm sig}^j\sim \kappa |\alpha_{\rm sig}|^2|G_{jI}(\omega_s)|^2$ is the number of coherent photons induced by a signal with photon flux $\alpha_{\rm sig}$ and its full expression can be found in the Appendix~\ref{app:amplifier_properties}.

Finally, our theory also allows us to evaluate any circuit quantity such as expectation values or correlations of flux $\phi_j(t)$ or charge $q_j(t)$ operators. Especially important is to check the validity of the low phase drop approximation $\langle \phi_j^2 \rangle/\Phi_0^2\ll 1$, for which we evaluate
\begin{align}
    \frac{\langle \phi_j^2 \rangle}{\Phi_0^2} ={}&2\left(\frac{\phi_{\rm zpf}}{\Phi_0}\right)^2(\langle a_j^\dag a_j\rangle(t) + {\rm Re}\langle a_j^2\rangle(t) +1/2 ),\label{lowflux2}
\end{align}
where the total occupation $\langle a_j^\dag a_j\rangle(t)$ and the second moment $\langle a_j^2\rangle(t)$ can also be expressed in terms of the Green's function as shown in Appendix~\ref{app:amplifier_properties}.

In the remainder of the paper, we evaluate all these figures of merit (\ref{gainjI})-(\ref{lowflux2}) to characterize the basic processes and the performance of the topological JPA array.

\subsection{Characterizing the topological amplification regime}\label{Characterizing_TopAmp}

In a topologically non-trivial steady-state phase, the Green's function components become asymmetric, $G_{jl}\neq G_{lj}$, and therefore the amplification is \tr{directional}. To see for which parameters this indeed occurs, we rely on a connection between open quantum systems and topological band theory that we developed in Refs.~\cite{porras_topological_2019,ramos_topological_2021,gomezleon_bridging_2021}. This consists in constructing \tr{an extended} Hamiltonian ${\cal H}(\omega)$ from the \tr{Green's function matrix $G(\omega)$} as,
\begin{align}
{\cal H}(\omega)=\begin{pmatrix}
    0 & G^{-1}(\omega) \\
    [G^{-1}(\omega)]^\dag & 0
\end{pmatrix},\label{Hext}
\end{align}
and relating the eigenvalue problem of ${\cal H}(\omega)$ to the steady-state \tr{of the quantum fluctuation $\delta a_j(t)$ in Eq.~(\ref{FullQLangevin})}. In essence, if the amplifier's parameters are such that ${\cal H}(\omega)$ is in a topologically non-trivial phase according to the ten-fold way of topological insulators \cite{ryu_topological_2010}, then the Green's function matrix $G(\omega)$ describing the physical response of the system develops an exponential spatial dependence. In particular, we show in Appendix~\ref{TopRelationExtHam} that the components $G_{jI}(\omega)$ can be well approximated by,
\begin{align}
    \tr{G_{jI}(\omega)\approx e^{(j-[I+1])\zeta(\omega)}G_{I+1,I}(\omega),\qquad j\geq I+1},\label{exponentialdep}
\end{align}
with small deviations due to boundary effects. Remarkably, the exponent $\zeta(\omega)$ of the Green's function corresponds to the inverse localization length of the edge states associated with the extended Hamiltonian ${\cal H}(\omega)$ \cite{ramos_topological_2021,GomezLeon2022}.

Directional amplification is a \tr{direct consequence of the exponential dependence in Eq.~(\ref{exponentialdep}) as this implies} that the gain of the topological amplifier grows exponentially from left to right: ${\cal G}_j=\kappa^2|G_{jI}|^2\sim e^{2j{\rm Re}(\zeta)}$, whereas reverse gain is exponentially suppressed: ${\cal G}_j^{(R)}=\kappa^2|G_{Ij}|^2\sim e^{-2j{\rm Re}(\zeta)}$. 

Figs.~\ref{Fig_directional_amplification}(a)-(b) demonstrate numerically this behaviour for a topological parametric amplifier array with effective parameters $\bar{\Delta}=0$, $\bar{\varphi}=\pi/2$, $\bar{J}/\kappa=0.52$, $\bar{g}_c/\kappa=0.23$, $\bar{g}_s/\bar{g}_c=1$, and the mean-field configuration in Fig.~\ref{Fig_inhomogeneities_auxiliary} with $\Omega_{\rm pw}=1.012\Omega^{\rm theo}_{\rm pw}$.
In particular, Fig.~\ref{Fig_directional_amplification}(a) displays the gain ${\cal G}_j$ (in dB) for a signal of frequency $\omega_s=\omega_p+0.3\kappa$ sent on input port $I=2$ and retrieved at site $j$. We confirm that at the signal frequency (blue circles) and idler frequency (red circles), the gain grows exponentially with distance, well-approximated by the theoretical prediction from Eq.~(\ref{exponentialdep}) (dashed line). For the same signal sent through site $j$ and propagating in the opposite direction to site $I=2$, Fig.~\ref{Fig_directional_amplification}(b) shows the reverse gain ${\cal G}^{(R)}_j$ (in dB), which indeed decreases exponentially as the estimation (dashed line) for both signal and idler fields. Note that deviations from the exponential dependence are due to finite-size effects (see Appendix~\ref{TopRelationExtHam}). 

In practice, the exponential scaling of the gain allows a topological amplifier array of a moderate size such as $N=\tr{17}$ can reach \tr{nearly $30$} dB of amplification and \tr{$-40$} dB of reverse attenuation. This large gain can lead to a large occupation of the quantum fluctuations of the device $\langle \delta a_j^\dag (t)\delta a_j(t)\rangle$, which also grows exponentially from left to right as shown in Fig.~\ref{Fig_directional_amplification}(c). The dashed curve corresponds to the case of no applied signal (only contribution from the vacuum noise $n_{\rm amp}$ in Eq.~(\ref{eq:occupation_amp})), whereas the solid curve also includes the coherent contribution $n_{\rm sig}$ from a signal of amplitude $|\alpha_{\rm sig}|^2=0.1\kappa$. To avoid this generated noise leads to unwanted saturation of the device, one must ensure that mean-field configuration has large enough displacements, so that $\langle \delta a_j^\dag (t)\delta a_j(t)\rangle\ll |\alpha_{\rm ss}^j|^2$. Here, we consider the mean-field configuration in Fig.~\ref{Fig_inhomogeneities_auxiliary} with $\Omega_{\rm pw}=1.012\Omega^{\rm theo}_{\rm pw}$, leading to a maximum occupation $|\alpha_{\rm ss}^j|^2\sim 3000$ at the central region. In this case, Fig.~\ref{Fig_directional_amplification}(c) shows that saturation is low enough for the applied signal $|\alpha_{\rm sig}|^2=0.1\kappa$. For stronger signals, saturation can be still controlled by having a larger number $M$ of JJs per site, but this is discussed further in Sec.~\ref{Peformance_ExpParameters}. Another important approximation to ensure the proper functioning of the amplifier array is the low phase drop condition $\langle \phi_j^2 \rangle/(\Phi_0)^2\ll 1$. We numerically computed this quantity using Eq.~(\ref{lowflux2}), and we confirmed that the amplifier array in the topological regime also respects this condition for the parameters considered [see Fig.~\ref{Fig_directional_amplification}(d)]. 

Looking for edge states in ${\cal H}(\omega)$ allows for a systematic study of the parameter conditions leading to topological amplification. As an example, Fig.~\ref{Fig_directional_amplification}\tr{(e)} displays a phase diagram of the steady-state of quantum fluctuations as a function of the effective parameters $\tr{\bar{g}_c/\kappa}$ and \tr{$\bar{J}/\kappa$}, for given $\bar{g}_s/\bar{g}_c=1$, \tr{$\omega=\omega_p+0.3\kappa$}, $\tr{\bar{\Delta}}=0$, $\tr{\bar{\varphi}}=\pi/2$, and $N=\tr{17}$. The colored region indicates the topological amplifying phase, which is defined by $E_0(\bar{g}_c,\bar{J})\leq 1/N$, with $E_0$ the lowest eigenvalue of ${\cal H}(\omega)$ (see Appendix~\ref{TopRelationExtHam}). Moreover, in this region, the steady state must be stable, which is guaranteed when all eigenvalues of $H_{\rm nh}$ have negative imaginary parts \cite{porras_topological_2019,ramos_topological_2021}. 

Inspecting the phase diagram, we see that a stable topological amplifying phase requires a balance between parametric drive $\bar{g}_c$ and decay $\kappa$: When $\bar{g}_c/\kappa$ is too small, the system becomes topologically trivial and there is no amplification, and when $\bar{g}_c/\kappa$ is too large, the amplifier becomes unstable. Given an intermediate value of $\bar{g}_c/\kappa\sim 0.2-0.4$, there is a minimum of $\bar{J}/\kappa$ to induce the topological regime. To be more precise, the color bar in Fig.~\ref{Fig_directional_amplification}\tr{(e)} displays the inverse localization length ${\rm Re}[\zeta]$ as a function of $\bar{g}_c/\kappa$ and $\bar{J}/\kappa$, which determines the exponent of the gain in the topological phase. The maximum amplification occurs at the center of the topological region where ${\rm Re}[\zeta]\sim 0.4-0.5$. However, we need to keep a balance between the gain and the saturation that the array can tolerate in practice, and thus we typically target more moderate working points with ${\rm Re}[\zeta]\sim 0.2$ such as indicated in Fig.~\ref{Fig_directional_amplification}(e). It is exactly the balance of parameters that we considered in Figs.~\ref{Fig_directional_amplification}(a)-(d). In the next section, we discuss in detail various parameter sets and how to obtain the best overall performance of the topological amplifier array. 

\begin{figure*}
    \centering
    \includegraphics[width=\textwidth]{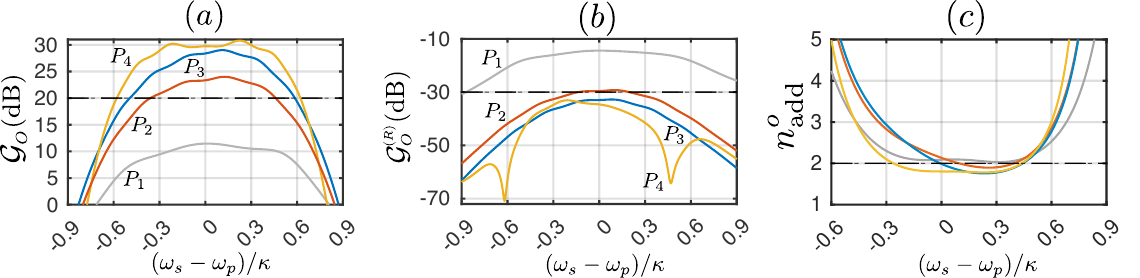}
    \caption{Broadband performance of the topological \tr{JPA array}. As a function of the frequency of the probe signal $\omega_s$, we calculate (a) Gain ${\cal G}_O$ (in dB), (b) reverse gain ${\cal G}^{(R)}_O$ (in dB), and (c) added noise $n^{\rm add}_O$ for 4 different parameter sets. \tr{For $P_1$ (grey), $P_2$ (red), and $P_3$ (blue) parameters are $\bar{J}=0.52\kappa$, $\bar{g}_c=\bar{g}_s= 0.23\kappa$, $\bar{\Delta}=0$, $\varphi=\pi/2$, with increasing sizes $N=8$, $N=15$, and $N = 17$, respectively. For $P_4$ (yellow), parameters are $\bar{J}=0.42\kappa$, $\bar{g}_c=\bar{g}_s/2=0.19\kappa$, $\bar{\Delta}=0$, $\bar{\varphi}=\pi/2$  with size $N=11$. See Table~\ref{tab:four_wave_mixing} for more details.}}
    \label{Fig_performance}
\end{figure*}

\section{Performance and experimental parameters}\label{Peformance_ExpParameters}

In this section, we first quantify the performance and frequency response of the topological JPA array at four different operation points (see Sec.~\ref{freq_response}). Then, we show the superconducting circuit parameters to realize those setups experimentally (see Sec.~\ref{analisis_experimental_parameters}).

\subsection{Frequency-dependent response}\label{freq_response}

To characterize the broadband performance of the directional amplification, we display in Figs.~\ref{Fig_performance}(a)-(c) the frequency dependence of the gain ${\cal G}_O$, reverse gain ${\cal G}^{(R)}_O$, and added noise $n^{\rm add}_O$ at the output site $O=N-1$ when a probe signal of frequency $\omega_s$ is sent at site $I=2$. We consider 4 parameter sets. On the one hand, $P_1$, $P_2$, and $P_3$ correspond to the same effective parameters as in Figs.~\ref{Fig_directional_amplification}(a)-(b), but with different array sizes: $P_1$ for $N=8$ (grey), $P_2$ for $N=15$ (red), and $P_3$ for $N=17$ (blue). On the other hand, parameter set $P_4$ (yellow) corresponds to a qualitatively different operation point, discussed below, which allows the device to reach high performance with a smaller size of $N=11$. All effective parameters are detailed in Table~\ref{tab:four_wave_mixing} and the associated tapering functions at the boundaries are indicated in Table~\ref{tab:tapering_f} of the Appendix.

\begin{table*}[!ht]
\center
\begin{tabular}{|c||c|c|c|c||c|c|c|c||c|c|c|c|c||c|c|c|c||c|c|c|}
\hline
    & $N$ & $N_L$ & $N_R$ & $M$ & $|\bar{\alpha}_{\rm ss}|^2$ & $\lambda$ & $\bar{J}/\kappa$ & $\bar{g}_c/\kappa$ & $ \bar{J}/2\pi$ & $\bar{g}_c/2\pi$ & $\bar{g}_s/2\pi$ & $\bar{\Delta}/2\pi$ & $\kappa/2\pi$ & ${\cal G}_O$ & $\Delta\omega/2\pi$ & ${\cal G}^{(R)}_O$ & $n_{\rm add}^O$ & $\frac{\langle\phi_j^2\rangle}{\Phi_0^2}$ & $\frac{\langle  \delta a^\dag_j\delta a_j\rangle}{|\alpha_{\rm ss}^j|^2}$ & $P_s$ \\
    & & & & & & & & & (MHz)  & (MHz)  & (MHz) & (MHz) & (MHz) & (dB) & (MHz) & (dB)  & & & & (dBm)\\
\hline
$P_1$ & 8 & 2 & 3 & 1 & 27 & 1 & 0.52 & 0.23 & 145 (152) & 64 (68) & 64 (74) & 0 (5) & 279 & 11 & 210 & -14 & 2.0 & 0.21 & 0.29 & -130 \\
\hline
$P_2$ & 15 & 2 & 3 & 16 & 680 & 1 & 0.52 & 0.23 & 145 (152) & 64 (68) & 64 (70) & 0 (18) & 279 & 23 & 232 & -29 & 1.9 & 0.20 & 0.28 & -122\\
\hline
$P_3$ & 17 & 2 & 3 & 25 & 2800  & 1 & 0.52 & 0.23  & 145 (154) & 64 (69) & 64 (70) & 0 (24) & 279 & 28 & 303 & -33 & 1.8 & 0.18 & 0.24 & -120 \\
\hline
$P_4$ & 11 & 2 & 2 & 30 & 2700 & 2 & 0.42 & 0.19 & 94 (98) &  43 (45) & 85 (89) & 0 (6) & 225 & 30 & 271  & -33 & 1.8 & 0.20  & 0.25 & -120 \\
\hline
\end{tabular}
\caption{Effective parameters and figures of merit of topological JPA array at four operation points $P_1-P_4$. The first seven columns show quantities that define each operation point $P_1-P_4$: number of sites $N$ of the JJA, number of sites in the left and right buffer regions $N_L$ and $N_R$, the number of JJs $M$ on each sub-array, target mean occupation of the JJA at the central region $|\bar{\alpha}_{\rm ss}|^2$, and ratios of effective amplifier parameters $\lambda=\bar{g}_s/\bar{g}_c$, $\bar{J}/\kappa$, and $\bar{g}_c/\kappa$. The next five columns show the effective hopping $\bar{J}$, non-local and local squeezing terms $\bar{g}_c$ and $\bar{g}_s$, and effective detuning $\bar{\Delta}$. Notice that values inside (outside) parenthesis correspond to the setup with (without) the auxiliary waveguide. In all cases, we additionally assume $\bar{\varphi}=\pi/2$. The next four columns indicate the figures of merit of gain, bandwidth, reverse gain, and added noise, for a signal at frequency $\omega_s=\omega_p+0.3\kappa$ sent on input $I=2$ and retrieved at output $O=N-1$. The last three columns show the indicators of low phase drop and controlled saturation for a probing signal of power $P_s$. The parameters of the superconducting circuit leading to these effective parameters are detailed in Tables~\ref{tab:circuit_param}-\ref{tab:auxiliar_waveguide}. Parameters of the tapering functions used ($f_j$, $h_j$, and $q_j$) are summarized in Table~\ref{tab:tapering_f}.}\label{tab:four_wave_mixing}
\end{table*}

For $P_1$, the smallest setup with $N=8$, the directional amplifier can reach above 10 dB of gain ${\cal G}_{O}$ over a bandwidth of $\Delta \omega_{\rm 10 dB}\approx 0.8\kappa$ [see Fig.~\ref{Fig_performance}(a), grey]. Protection to the quantum source is provided by a reverse isolation ${\cal G}_{O}^{(R)}<-14$ dB over all frequencies, and an added noise of one photon above the quantum limit: $n_{\rm add}^O\approx 2$ [see Fig.~\ref{Fig_performance}(b)-(c), grey]. This is a good performance for a proof-of-principle demonstration of the topological amplifier, but more competitive figures of merit can be obtained when increasing the array size. 

For $P_2$ ($N=15$), the device surpasses 20 dB of gain ${\cal G}_{O}$ over a bandwidth $\Delta \omega_{\rm 20 dB}\approx 0.8\kappa$, with reverse isolation ${\cal G}_{O}^{(R)}< -29$ dB, and added noise $n_{\rm add}^O\approx 1.9$ [see Fig.~\ref{Fig_performance}(a)-(c), red]. Further increasing the size to $N=17$, $P_3$ reaches a maximum gain of ${\cal G}_{O}\approx 29$ dB, leading to a larger bandwidth above 20 dB gain of $\Delta \omega_{\rm 20 dB}\approx 1.1\kappa$. Reverse gain is below ${\cal G}_{O}^{(R)}<-33$ dB over all frequencies, and added noise reaches closer to the quantum limit, $n_{\rm add}^O\approx 1.8$ [see Fig.~\ref{Fig_performance}(a)-(c), blue]. The idler response is nearly identical to the signal field shown above [see Fig.~\ref{Fig_Performance_idler} of Appendix for details].

The overall performance of the topological amplifier increases exponentially with array size $N$. Unfortunately, there is a practical limit that a real device can reach due to saturation. However, a compact device with $N\lesssim 17$ already leads to outstanding performance, and in the next Sec.~\ref{analisis_experimental_parameters}, we show that this is possible to realize with state-of-art superconducting parameters. 

Another route to reach strong directional amplification performance without a large array is to explore other operation points not shown in the phase diagram in Fig.~\ref{Fig_topology}(e). In particular, here we consider the parameter set $P_4$, characterized by a larger ratio $\lambda = \bar{g}_c/\bar{g}_s= 2$ between local and non-local squeezing terms ($P_1-P_3$ assume $\lambda=1$). Other parameters are $\bar{J}/\kappa = 0.42$, $\bar{g}_c/\kappa = 0.19$, $\bar{\Delta}=0$, and $\bar{\varphi}=\pi/2$ (see Table~\ref{tab:four_wave_mixing}). In this case, we predict that a JJA with only $N=11$ sites, including left and right buffer regions of two sites ($N_L=N_R=2$), can reach a stronger gain and reverse gain than $P_3$ with $N=17$. As shown in Figs.~\ref{Fig_performance} (yellow), $P_4$ reaches a gain above ${\cal G}_{O}\approx 30$ dB, a reverse gain ${\cal G}_{O}^{(R)}\approx -33$ dB, and low added noise $n_{\rm add}^O=1.8$. The bandwidth above 20 dB of amplification is also increased to $\Delta \omega_{\rm 20 dB}\approx 1.23\kappa$. 

To respect all the conditions for topological amplification, we show below that the parametric JPA array can have effective parameters up to the range $\bar{g}_{c,s}/2\pi \sim 40-90$ MHz, $\bar{J}/2\pi\sim 100-150$ MHz, and $\kappa/2\pi \sim 230-280$ MHz (see Table~\ref{tab:four_wave_mixing}). This leads to an amplification bandwidth above 20 dB of order $\Delta\omega_{\rm 20dB}/2\pi\sim 230-300$ MHz in practical implementations of the device. For a directional amplifier, this bandwidth is already one order of magnitude larger than what has been achieved for JPAs \cite{abdo_directional_2013,sliwa_reconfigurable_2015,lecocq_nonreciprocal_2017}, and is on the same order as achieved by TWPAs with isolation via frequency-conversion \cite{ranadive_traveling_2024,malnou_traveling-wave_2024}. Comparative advantages of the topological JPA array include the suppression of gain ripples (see Fig.~\ref{Fig_performance}) due to the intrinsic protection to disorder and inhomogeneities in the JJA, and that the strong reverse isolation ${\cal G}_O^{(R)}<-30$ dB acts on all frequencies, not only on the amplification band. 

In Appendix~\ref{app:quarton}, we explore the possibility of further increasing the bandwidth of the topological JPA array by using Quarton non-linearities \cite{ye_engineering_2021} instead of simple JJs. Quartons have been recently demonstrated to be useful in increasing the coupling between qubits and non-linear resonators \cite{quarton_exp} leading, for instance, to ultra-fast dispersive qubit readout \cite{quarton_readout}. In the present case of the JJA, we show that Quartons can induce larger Kerr non-linearities without reducing capacitance $\bar{C}_{\rm eq}$. Using this property, we estimate in Appendix~\ref{app:quarton} that the operation point $P_2$ can be realized with all effective quantities in Table~\ref{tab:four_wave_mixing} scaled by a factor $\chi\approx 4.4$, and thus operate the amplifier with the same high performance shown in Fig.~\ref{Fig_performance}(red), but with a larger bandwidth above 20 dB gain of order $\Delta\omega_{\rm 20dB}\approx 0.83\kappa \approx 2\pi \cdot 1$ GHz. 

\subsection{Experimental paramaters}\label{analisis_experimental_parameters}

We now analyze the superconducting circuit parameters to realize the four operation points $P_1-P_4$ shown in Table~\ref{tab:four_wave_mixing}. To allow for the highest gain and still have saturation under control $\langle \delta a_j^\dag \delta a_j\rangle/|\alpha_{ \rm ss}^j|^2\ll 1$, one would like to stabilize an amplifier configuration with central mean occupation $|\bar{\alpha}_{\rm ss}|^2\gg 1$ as large as possible. However, the main practical restriction in a JJA implementation is the low phase drop condition (\ref{upperbound}) which limits the maximum achievable $|\bar{\alpha}_{\rm ss}|^2$. A feasible operation point of the amplifier array with gain ${\cal G}_j$ must thus respect all these conditions:
\begin{align}
    {\cal G}_j \lesssim \langle \delta a_j^\dag \delta a_j\rangle\ll |\bar{\alpha}_{\rm ss}|^2\ll M^2(\hbar\bar{\omega}_b \bar{C}_{\rm eq})/(8e^2).\label{summary_lowsat_conditions}
\end{align}
For a given frequency of the JJA, which we fix here to $\bar{\omega}_b/2\pi = 8.2$ GHz, it is, therefore, a requisite to build a JJA with large total capacitance $\bar{C}_{\rm eq}$ or with large sub-arrays of JJs ($M\gg 1$) to allow for the maximum gain and dynamic range.

We first discuss the realization of the operation point $P_1$, which due to the moderate gain ${\cal G}_O\sim 10$dB, is an example where saturation can be controlled without sub-arrays of JJs, i.e.~for $M=1$. Because of this simplification and the small size $N=8$, we think $P_1$ is a good candidate for a proof-of-principle demonstration of topological amplification. To realize a JJA with large capacitance, we propose circuit parameters similar to the experiment reported in Ref.~\cite{mutus_strong_2014}, where an amplifier with large capacitance $\bar{C}_{\rm eq}\approx 4$ pF and similar frequency $\bar{\omega}_b/2\pi\sim 8$GHz has been demonstrated. Following the procedure in Appendix~\ref{sec:inverse_procedure}, we obtain all the required quantities to realize $P_1$ with the above conditions. These are detailed in Table~\ref{tab:circuit_param} of the Appendix. Here, we summarize the order of magnitude of the main circuit parameters. We require on-site and coupling capacitances in the range $\bar{C},\bar{C}',\bar{C}_{\rm TL}\sim 0.5-1.5$ pF, Josephson inductances $L_J=L_J'/2\sim 200$ pH, and external pumps of power $P_p\sim -100$ dBm and detuning $\bar{\Delta}_{b}/2\pi\sim -400$ MHz. This leads to Kerr non-linearities $K_c=K_s/2\sim 2$ MHz and a central JJA occupation of $|\bar{\alpha}_{\rm ss}|^2\sim 30$, which allows to control saturation $\langle  \phi_j^2\rangle/\Phi_0^2< 0.2$ and $\langle\delta a_j^\dag \delta a_j \rangle/|\alpha_{\rm ss}^j|^2< 0.3$, when ampliying signals up to power $P_s\sim -130$ dBm.

\begin{figure*}
    \centering
    \includegraphics[width=0.95\textwidth]{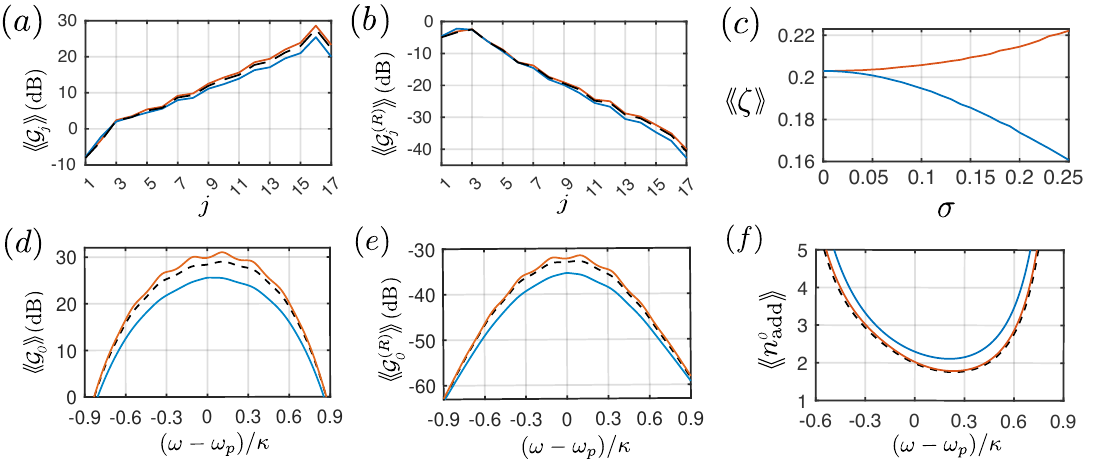}
    \caption{Directional amplification under the influence of disorder. (a) Average gain $\langle\!\langle{\cal G}_O\rangle\!\rangle$ vs site index, (b) average reverse gain $\langle\!\langle{\cal G}^{(R)}_O\rangle\!\rangle$ vs site index, for disorder of strength $\sigma=0.15$ in $\Delta_j$ (blue) and in $g_c^j$ (red). Dashed black lines correspond to $\sigma=0$ as reference. (c) Average inverse localization length $\langle\!\langle\zeta\rangle\!\rangle$ vs disorder strength vs disorder in $\Delta_j$ (blue) and in $g_c^j$ (red). (d) Average gain $\langle\!\langle{\cal G}_O\rangle\!\rangle$, (e) reverse gain $\langle\!\langle{\cal G}^{(R)}_O\rangle\!\rangle$, and (f) added noise $\langle\!\langle n_{\rm add}^O\rangle\!\rangle$ as function of signal frequency $\omega$, for disorder $\sigma = 0.15$ in $\Delta_j$ (blue), and in $g_c^j$ (red), as well as $\sigma=0$ as reference (dashed black). Quantities are averaged over $10^4$ realizations around the mean values defined by parameters in $P_3$, see Table~\ref{tab:four_wave_mixing}.}
    \label{fig:disorder_N_freq}
\end{figure*}

Setups $P_2-P_4$ provide much higher performance than $P_1$, allowing the directional amplification of stronger signals up to $P_s\sim -120$ dBm with gains ${\cal G}_O$ above 20 dB. The higher gain implies that satisfying the inequalities (\ref{summary_lowsat_conditions}) requires increasing each term by more than an order of magnitude compared to $P_1$. A naive solution to this would be to increase the capacitance $\bar{C}_{\rm eq}$ accordingly, but having too large capacitances is problematic in JJ arrays as it generates parasitic geometrical inductances larger than the Josephson inductances themselves \cite{mutus_strong_2014,eichler_quantum-limited_2014}. Therefore, we take the route of sub-arrays of $M\gg 1$ JJs in series to implement all non-linear inductances of the circuit (see Sec.~\ref{sec:saturation_main}). This technique has been successfully implemented to reduce saturation in standard JPAs (with $M=80$) \cite{planat_understanding_2019} and even in an array of two linearly coupled JPAs (with $M=30$) \cite{eichler_quantum-limited_2014}. Due to the similarity to our setup, for $P_2-P_4$ we propose circuit parameters similar to Ref.~\cite{eichler_quantum-limited_2014} with sub-arrays of size $M\sim 16-30$, and standard capacitances in the range $\bar{C},\bar{C}',\bar{C}_{\rm TL}\sim 50-170$ fF, Josephson inductances $L_J/M, L_J'/M\sim 40-230$ pH, as well as pumps of power $P_p\sim -83/-77$ dBm and detuning $\bar{\Delta}_b\sim -350$ MHz. Note that the ratio between local and non-local squeezing terms $\lambda=\bar{g}_s/\bar{g}_c$ is controlled via the ratio of Josephson inductances $\lambda = L_J'/(2L_J)$. We refer to Table~\ref{app:Table_Parameters} for a detailed description of each operation point $P_2-P_4$.

In general, these configurations lead to effectively small Kerr-nonlinearities in the JJA in the range $K_{c,s}/(2\pi M^2) \sim 10-200$ kHz and to large JJA mean occupations $|\bar{\alpha}_{\rm ss}|^2\sim 700-3000$, which allows the amplifier to control saturation effectively with $\langle  \phi_j^2\rangle/\Phi_0^2\lesssim 0.2$, and  $\langle  \delta a_j^\dag \delta a_j \rangle/|\alpha_{\rm ss}^j|^2\lesssim 0.25$. This prediction considers the application of signals up to $P_s\approx -120$ dBm and for stronger signals, one would need to increase the dynamic range further by using sub-arrays with larger $M$.

Regarding the implementation of the auxiliary waveguide, the required circuit parameters are very similar for all parameter sets $P_1-P_4$. Here we provide the typical order of magnitude, but in Table~\ref{tab:auxiliar_waveguide} of the Appendix they can be found with precision for each operation point. In general, we require waveguide inductances $L_{\rm w}\sim 400$ pH, on-site capacitances $\bar{C}_{\rm w}\sim 900$ fF, inter-site capacitances $C_{\rm w}'\sim 50$ fF, intra-array capacitances $\bar{C}''\sim 3-10$ fF, and pump transmission lines with coupling capacitance $C_{\rm pw}\sim 300$ fF, and impedance $Z_{\rm pw} = 10\Omega$. This leads to a bare waveguide frequency $\omega_{\rm w}/2\pi\sim 7.8$ GHz, intra-array hopping $J_{\rm w}/2\pi\sim 180$ MHz, inter-array hopping $J_{\rm w}'/2\pi \sim 20$ MHz, and decay rate on the boundaries $\kappa_{\rm w}^{1,N}/2\pi\sim 360$ MHz. For these parameters, the waveguide is on-resonance with the pump $\omega_p/2\pi\sim 7.8$ GHz and is weakly and off-resonantly coupled to the JJA, $\bar{J}'_{\rm w}\ll J_{\rm w}, |\Delta_b|$. The non-local decay induced on the JJA is highly suppressed, $\bar{\gamma}/2\pi\lesssim 1$ MHz, and can be neglected compared to all other effective quantities. Finally, the pump powers $P_{\rm pw}$ are chosen in the range between $-66$ to $-46$ dBm [cf.~Table~\ref{tab:auxiliar_waveguide}], which is relatively strong but routinely used in the field \cite{esposito_perspective_2021}.

In the case of implementing decay $\kappa$ of the JJA via local resistors, at sites $j\neq I,O$ we must consider resistances on order $R_j\sim 140 \Omega$ for $P_1$ and $R_j\sim 850-1500 \Omega$ for $P_2-P_4$ as the total capacitance is smaller (see Table~\ref{tab:auxiliar_waveguide}). We show below that the topological amplifying phase is resilient up to $15\%$ of spatial variation in local dissipation rates $\kappa_j$ which implies a similar tolerance to variation in these resistances $R_j$ in the $15\%$ range.

Finally, the specific tapering functions $f_j$, $h_j$, and $q_j$ optimized for the different operation points $P_1-P_4$ are given in Table~\ref{tab:tapering_f} of the Appendix. From these functions, the parameters at the central region given in Tables \ref{tab:four_wave_mixing}-\ref{tab:auxiliar_waveguide}, and the relations in Secs.~\ref{Sec:4WM_tot}-\ref{auxiliaryarray}, one can determine all inhomogeneous parameters of the circuit.

\begin{figure*}
    \centering
    \includegraphics[width=\textwidth]{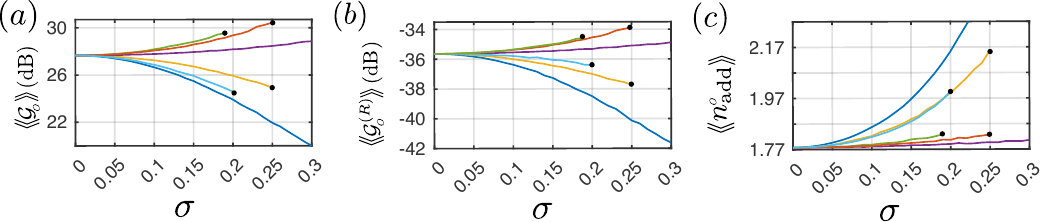}
    \caption{{Robustness against disorder in all system parameters}. (a) Average gain $\langle\!\langle{\cal G}_O\rangle\!\rangle$, (b) reverse gain $\langle\!\langle{\cal G}^{(R)}_O\rangle\!\rangle$, and (c) added noise $\langle\!\langle n_{\rm add}^O\rangle\!\rangle$ as function of the dimensionless disorder strength $\sigma$, defined for each system parameter as $\sigma=\delta\Delta/\tr{\kappa}$ (blue), $\sigma=\delta\kappa/\tr{\kappa}$ (green), $\sigma=\delta J/\tr{\bar{J}}$ (yellow), $\sigma=\delta g_s/\tr{\bar{g}_s}$ (purple), $\sigma=\delta g_c/\tr{\bar{g}_c}$ (red), and $\sigma=\delta \tr{\varphi/\varphi}$ (cyan). All quantities are averaged over \tr{$10^4$} realizations for each type of disorder around the mean values defined by parameters in $P_3$, see Table~\ref{tab:four_wave_mixing}. The signal frequency is $\omega_s=\tr{\omega_p+0.3\kappa}$. The black circle at the end of some curves indicates the disorder $\sigma$ from which the steady state becomes unstable.}
    \label{fig:disorder_all}
\end{figure*}

\section{Topological protection against disorder}\label{topProtectionP1}

The last remarkable property of the topological \tr{JPA array} is the robustness to disorder, which can highly facilitate its experimental realization with current superconducting circuit technology. The origin of the topological protection lies in the presence of an intrinsic chiral symmetry of the extended Hamiltonian ${\cal H}(\omega)$ in Eq.~(\ref{Hext}), and in the robustness of the breaking of time-reversal symmetry (TRS) via $\tr{\bar{\varphi}}=\pi/2\neq 0$. Smooth changes due to disorder cannot modify the symmetry class of the edge states of ${\cal H}(\omega)$, and thus the topologically amplifying steady-state persists as long as the disorder strength \tr{is not larger than} the gap, $\delta x < \Delta_{\rm top}\sim 0.3\kappa$ \cite{porras_topological_2019,ramos_topological_2021,wanjura_correspondence_2021}. Here, $\delta x$ denotes the standard deviation of any system quantity $x$, assumed to be normally distributed around the mean values discussed in previous sections. Note that for robust breaking of TRS, we also require that the disorder in the phase fulfills \tr{$\delta\varphi \ll \bar{\varphi}$.}

To quantify the impact of the disorder on the topological amplification properties, we compute in Figs.~\ref{fig:disorder_N_freq}(a)-(b) the spatial dependence of the average gain $\langle\!\langle{\cal G}_j\rangle\!\rangle$ and reverse gain $\langle\!\langle{\cal G}^{(R)}_j\rangle\!\rangle$ for disorder in the squeezing term $x=g_c^j$ (red) and in effective detunings $x=\Delta_j$ (blue). In both cases, the disorder is relatively strong with standard deviation $\delta x = 0.15 \kappa$. For comparison, the black dashed line represents the case of no disorder given in Figs.~\ref{Fig_topology}(a)-(b). Remarkably, we see that the exponential scalings are preserved and that the disorder in $x=g_c^j$ can even enhance the gain, while the disorder in $x=\Delta_j$ can suppress it. To understand this behavior, we fit an exponential function of the same form as in Eq.~(\ref{exponentialdep}), and extract the exponent $\langle\!\langle \xi \rangle\!\rangle$ of the average as a function of the dimensionless disorder strength $\sigma=\delta x/\kappa$. As shown in Fig.~\ref{fig:disorder_N_freq}(c), the average exponent $\langle\!\langle \xi \rangle\!\rangle$, effectively grows with the disorder in $g_c^j$ (red) and reduces with disorder in $\Delta_j$. This noise-induced effect is reminiscent of topological Anderson insulators observed in photonic systems \cite{stutzer_photonic_2018}. Here, the noise re-normalizes the system parameters and can even trigger topological phase transitions by modifying the localization properties of edge states $\langle\!\langle \xi \rangle\!\rangle$ on average. 

We also checked the effect of disorder on the frequency dependence of the average gain $\langle\!\langle{\cal G}_O\rangle\!\rangle$, reverse gain $\langle\!\langle{\cal G}_O^{(R)}\rangle\!\rangle$, and added noise $\langle\!\langle n^O_{\rm add}\rangle\!\rangle$ at the output site $O=N-1$ [see Fig.~\ref{fig:disorder_N_freq}(d)-(f)]. We see that the frequency response of the directional amplification is also robust to strong disorder $\sigma=0.15\kappa$. For disorder in detuning $\Delta_j$ (blue), the average gain and reverse gain reduce and added noise increases but the overall behavior remains. For disorder in $g_c^j$ (red), the enhanced gain induces small gain ripples but added noise is nearly unchanged.

Finally, we evaluate the impact of disorder in all possible effective parameters of the topological amplifier array dynamics. For this, Figs.~\ref{fig:disorder_all}(a)-(c), show the average gain $\langle\!\langle{\cal G}_O\rangle\!\rangle$, reverse gain $\langle\!\langle{\cal G}^{(R)}_O\rangle\!\rangle$, and added noise $\langle\!\langle n^O_{\rm add}\rangle\!\rangle$ as function of the dimensionless disorder strength $\sigma=\delta x/\tr{\kappa}$ for disorder in effective detuning $x=\bar{\Delta}$ (blue), decay rate $x=\kappa$ (\tr{green}), hopping $x=\bar{J}$ (yellow), local squeezing $x=\bar{g}_s$ (purple), non-local squeezing $x=\bar{g}_c$ (\tr{red}), as well as for the phase $\sigma=\delta \varphi/\bar{\varphi}$ (cyan). The frequency of the input signal is set to $\omega_s=\omega_p+\tr{0.3\kappa}$. We see that for $\sigma\lesssim 5\%$, none of the figures of merit are appreciably affected. For a larger disorder between $\sim 5\%$ and $\sim 20\%$ the topological amplification is preserved but its performance is modified. For all types of disorder, the amplifier's noise $\langle\!\langle n_{\rm add}^O \rangle\!\rangle$ slightly increases. Gain is enhanced for disorder in $\delta g_{c,s}$ and $\delta \kappa$ at the expense of reducing directionality in $\langle\!\langle{\cal G}^{(R)}_O \rangle\!\rangle$, while disorder in $\Delta_j$, $J_j$, and $\varphi_j$ has the opposite effect. For even larger disorder $\sigma\simeq 1$, the gain always reduces, and directional topological amplification is lost.

Notice that a strong enough disorder can destabilize the amplifying steady state depending on the operation point and/or disorder type. For $P_3$, the early termination of certain curves in Figs.~\ref{fig:disorder_all}(a)-(c) indicates the specific disorder strength at which the system becomes unstable. Including this stability analysis, we predict an overall tolerance to the disorder of at least $\sigma\sim \tr{15}\%$ in all effective system parameters, while retaining the excellent performance of the directional amplification. A practical implementation of the topological \tr{JPA array} will be affected by disorder and inhomogeneities inherent to the fabrication \tr{of capacitances} and JJs \cite{kreikebaum_improving_2020} and this constitutes a natural protection against it.

\section{Conclusions and outlook}\label{conclusions}

We have proposed a new design of an array of non-linearly coupled JPAs that exploits non-Hermitian topological effects to work collectively in a regime of high directional gain, near quantum-limited noise, and broad bandwidth. This topological amplifier can be immediately implemented with state-of-art superconducting technology as it only combines conventional elements such as a few coupled JPAs, 50$\Omega$ input/output ports, microwave pumps with inhomogeneous profiles of amplitudes and phases, and, optionally, an auxiliary resonator waveguide to distribute a single pump with phase gradient \cite{PatentApp}. This quantum device can be integrated on-chip and used as a directional broadband pre-amplifier, alleviating the need for bulky isolators to protect the quantum source, and facilitating the control of large-scale quantum processors \cite{esposito_perspective_2021}.

The primary limitation of the current device is saturation and dynamic range, which also constrains the achievable bandwidth. This limitation arises from the low phase-drop condition required to implement Kerr nonlinearities in Josephson junctions (JJs). To overcome this, we can build the Kerr resonator array using high kinetic inductance nonlinearities, as demonstrated in \cite{parker_degenerate_2022,Frasca2024,jouanny_band_2024,ho_eom_wideband_2012}. This approach has the potential to enhance both dynamic range and bandwidth while maintaining the same performance metrics discussed herein.

Our work opens several routes for exploring further technological applications, as well as the fundamental physics of topologically driven-dissipative systems. The topological JPA array can be applied, for instance, for efficient multiplexed readout of multiple superconducting qubits \cite{pereira_parallel_2022,heinsoo_rapid_2018} or broadband detection of itinerant single-photons \cite{grimsmo_quantum_2021}. Saturation effects appearing in the device \cite{eichler_controlling_2014,planat_understanding_2019,remm_intermodulation_2022} can lead to novel regimes of operation, which can be systematically addressed by approximations beyond mean-field like Gaussian ansatz \cite{menu_gaussian-state_2023}, truncated-Wigner approximation \cite{Vicentini18}, or Matrix Product State techniques \cite{daley_quantum_2014}. 

From a more fundamental perspective, our proposal is also a versatile platform for the quantum simulation of open quantum systems \cite{kounalakis_tuneable_2018,guimond_unidirectional_2020}, non-Hermitian physics \cite{Viola3,PhysRevX.9.041015}, and novel topological driven-dissipative phases of matter \cite{GomezLeon2022,McDonaldPRB2022}, including those with disorder \cite{tangpanitanon_topological_2016} and strong photon-photon interactions \cite{hafezi_non-equilibrium_2013, jin_photon_2013}. For quantum simulation, a particularly convenient aspect of our design is that the strength of the synthetic gauge field \cite{Koch2010} and the nonlinear interactions can be conveniently controlled via the external pump, without requiring Floquet engineering \cite{roushan_chiral_2017} or external magnetic fields.

\section*{Acknowledgments}
\tr{We thank Dian Tan, Eunjong Kim, and Adri\'an Parra-Rodr\'iguez for helpful discussions.} This work has been supported by funding from Spanish project PGC2018-094792-B-I00 (MCIU/AEI/FEDER, UE), CSIC Interdisciplinary Thematic Platform (PTI+) on Quantum Technologies (PTI-QTEP+), and Proyecto Sinergico CAM 2020 Y2020/TCS-6545 (NanoQuCo-CM). T.R. further acknowledges support \tr{from the Ramón y Cajal program RYC2021-032473-I, financed by MCIN/AEI/10.13039/501100011033 and the European Union NextGenerationEU/PRTR.}

\appendix

\section{Hamiltonian of the Josephson junction array}\label{TotalH}

In this Appendix, we derive the quantum Hamiltonian $H=H_{\rm JJA}+H_{\rm TL}$ of the JJA coupled to transmission lines, starting from the classical Lagrangian (see Sec.~\ref{app:classical_Ham}). We then quantize the excitations in the system and derive the quantum Hamiltonians in Eqs.~(\ref{Hexpand})-(\ref{TRHam}) of the main text (see Sec.~\ref{app:quantum_HamJJA}).

\subsection{Classical Hamiltonian}\label{app:classical_Ham}

We determine the total Hamiltonian of the system by a standard Legendre transformation, $H = \sum_j q_j \dot{\phi}_j - {\cal L}$, with the canonical charge variables $q_j=\partial {\cal L}/\partial \dot{\phi}_j$ given by
\begin{align}
    q_j ={}&C^j_{\rm eq}\dot{\phi}_j-C'_{j-1}\dot{\phi}_{j-1}-C'_{j}\dot{\phi}_{j+1}-C^j_{\rm TL}V_{\rm TL}^j(t).\label{qjJJA}
\end{align}
Here, $\delta_{jl}$ denotes the Kronecker delta, with $j,l=1,\dots,N$, and the equivalent total capacitance $C_{\rm eq}^j$ at site $j$, is given in Eq.~(\ref{Ceq}). The resulting Hamiltonian can be decomposed as $H=H_{\rm JJA}+H_{\rm TL}$, where the individual Hamiltonians in the limit of low coupling capacitances $C'_j, C^j_{\rm TL}\ll C^j_{\rm eq}$ take the form, 
\begin{align}
    H_{\rm JJA}={}&\sum_{j=1}^N\frac{1}{2C^j_{\rm eq}}q_j^2+\sum_{j=1}^{N-1}\frac{C'_j}{C^j_{\rm eq}C^{j+1}_{\rm eq}}q_jq_{j+1}\label{Htot}\\
    -{}&\sum_{j=1}^N E_{J}^j\cos\left(\frac{\phi_j}{\Phi_0}\right)-\sum_{j=0}^{N}E'_{J}{}^j\cos\left(\frac{\phi_{j+1}-\phi_j}{\Phi_0}\right),\nonumber\\
    H_{\rm TL}={}&\sum_{j=1}^N\frac{C^j_{\rm TL}}{C^j_{\rm eq}}q_j V_{\rm TL}^j(t).\label{TRHam2}
\end{align}
The Hamiltonian $H_{\rm TL}$ of the coupling to transmission lines is obtained by further neglecting the terms $\sim (V_{\rm TL}^j)^2$ as these do not couple dynamical variables of the JJA. Notice that Eq.~(\ref{Htot}) considers the possibility of inhomogeneous Josephson energies $E_J^j$, $E'_J{}^j$ in addition to inhomogeneous $C_j$, $C'_j$, and $C_{\rm TL}^j$. Although this is not required in our setup, having the most general expressions is useful if one prefers to taper the frequencies via inductances instead of capacitances as explained in Sec.~\ref{Sec:Stabilization_inhomogeneities}.

The Hamiltonian of the JJA can be further simplified in the limit of low phase drop along the JJs, i.e. $\langle \phi_j^2 \rangle\ll \Phi_0^2$ such that it is a good approximation to expand the cosine potentials in Eq.~(\ref{Htot}) as $\cos(x)=1-x^2/2+x^4/24+{\cal O}(x^6)$. In this way, the Hamiltonian of the JJA up to the fourth order in the flux reads
\begin{align}
    H_{\rm JJA}={}&\sum_{j=1}^N\frac{1}{2C^j_{\rm eq}}q_j^2+\sum_{j=1}^{N-1}\frac{C'_j}{C^j_{\rm eq}C^{j+1}_{\rm eq}}q_jq_{j+1}\label{Hexpand2}\\
    +{}&\sum_{j=1}^N\frac{1}{2L^j_{\rm eq}}\phi_j^2-\sum_{j=1}^{N-1}\frac{1}{L'_{J}{}^j}\phi_j\phi_{j+1}\nonumber\\
    -{}&\sum_{j=1}^N \frac{(E_J^j+E'_J{}^j+E'_J{}^{j-1})}{24(\Phi_0)^4}\phi_j^4-\sum_{j=1}^{N-1}\frac{E'_{J}{}^j}{4(\Phi_0)^4}\phi_{j+1}^2\phi_j^2\nonumber\\
    +{}&\sum_{j=1}^{N-1}\frac{E'_J{}^j}{6(\Phi_0)^4}(\phi_{j+1}^3\phi_j+\phi_{j+1}\phi_j^3)+{\cal O}(\phi_j^6),\nonumber
\end{align}
where $L_J^j=(\Phi_0)^2/E_J^j$ and $L'_J{}^j=(\Phi_0)^2/E'_J{}^j$ are the Josephson inductances from on-site and inter-site JJs, respectively. Using these definitions, the inverse of the linear equivalent inductance at node $j$ of the circuit reads 
\begin{align}
1/L^j_{\rm eq}=1/L_{J}^j+1/L'_{J}{}^j+1/L'_{J}{}^{j-1}.\label{LeqInhomo}
\end{align}
Notice that equations (\ref{Hexpand2})-(\ref{LeqInhomo}) reduce to Eqs.~(\ref{Hexpand}) and (\ref{Leq}) of the main text, in the case of homogeneous Josephson inductances.

\subsection{Quantum Hamiltonian}\label{app:quantum_HamJJA}

Since the Hamiltonians are now expressed only in terms of conjugate dynamical variables $\phi_j$ and $q_j$, we can apply the standard quantization procedure, which consists in promoting flux $\phi_j$ and charge $q_j$ variables to operators that satisfy canonical commutation relations, namely $[\phi_j,q_l]=i\hbar \delta_{jl}$, for $j,l=1,\dots, N$. Note that in Eq.~(\ref{Htot}) we have $\phi_0=\phi_{N+1}=0$ so they are not dynamic variables (no associated charge).

The quadratic local part $H_{\rm JJA}^{(0)}=(1/2)\sum_{j=1}^N(q_j^2/C^j_{\rm eq}+\phi_j^2/L^j_{\rm eq})$ in Eq.~(\ref{Hexpand}) or (\ref{Hexpand2}) can be diagonalized using the standard quantum harmonic oscillator procedure \cite{blais_circuit_2021} as $H_{\rm JJA}^{(0)}=\sum_{j=1}^N\hbar\omega_b^ja_j^\dag a_j$. Here, $\omega_b^j = (L_{\rm eq}^j C_{\rm eq}^j)^{-1/2}$ is the bare frequency of each local oscillator, and $a_j^\dag$, $a_j$ are ladder operators satisfying canonical commutation relations $[a_j,a_l^\dag]=\delta_{jl}$. Conveniently, we can express flux $\phi_j$ and charge $q_j$ operators in terms of $a_j$ and $a_j^\dag$ as in Eqs.~(\ref{phia}) of the main text, and then $H_{\rm JJA}$ in Eq.~(\ref{Hexpand2}) can be recast as
\begin{align}
    \frac{H_{\rm JJA}}{\hbar} ={}& \sum_{j=1}^N \omega'_b{}^j a_j^\dag a_j - \sum_{j=1}^{N-1} J^j_L(a_{j+1}+a_{j+1}^\dag)(a_{j}+a_{j}^\dag)\nonumber\\
    -{}&\sum_{j=1}^{N-1} J^j_C(a_{j+1}-a_{j+1}^\dag)(a_{j}-a_{j}^\dag)-\sum_{j=1}^N\frac{K^j_s}{12}(a_j + a_j^\dag)^4\nonumber\\ 
    -{}&\sum_{j=1}^{N-1}\frac{K^j_c}{4} (a_{j}+a_{j}^\dag)^2(a_{j+1}+a_{j+1}^\dag)^2,\nonumber\\
    +{}&\sum_{j=1}^{N-1} \frac{T_j}{6}(a_{j+1}+a_{j+1}^\dag)^3(a_{j}+a_{j}^\dag)\nonumber\\
    +{}&\sum_{j=1}^{N-1} \frac{T_j'}{6}(a_{j}+a_{j}^\dag)^3(a_{j+1}+a_{j+1}^\dag)\label{H1}.
\end{align}
Here, the linear inductive $J_L^j$ and capacitive $J_C^j$ couplings read
\begin{align}
    J_C^j ={}& \frac{C'_j(Z_jZ_{j+1})^{-1/2}}{2C_{\rm eq}^{j}C_{\rm eq}^{j+1}},\qquad J^j_L = \frac{(Z_jZ_{j+1})^{1/2}}{2L'_J{}^j}.
\end{align}
The non-linear self-Kerr and cross-Kerr couplings $K_s^j$, $K_c^j$, $T_j$, and $T_j'$ are given by 
\begin{align}
    K_s^j = {}&\frac{\hbar(E_{J}^j+E'_J{}^{j-1}+E'_{J}{}^j)}{8(\Phi_0)^4}(Z_j)^2,\\
    K_c^j = {}&\frac{\hbar E'_{J}{}^j}{4(\Phi_0)^4}Z_jZ_{j+1},\\
    T_j = {}& K_c^j \sqrt{Z_{j+1}/Z_j},\qquad T_j' = K_c^j \sqrt{Z_{j}/Z_{j+1}},
\end{align}
with $Z_j=\sqrt{L_{\rm eq}^j/C_{\rm eq}^j}$ the impedance at site $j$. Finally, to express $H_{\rm JJA}$ as in Eq.~(\ref{H2}) of the main text, we consider homogeneous Josephson energies $E_J^j=E_J$, $E_J'{}^j=E_J'$, and apply RWA, neglecting all highly off-resonant processes provided $J_L^j, J_C^j, K_s^j, K_c^j, T_j, T_j'\ll \omega_b^j$. The effective bare frequency $\omega_b^j$ appearing in Hamiltonian (\ref{H2}) includes a small renormalization from Kerr couplings, $\omega_b^j\rightarrow \omega_b^j - K^j_s - (K^j_c + K^{j-1}_c)/2$, as well as the total bare linear hopping $J_b^j$, which reads $J_b^j=J_C^j-J_L^j+(T_j+T_j')/2$.

\section{Modelization of coupling to transmission lines, dissipation, and input signal via input-output theory}\label{FulldissipationModel}

In this Appendix, we first derive the coherent pump coherent Hamiltonian $H_p$ and describe the effects of dissipation and coupling to the transmission line via input-output theory (see Sec.~\ref{app:transmission_lines}). We then show how to model a coherent input signal within the formalism (see Sec.~\ref{app:input_signal}). 

\subsection{Coherent pump Hamiltonian and dissipation via transmission lines}\label{app:transmission_lines}

To describe the effects of coherent pump and dissipation in the JJA, it is convenient to decompose the transmission line Hamiltonian $H_{\rm TL}$ in Eq.~(\ref{TRHam}) as $H_{\rm TL}=H_{p}+H_{\rm diss}$, where $H_p$ is associated to the classical part of the voltage $V_{\rm TL}^j$ in Eq.(\ref{Vpj}) and $H_{\rm diss}$ depends exclusively on the voltage vacuum fluctuations $\delta V_{\rm TL}^j$. 

On the one hand, the explicit form of $H_p$ is obtained when replacing Eqs.~(\ref{phia}) and the classical part of the voltage (\ref{Vpj}) into Eq.~(\ref{TRHam}), namely
\begin{align}
    H_p/\hbar = {}& 2i\sum_{j=1}^N\Omega_p^j\cos(\omega_p^j t+\theta_p^j)(a_j^\dag-a_j)\label{Hp}.
\end{align}
Here, the strength $\Omega_p^j$ of the pump at site $j$ is given in Eq.~(\ref{Omegap_Circuit}). Note that Eq.~(\ref{HpRWA}) of the main text is obtained when applying the RWA to Eq.~(\ref{Hp}), provided $\Omega_p^j\ll \omega_p,\omega_b^j$, and going to a rotating frame with the pump frequency $\omega_p$.

On the other hand, the dissipation Hamiltonian $H_{\rm diss}$ is obtained when replacing Eqs.~(\ref{phia}) and the quantum fluctuation part of the voltage (\ref{Vpj}) into Eq.~(\ref{TRHam}), namely
\begin{align}
H_{\rm diss} ={}&\sum_{j=1}^N\frac{C^j_p}{C^j_{\rm eq}}q_j \delta V_{\rm TL}^j.\label{interaction_Ham}
\end{align}
Assuming semi-infinite transmission lines, the quantum voltage fluctuations at site $j$ can be expressed as \cite{blais_circuit_2021},
\begin{align}
    \delta V_{\rm TL}^j ={}& i\int d\omega \sqrt{\frac{\hbar\omega Z_{\rm TL}}{4\pi}}(d_{j}{}^\dag(\omega)-d_{j}(\omega)).\label{quantFluct1}
\end{align}
Here, $Z_{\rm TL}$ is the impedance of the transmission lines, and $d_j(\omega)/d_j^\dag(\omega)$ are annihilation/creation operators for excitations of frequency $\omega$ propagating in the transmission line $j$, satisfying continuum canonical commutation relations, $[d_j(\omega),d_{l}{}^\dag(\omega')]=\delta_{jl}\delta(\omega-\omega')$.

We describe the driven-dissipative JJA as an open quantum system with weak coupling to the continuum of photons at each transmission line in Eq.~(\ref{interaction_Ham})-(\ref{quantFluct1}) as the environment. We describe the system-environment interaction in the Markov approximation and use standard input-output theory \cite{QuantumNoise}. This approximation requires the coupling to all transmission to be weak compared to the free transition energies. Following the standard procedure, we obtain quantum Langevin equations describing the coherent and dissipative processes taking place in the JJA, which read
\begin{align}
    \dot{a}_j={}&\frac{i}{\hbar}[H_{\rm JJA}+H_p,a_j]-\frac{\kappa_j}{2}a_j+\sqrt{\kappa_j}a_{\rm in}^j(t).\label{QLE1disp_App}
\end{align}
where $\kappa_j$ is the decay rate into the transmission line at site $j$, 
\begin{align}
    \kappa_j=(C_{\rm TL}^j/2C_{\rm eq}^j)^2(Z_{\rm TL}/Z_j)\omega_b^j.\label{kappaj_expression}
\end{align} 
In the main text, we assume a homogeneous decay $\kappa_j=\kappa$ to all transmission lines, being a special case of the above equations. The input operators $a_{\rm in}^j(t)$ describe the effect of quantum noise fluctuations entering the JJA at site $j$, which is non-zero even if these channels are in vacuum at zero temperature. The output fields $a_{\rm out}^j(t)$ describe the emission of microwave excitations in the outward direction of each pump channel $j$. They are related to the input operators and the local modes $a_j$ by standard input-output relations \cite{QuantumNoise}:
\begin{align}
    a^j_{\rm out}(t) ={}& a^j_{\rm in}(t) - \sqrt{\kappa_j} a_j(t),\quad j=1,\dots, N,
\end{align}
Input and output operators also satisfy canonical commutation relations $[a^j_{\rm in/out}(t),a^{l}_{\rm in/out}{}^\dag(t')]=\delta_{jl}\delta(t-t')$. 

\subsection{Coherent input signal}\label{app:input_signal}

Sending a coherent signal on transmission line $j=I$, correspond to applying an extra classical voltage $V_s(t)=A_s\cos(\omega_s t)$ on that line, namely
\begin{align}
    V^j_{\rm TL}(t) \rightarrow V^j_{\rm TL}(t) + \delta_{jI} V_s(t),\label{Vsignal}
\end{align}
with $A_s=\sqrt{2Z_{\rm TL}P_s}$ the voltage amplitude of the signal determined by its power $P_s$. Following the same treatment as for the pump, replacing this extra voltage in Eq.~(\ref{TRHam}) induces a coherent drive Hamiltonian for the signal of the form:
\begin{align}
\frac{H_s}{\hbar}= {}& 2i\Omega_s\cos(\omega_s t)(a_I^\dag-a_I)\label{Hs}\\
    \approx {}& i\Omega_s (e^{-i\omega_s t}a_I^\dag-e^{i\omega_s t}a_I),\label{HsRWAapp}
\end{align}
where in the last line we have applied the RWA provided the signal is weak, $\Omega_s\ll \omega_s,\omega_b^j$. The strength of the signal reads 
\begin{align}
\Omega_s =(C_{\rm TL}^I/2C_{\rm eq}^I)[(Z_{\rm TR}/Z_I)P_s/\hbar)]^{1/2}.\label{Omegas_expression}
\end{align}
Instead of adding the signal Hamiltonian $H_s$ to the equations of motion (\ref{QLE1disp_App}) or (\ref{QLE1disp}) in the main text, it is convenient to apply a displacement transformation so that the signal is absorbed in the input operator as
\begin{align}
    a^j_{\rm in}(t) = \langle a^j_{\rm in}(t) \rangle + \delta a^j_{\rm in}(t).\label{decompdin}
\end{align}
Here, $\delta a^j_{\rm in}(t)$ is the vacuum noise contribution, and the non-zero coherent part depends on the signal strength $\Omega_s$ as, 
\begin{align}
    \langle a^j_{\rm in}(t) \rangle = \delta_{jI} \frac{\Omega_s}{\sqrt{\kappa_I}}e^{-i\omega_s t},\label{signal_input_field_app}
\end{align}
Finally, writing Eq.~(\ref{signal_input_field_app}) in a rotating frame with the pump frequency $\omega_p$, re-expressing $\Omega_s$ and $\kappa_I$ via Eqs.~(\ref{kappaj_expression}) and (\ref{Omegas_expression}), and approximating $\omega_b^j\approx  \omega_s$ due to the RWA, we obtain Eq.~(\ref{inputsignal}) of the main text for $a^j_{\rm in}(t)$ in terms of the signal input flux, $|\alpha_{\rm sig}|^2 = P_s/\hbar\omega_s$. 

\section{Control of saturation via sub-arrays of JJs in series}\label{ControlSaturation}

In this Appendix, we explain a procedure to reduce the effective non-linearity of the JJA while keeping the same linear properties of the circuit, i.e. without changing the linear inductances. By doing so, we can effectively reduce the saturation effects in the device and populate it with a larger number of photons. This method has already been experimentally demonstrated, see for instance \cite{eichler_controlling_2014,eichler_quantum-limited_2014,planat_understanding_2019}.

\begin{figure}
    \centering
    \includegraphics[width=0.8\columnwidth]{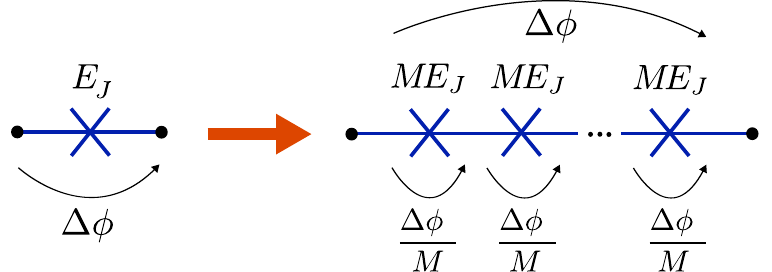}
    \caption{Control of saturation in the topological parametric amplifier array via sub-arrays of $M$ identical JJs in series with $M$ times larger Josephson energy. The phase drop $\Delta\phi$ on each JJ is reduced by $M$.}
    \label{fig:control_saturation}
\end{figure}

As sketched in Fig.~\ref{fig:control_saturation}, the idea consists of replacing all JJs in the setup by a sub-array of $M$ identical JJs in series, with Josephson energies $M$ times larger. In the limit of low phase drop $\Delta \phi/\Phi_0\ll 1$, the Lagrangian ${\cal L}_{\rm JJ}=E_J\cos(\Delta \phi/\Phi_0)$ of a single JJ can be expanded as,
\begin{align}
{\cal L}_{\rm JJ} ={}& -\frac{E_J}{2} \left(\frac{\Delta \phi}{\Phi_0}\right)^2 \!+ \frac{E_J}{24}\left(\frac{\Delta\phi}{\Phi_0}\right)^4 \!+ {\cal O}^6.\label{LagJJ_App}
\end{align}
For $M$ of the above JJs in series with $ME_J$ Josephson energy, the flux drop per element reduces to $\Delta\phi\rightarrow\Delta\phi/M$ and the resulting Lagrangian reads,
\begin{align}
{\cal L}_{\rm JJ}^{(M)} ={}& -\sum_{i=1}^M \left[\frac{ME_J}{2} \left(\frac{\Delta \phi}{M}\right)^2 \!+ \frac{ME_J}{24} \left(\frac{\Delta\phi}{M}\right)^4 \!+ {\cal O}^6\right]\nonumber\\
={}&-\frac{E_J}{2} \left(\frac{\Delta \phi}{\Phi_0}\right)^2 \! + \frac{E_J}{24M^2} \left(\frac{\Delta\phi_j}{M}\right)^4 + {\cal O}^6.
\end{align}
We see that the quadratic term is unchanged and thus the linear Josephson inductance $L_J=(\Phi_0)^2/E_J$ of the device is the same as for a single JJ. However, the quartic non-linearity is reduced by $M^2$, leading to a quadratic reduction of the Kerr-nonlinearity $K\rightarrow K/M^2$.

As discussed in Sec.~\ref{sec:saturation_main}, if we replace all JJs of the setup with these sub-arrays of JJs in series, the $M$ times smaller non-linearity increases the bound on the maximum number of photons that the JJA can sustain at each site $j$, obtaining $|\alpha_{\rm ss}^j|^2\ll M^2(\hbar\omega_b^jC_{\rm eq}^j)/(8e^2)$. This increases the dynamic range of the amplifier since it can sustain $M^2$ more noise photons and larger power of the signal and still have negligible saturation, $\langle \delta a_j^\dag \delta a_j\rangle / |\alpha_{\rm ss}^j|^2\ll 1$. Since $K_{c,s}^j$ reduce by a factor $M^2$, but $|\alpha_{\rm ss}^j|^2$ increase by the same factor, the effective squeezing terms $g_c$ and $g_s^j$ in Eqs.~(\ref{effectivegc})-(\ref{effectivegs}) are unchanged. Notice that increasing $|\alpha_{\rm ss}^j|^2$ by factor $M^2$ requires increasing the pump strength as $\Omega_p^j\rightarrow M\Omega_p^j$.

\section{Duffing oscillator equation for stabilizing a quasi-homogeneous steady-state compatible with topological amplification}\label{central_parameters}

In this Appendix, we show that a fully homogeneous array with open boundary conditions does not admit a homogeneous solution of the mean-field displacements due to boundary effects (see Sec.~\ref{sec:no_hom_sol}). However, as discussed in the main text, it is possible to engineer the boundaries to induce a quasi-homogenous steady-state configuration. For this configuration, we show how to solve the third-order non-linear Duffing equation at the homogeneous central region of the JJA. (see Sec.~\ref{sec:sol_duffing}). Finally, we show how to obtain the circuit parameters that fulfill target effective values compatible with topological amplification (see Sec.~\ref{sec:inverse_procedure}).  

\subsection{Fully homogenous steady-state solution is prohibited by boundary effects}\label{sec:no_hom_sol}

First, to look for steady-state solutions we set $\dot{\alpha_j}=0$ on the left-hand side of Eq.~(\ref{classicalTime}). In addition, we impose the homogeneous condition by using a plane-wave ansatz,
\begin{align}
    \alpha_{\rm ss}^j = |\bar{\alpha}_{\rm ss}| e^{-i\bar{\varphi} j},\label{planewave}
\end{align}
obtaining a set of $N$ independent nonlinear algebraic equations:
\begin{align}
    {}&\left\lbrace\kappa/2-\Im(X_j)+\Im(Y_j)|\bar{\alpha}_{\rm ss}|^2\right.\label{SSgeneral}\\
    {}&\left.-i(\Re(X_j)+\Re(Y_j)|\bar{\alpha}_{\rm ss}|^2)\right\rbrace |\bar{\alpha}_{\rm ss}|e^{-i\varphi j}= \bar{\Omega}_p e^{-i\theta^j_p}.\nonumber
\end{align}
First, note that setting $\theta_p^j = \bar{\varphi} j$ indeed allows one to imprint this same phase on the steady-state solution since this global phase can be canceled at both sides of all equations. We are thus left with $N$ equations for the unique amplitude $|\bar{\alpha}_{\rm ss}|$, all of which need to be simultaneously satisfied. The coefficients $X_j$ and $Y_j$ in Eqs.~(\ref{SSgeneral}) take three different values depending if they are evaluated at the left boundary $j=1$, right boundary $j=N$, or in the central region $1<j<N$. Explicitly they are given by 
\begin{align}
X_j={}&
    \bar{\Delta}_b-\bar{J}_b\begin{cases}
        e^{-i\bar{\varphi}}, & j = 1\\
        e^{i\bar{\varphi}}, & j = N\\
        2\cos(\bar{\varphi}), & \text{else }
    \end{cases}\\
Y_j={}&\bar{K}_s + \bar{K}_c
    \begin{cases}
        1+e^{-2i\bar{\varphi}}/2-3e^{-i\bar{\varphi}}/2-e^{i\bar{\varphi}}/2, & j = 1\\
        1+e^{2i\bar{\varphi}}/2-3e^{i\bar{\varphi}}/2-e^{-i\bar{\varphi}}/2, & j = N\\
        2+\cos(2\bar{\varphi})-4\cos(\bar{\varphi}), & \text{else }
    \end{cases}
\end{align}
Notice that these three different values come from the fact that sites at the boundary, $j=1$ and $j=N$, have only one neighbor, while all other sites in the bulk have two. Consequently, Eqs.~(\ref{SSgeneral}) reduce to three different equations which can only be simultaneously satisfied when $\bar{J}_b=0$ and $\bar{K}_c=0$. However, in that case, $\bar{g}_c=\bar{J}=0$ and the effective parametric amplifier array model in Eqs.~(\ref{QLE2})-(\ref{effH}) cannot reach a topological amplifying steady state.

The only way that all equations (\ref{SSgeneral}) are equal is to have a closed loop configuration, i.e. periodic boundary conditions in the system instead of an open chain configuration, but in this steady-state topological amplification is unstable as the number of photons grows without limit. Consequently, a stable realization of the device must necessarily have open boundaries and be inhomogeneous in the circuit and controls as discussed in Sec.~\ref{Sec:Stabilization_inhomogeneities}.

\subsection{Solving the Duffing equation to estimate circuit parameters at central region}\label{sec:sol_duffing}

In Sec.~\ref{Sec:Stabilization_inhomogeneities} of the main text, we showed that the estimation $|\bar{\alpha}_{\rm ss}|$ for the homogeneous steady-state displacement at the central region must satisfy the Duffing equation (\ref{SSperiodic}), which is obtained when considering periodic boundary conditions in Eq.~(\ref{SSgeneral}). In this subsection, we show how to solve this equation to obtain a solution compatible with topological amplification.

The most important constraint is that the mean-field solution $|\bar{\alpha}_{\rm ss}|$ should originate the desired value of effective detuning $\bar{\Delta}$ in Eq.~(\ref{DeltaHom}) so that the quantum fluctuations with dynamics given by Eqs.~(\ref{QLE2}) can reach a topological steady-state. To incorporate this extra constraint, we replace $\bar{\Delta}_b$ in Eq.~(\ref{SSperiodic}) by 
\begin{align}
 \bar{\Delta}_b ={}& \bar{\Delta} - 2(\bar{K}_s+\bar{K}_c[1-2\cos(\bar{\varphi})])|\bar{\alpha}_{\rm ss}|^2.\label{eq:inverse_deltab}
\end{align}
In this way, we obtain a new Duffing equation of the same form:
\begin{align}
    \left\lbrace\kappa/2-i(\bar{X}'-\bar{Y}'|\bar{\alpha}_{\rm ss}|^2)\right\rbrace |\bar{\alpha}_{\rm ss}|e^{-i\bar{\varphi} j} 
    = \bar{\Omega}_p e^{-i\theta^j_p},\label{SSperiodic_constraint} 
\end{align}
but with modified coefficients $\bar{X}' = \bar{\Delta} - 2\bar{J}_b\cos(\bar{\varphi})$, and $\bar{Y}' = \bar{K}_s - \bar{K}_c\cos(2\bar{\varphi})$.

Choosing $\theta_p^j = \bar{\varphi}j$, we can compensate for the phase dependence. We then square Eq.~(\ref{SSperiodic_constraint}) on both sites and manipulate them to obtain a third-order equation for the mean-field displacement, namely
\begin{align}
    n^3-2\xi n^2 + (\xi^2+1/4)n  = \eta,\label{thirdorderEq}
\end{align}
where $n=(\bar{Y}'/\kappa)|\bar{\alpha}_{\rm ss}|^2$ is the normalized displacement, $\eta = (\bar{\Omega}_p)^2\bar{Y}'/\kappa^3$ the normalized effective pump, and $\xi = \bar{X}'/\kappa$ the normalized effective detuning.

The third-order non-linear equation (\ref{thirdorderEq}) admits analytical solutions. In particular, for \tr{$\bar{\Delta}=0$ and $\bar{\varphi}=\pi/2$} as it is of interest in this work, we have $\tr{\xi=0}$, $|\tr{\bar{\alpha}_{\rm ss}}|^2=[\kappa/(\bar{K}_s+\bar{K}_c)]n$ and $\tr{\bar{\Delta}_b}=-2\kappa n$, \tr{with $n$ given by}
\begin{align}
n(\eta)=\frac{3^{1/3}(36 \eta + \sqrt{3 + 1296\eta^2})^{2/3}-3^{2/3}}{6(36 \eta + \sqrt{3 + 1296\eta^2})^{1/3}}.
\end{align}
In contrast to a single-site JPA \cite{eichler_controlling_2014}, the topological parametric amplifier array has no instability at $\eta>1/\sqrt{27}$. Therefore, quasi-homogeneous solutions with much larger $n(\eta)$ can be stabilized by engineering the suitable inhomogeneous boundaries as explained in Sec.~\ref{Sec:Stabilization_inhomogeneities}.

\subsection{Inverse procedure to determine the circuit parameters compatible with the effective amplifier array model}\label{sec:inverse_procedure}

In practice, to find suitable parameters at the center of the JJA we do not need to find an explicit solution for $|\bar{\alpha}_{\rm ss}|$, but we start from a target $|\bar{\alpha}_{\rm ss}|$ that we want to stabilize at the center, and use the Duffing equation (\ref{thirdorderEq}) to find the parameters of the system that are compatible with it. The steps of this inverse procedure are the following:
\begin{enumerate}
    \item We set $\bar{C}_{\rm eq}$ and $\bar{\omega}_b$ at the central region. This in turns determines $\bar{E}_C=e^2/(2\bar{C}_{\rm eq})$, $\bar{K}_s=\bar{E}_C/\hbar$, and $|\bar{\alpha}_{\rm ss}|$ by having the condition $|\bar{\alpha}_{\rm ss}|\ll \hbar\bar{\omega}_b/(16\hbar\bar{{E}}_C)$.
    \item Then, we target a solution with given effective parameters at the center: $|\bar{\alpha}_{\rm ss}|$, $\bar{\varphi}$, $\bar{\Delta}/\kappa$,  $\bar{J}/\kappa$,  $\bar{g}_c/\kappa$, and  $\bar{g}_s/\bar{g}_c$.
    \item We use Eqs.~(\ref{DeltaHom})-(\ref{gsHom}) to determine $\bar{K}_c$, $\bar{g}_c$, $\bar{g}_s$, $\kappa$, $\bar{\Delta}$, $\bar{J}$, and $\bar{J}_b$.
    \item We compute $\bar{X}' = \bar{\Delta} - 2\bar{J}_b\cos(\bar{\varphi})$, $\bar{Y}' = \bar{K}_s - \bar{K}_c\cos(2\bar{\varphi})$, $n=(\bar{Y}'/\kappa)|\bar{\alpha}_{\rm ss}|^2$, and $\xi = \bar{X}'/\kappa$. Then, we use the Duffing Eq.~(\ref{thirdorderEq}) to determine $\eta$.
    \item We calculate $\bar{\Omega}_p = \sqrt{\eta \kappa^3/ \bar{Y}'}$ and determine $\bar{\Delta}_b$ via Eq.~(\ref{eq:inverse_deltab}).
    \item Finally, we use Eqs.~(\ref{omegabFull}), (\ref{Ks_full}), (\ref{bareHopping}), (\ref{fullKc}), (\ref{Omegap_Circuit}), and (\ref{kappap}), evaluated in the homogenous region, to solve for circuit parameters $\bar{C}$, $\bar{C}'$, $\bar{C}_{\rm TL}$, $L_J$, $L_J'$, $\bar{P}_p$, and $\bar{\omega}_p$.
\end{enumerate}

This procedure determines univocally the circuit parameters at the homogenous central region. Then, we use the procedure detailed in Sec.~\ref{Sec:Stabilization_inhomogeneities} to engineer the parameters on the boundary regions and induce the target quasi-homogeneous steady-state configuration. Finally, we confirm that the system indeed stabilizes to this unique target steady state by evolving the full numerical non-linear mean-field equations (\ref{classicalTime}). This works, provided the boundary effects are properly suppressed as discussed in Sec.~\ref{Sec:Stabilization_inhomogeneities}.

\section{Distributing a coherent pump and 4WM via auxiliary array}\label{auxWaveguide}

In this Appendix, we derive the quantum Hamiltonian for the auxiliary waveguide, including the interaction with the JJA and the external transmission lines to provide pump and dissipation (see Sec.~\ref{app:Ham_Aux}). We then describe details to perform adiabatic elimination of the waveguide fluctuations, obtaining effective linearized equations for the JJA including residual effects of the auxiliary array (see Sec.~\ref{app:adiabatic_elimination}). Finally, we solve for the non-linear mean-field coupled dynamics of JJA and waveguide including non-RWA terms, and show that one obtains quantitatively the same results when slightly tuning the waveguide driving $\Omega_{\rm pw}$ (see Sec.~\ref{SimNonRWA}). 

\subsection{Hamiltonian for auxiliary waveguide}\label{app:Ham_Aux}

Following the same procedure as for the JJ array, one can obtain the total Hamiltonian of the design with an auxiliary waveguide from the Legendre transformation, $H'=\sum_j(\dot{\phi}_jq_j+\dot{\psi}_jq^j_{\rm w})-{\cal L}'$. Here, the modified total Lagrangian reads ${\cal L}'={\cal L}_{\rm JJA}+{\cal L}_{\rm w}+{\cal L}_{\rm TL}'+{\cal L}_{\rm pw}$ with each term defined in the main text, whereas $\phi_j$, $q_j=\frac{\partial {\cal L}'}{\partial \dot{\phi}_j}$, and $\psi_j$, $q^j_{\rm w}=\frac{\partial {\cal L}'}{\partial \dot{\psi}_j}$ are the flux and charge variables for the JJA and waveguide, respectively. We decompose the resulting Hamiltonian as $H'=H_{\rm JJ}+H_{\rm w}+H_{\rm TL}'$, where $H_{\rm JJA}$ is as given in Eq.(\ref{Htot}), but the new Hamiltonian $H_{\rm w}$ describing the waveguide and its coupling to the JJ array as well as $H_{\rm TL}'$ describing the coupling to the transmission lines and voltage sources read
\begin{align}
H_{\rm w}={}&\sum_{j=1}^N\frac{1}{2C^j_{\rm eq,w}}(q^j_{\rm w})^2+\sum_{j=1}^N\frac{1}{2L_{\rm w} }\psi_j^2\label{Hwaveguide}\\
+{}&\sum_{j=1}^{N-1}\frac{{C}_{\rm w}'}{C^j_{\rm eq,w}C^{j+1}_{\rm eq,w}}q^j_{\rm w}q^{j+1}_{\rm w}+\sum_{j=1}^{N}\frac{C''_j}{C^j_{\rm eq}C^{j}_{\rm eq,w}}q^jq^{j}_{\rm w},\nonumber\\
H_{\rm TL}'={}&\sum_{j=1,N}\frac{C_{\rm pw}}{C^j_{\rm eq,w}}q^j_{\rm w} V_{\rm w}^j(t)+\sum_{j}\frac{C^j_{\rm TL}}{C^j_{\rm eq}}q_j V_{\rm TL}^j(t).\label{HTLprime}
\end{align}
Notice that in the setup with local resistors and JJA transmission lines only at sites $j=I$ and $j=O$, we have $C^j_{\rm TL}=\delta_{jI}C^I_{\rm TL}+\delta_{jO}C^O_{\rm TL}$, and otherwise these couplings are non-zero at all sites. In addition, notice that we have neglected long-range couplings provided the coupling capacitances are much weaker than the total equivalent capacitance per site, i.e. $C_{\rm w}', C_j'', C_{\rm pw}, C_{\rm TL}^j\ll C^j_{\rm eq,w}, C^j_{\rm eq}$. The charge variable $q^j_{\rm w}$ at site $j$ of the auxiliary waveguide reads,
\begin{align}
    q^j_{\rm w} ={}&\frac{\partial {\cal L}'}{\partial \dot{\psi}_j}=C^j_{\rm eq,w}\dot{\psi}_j-C'_{\rm w}\dot{\psi}_{j-1}-{C}'_{\rm w}\dot{\psi}_{j+1}\nonumber\\
    {}&-\delta_{j1}C_{\rm pw}V_p^1(t)-\delta_{jN}C_{\rm pw} V_p^N(t)-C_j''\dot{\phi}_j,
\end{align}
with the equivalent waveguide capacitance $C_{\rm eq,w}^j$ given by,
\begin{align}
    C^j_{\rm eq,w}={}&C_{\rm w}^j+2C_{\rm w}'+C_j''+\delta_{j1}C_{\rm wp}^1+\delta_{jN}C^N_{\rm wp}.\label{Ceqw}
\end{align}
The charge variable $q_j$ of the JJA is given in Eq.~(\ref{qjJJA}) when adding the extra term $-C_j''\dot{\psi}_j$. Applying the standard quantization procedure to the modified total Hamiltonian $H'$, we promote the waveguide charge and flux variables as operators and express them as
\begin{align}
    \psi_j={}&\sqrt{\frac{\hbar Z^j_{\rm w}}{2}}(b_j^\dag+b_j),\qquad q^j_{\rm w} =i\sqrt{\frac{\hbar }{2 Z^j_{\rm w}}}(b_j^\dag-b_j),\label{psi}
\end{align}
where $b_j$ and $b_j^\dag$ destroy and create an excitation at site $j$ of the waveguide, and they satisfy canonical commutation relations, $[b_j,b_{l}^\dag]=\delta_{jl}$ such that $[\psi_j,q^{l}_{\rm w}]=i\hbar\delta_{jl}$. In addition, the effective impedance at site $j$ reads $Z^j_{\rm w}=\sqrt{L_{\rm w}/C_{\rm eq,w}^j}$. In terms of the ladder operators $b_j$ and $a_j$ introduced before for the JJA, the Hamiltonian $H_{\rm w}$ in Eq.~(\ref{Hwaveguide}) takes the form,
\begin{align}
    \frac{H_{\rm w}}{\hbar} ={}& \sum_{j=1}^N \omega_{\rm w}^j b_j^\dag b_j + \sum_{j=1}^{N-1} J_{\rm w}^j (b_{j} + b_{j}^\dag)(b_{j+1}+b_{j+1}^\dag)\nonumber\\
{}&+\sum_{j=1}^N J'_{\rm w}{}^j(a_{j}+a_{j+1}^\dag)(b_{j}+b_{j+1}^\dag),\label{Hb_app}
\end{align}
where $\omega_{\rm w}^j=(C_{\rm eq,w}^{j}L_{\rm w})^{-1/2}$ are the local frequencies of the auxiliary resonators at site $j$. The waveguide Hamiltonian in Eq.~(\ref{Hb}) of the main text is obtained when applying RWA to Eq.~(\ref{Hb_app}), provided the couplings are much weaker than transition frequencies. Here, the intra- and inter-array coupling rates $J^j_{\rm w}$ and $J'_{\rm w}{}^j$ read
\begin{align}
J^j_{\rm w} =\frac{C'_{\rm w}(Z^j_{\rm w}Z^{j+1}_{\rm w})^{-1/2}}{2C_{\rm eq,w}^jC_{\rm eq,w}^{j+1}},\quad
J'_{\rm w}{}^j = \frac{C_j''(Z_jZ^{j}_{\rm w})^{-1/2}}{2C_{\rm eq}^jC_{\rm eq,w}^{j}},
\end{align}
which reduce to Eqs.~(\ref{Jwcouplings_hom}) of the main text in the case of homogeneous equivalent capacitance $C_{\rm eq,w}^j=C_{\rm eq}^{\rm w}$. From now on, we also consider this homogeneous case in this Appendix, which leads to homogeneous waveguide frequencies $\omega_{\rm w}^j=\omega_{\rm w}$ and couplings $J_{\rm w}^j=J_{\rm w}$. 

To model coherent driving and dissipation in the auxiliary waveguide, we further decompose the transmission line Hamiltonian as $H_{\rm TL}'=H_{\rm pw}+H_{\rm diss}'$. First, the Hamiltonian $H_{\rm pw}$ describing the coherent drive on the waveguide is obtained when replacing Eq.~(\ref{psi}) and the classical part of the waveguide voltage (\ref{pump_voltage}) in Eq.~(\ref{HTLprime}). Explicitly, we have 
\begin{align}
    H_{\rm pw}/\hbar = {}&2i\Omega_{\rm pw}\cos(\omega_p t)(b_1^\dag - b_1),\label{HpRWA2}
\end{align}
where $\Omega_{\rm pw}= (C_{\rm pw}/2C_{\rm eq}^{\rm w})\sqrt{(Z_{\rm pw}/Z_{\rm w})(P_{\rm pw}/\hbar)}$ is the waveguide driving strength on site $j=1$ and $Z_{\rm pw}$ is the impedance of the line. This Hamiltonian reduces to Eq.~(\ref{HpRWA2_main}) of the main text when applying the RWA, $\Omega_{\rm pw}\ll \omega_{\rm w}^j, \omega_b^j$, and going to a rotating frame with the pump frequency $\omega_p$.

The Hamiltonian describing the dissipation $H_{\rm diss}'$ contains the voltage quantum fluctuations $\delta V_{\rm w}^j(t)$ and $\delta V_{\rm TL}^j(t)$ from transmission lines coupled to waveguide and JJA, respectively. Assuming semi-infinite transmission lines, we can express the quantum fluctuations $\delta V_{\rm w}^j(t)$ and $\delta V_{\rm TL}^j(t)$ in terms of annihilation/creation operators of local environments in the same form of Eq.~(\ref{quantFluct1}). Applying standard Markovian input-output theory to the interaction Hamiltonian $H_{\rm diss}'$, we obtain modified quantum Langevin equations describing the coherent and dissipative processes that take place in the extended device including JJA and auxiliary waveguide. Explicitly, the coupled equations for $\dot{a}_j$ and $\dot{b}_j$ are given in Eqs.~(\ref{QLEw1})-(\ref{QLEw2}) of the main text, where coherent processes are described by the total Hamiltonian $H'=H_{\rm JJA}+H_{\rm w}+H_{\rm pw}$, whereas dissipation with rates $\kappa$ and $\kappa_{\rm w}^j$, given in Eqs.~(\ref{kappap}) and (\ref{kappa_waveguide}) for JJA and waveguide, respectively. 

Finally, the signal sent in port $j=I$ is modeled in the same form as in Eq.~(\ref{QLEw1}), performing the same displacement transformation in the noise operators as in Eq.~(\ref{inputsignal}) or (\ref{decompdin}).

\subsection{Adiabatic elimination of waveguide and effective pump}\label{app:adiabatic_elimination}

In this subsection, we obtain the analytical solution for the steady-state mean displacement $\beta_{\rm ss}^j$ of the auxiliary waveguide displacement in the case of perfect absorbing boundary conditions. We then show how to perform adiabatic elimination of the waveguide fluctuations $\delta b_j(t)$ to obtain effective linearized dynamics of the JJA fluctuations $\delta a_j(t)$. We also solve for the dynamics of the quantum fluctuations including the effect of $\gamma_j$, and we show this can be neglected provided $\gamma_j\ll \kappa$.

First, we recast the classical equation for $\dot{\beta}_j$ in Eq.~(\ref{betaEq}) as,
\begin{align}
       \dot{\beta}_j={}&-\sum_l {\cal I}_{jl}\beta_l+\Omega_{\rm pw}\delta_{j1}-i J'_{\rm w}{}^j\alpha_j,
\end{align}
with the components of the ${\cal I}$ matrix given by,
\begin{align}
    {\cal I}_{jl}=(\frac{\kappa_{\rm w}^j}{2}-i\Delta_{\rm w})\delta_{jl}+iJ_{\rm w}\delta_{l,j-1}+iJ_{\rm w}\delta_{l,j+1}.\label{Imatrix}
\end{align}
To look for the steady state, we set $\dot{\beta}_j=0$ and the resulting algebraic equation for $\beta_{\rm ss}^j$ can be formally solve as
\begin{align}
    \beta_{\rm ss}^j=[{\cal I}^{-1}]_{j1}\Omega_{\rm pw}-i\sum_{l=1}^N[{\cal I}^{-1}]_{jl}J'_{\rm w}{}^l\alpha^l_{\rm ss},\label{betaSS}
\end{align}
with $[{\cal I}^{-1}]_{jl}$ components of the inverse of the matrix ${\cal I}$. Remarkably, in the case $\Delta_{\rm w}^j=0$ and $\kappa_{\rm w}^j=2J_{\rm w}(\delta_{j1}+\delta_{jN})$, the inverse of matrix ${\cal I}$ has an exact solution \cite{ramos_non-markovian_2016}, which reads
\begin{align}
    [{\cal I}^{-1}]_{jl}=\frac{e^{-i(\pi/2)|j-l|}}{2J_{\rm w}}.\label{ExactSol}
\end{align}
Replacing (\ref{ExactSol}) into Eq.~(\ref{betaSS}), we see that under the above special conditions, the steady-state waveguide displacements correspond to a plane-wave with phase $\bar{\varphi}=\pi/2$, 
\begin{align}
    \beta_{\rm ss}^j=\frac{i\Omega_{\rm pw}}{2J_{\rm w}}e^{-i\bar{\varphi} j}-i\sum_{l=1}^N \frac{J'_{\rm w}{}^l}{2J_{\rm w}}e^{-i\bar{\varphi}|j-l|}\alpha^l_{\rm ss},\label{betaSS2}
\end{align}
up to a correction on order $\sim (J'_{\rm w}{}^l/2J_{\rm w})\alpha^l_{\rm ss}$, which is small in the strong pump and weak coupling limit, $\Omega_{\rm pw}\gg J_{\rm w}\gg J'_{\rm w}{}^j$ required for the four-wave mixing process, as discussed in the main text. The solution (\ref{betaSS2}) exists due to perfect impedance matching conditions and is robust to small deviations in the exact conditions.

To find the effective algebraic equations for the steady state JJA displacements $\alpha_{\rm ss}^j$, we set $\dot{\alpha}_j=0$ in Eq.~(\ref{classicalTime_coupled}) and replace expression (\ref{betaSS2}), obtaining  
\begin{align}
      {}&\left(\frac{\kappa}{2}-i\Delta_b^j\right)\alpha_j -iK^j_{s}|\alpha_j|^2\alpha_j+i(J_b^j\alpha_{j+1}+J_b^{j-1}\alpha_{j-1})\nonumber\\
    {}&- F_{c}^j(\alpha_j,\alpha_{j\pm 1})
    +\sum_{l=1}^N \sqrt{\gamma_j\gamma_l}e^{-i\bar{\varphi}|j-l|}\alpha^l_{\rm ss}=\Omega_p^j e^{-i\bar{\varphi} j},\label{alphaEq2}
\end{align}
with $\Omega_p^j=(J'_{\rm w}{}^j/2J_{\rm w})\Omega_{\rm pw}$ the effective pump induced by the auxiliary array on the JJA. The term $\sim \sqrt{\gamma_j\gamma_l}$ in Eq.~(\ref{alphaEq2}) describes collective decay with rate $\gamma_j=(J'_{\rm w}{}^j)^2/(2J_{\rm w})$ leading to long-range dissipative couplings between JJA sites. Fortunately, this effect is very small in the weak coupling regime we consider and we neglect it provided
$\gamma_j\ll\kappa, \Omega_p^j, |\Delta_b^j|, K_s^j|\alpha_{\rm ss}^j|^2,K_c^j|\alpha_{\rm ss}^j|^2$. In this case, we effectively obtain the same equation for $\alpha_{\rm ss}^j$ as without the waveguide given by setting $\dot{\alpha}_j=0$ in Eq.~(\ref{classicalTime}). Note that the presence of additional local decay on the bulk of the auxiliary waveguide, i.e. $\kappa_{\rm w}^j\neq 0$ for $1<j<N$, deteriorates the perfect absorbing conditions, but the topological steady state phase is robust as long as this unwanted decay is much smaller than $J_{\rm w}$.

Regarding the dynamics of the quantum fluctuations $\delta a_j$ and $\delta b_j$, we can also linearize the dynamics provided $|\alpha_{\rm ss}^j|^2\gg \langle \delta a_j^\dag \delta a_j\rangle$ as done for the design without auxiliary waveguide. Starting from Eqs.~(\ref{QLEw1})-(\ref{QLEw2}) and going to a rotating frame with the pump frequency $\omega_p^1$, the resulting coupled quantum Langevin equations for fluctuations in the JJ array and the waveguide read,
\begin{align}
    \delta\dot{a}_j={}&i[H_{\rm pa},\delta a_j]-iJ'_{\rm w}{}^j\delta b_j-\frac{\kappa}{2}\delta a_j+\sqrt{\kappa} a_{\rm in}^j(t),\label{lineardeltaa}\\
    \delta\dot{b}_j={}&-\sum_{l=1}^N{\cal I}_{jl}\delta b_l-iJ'_{\rm w}{}^j\delta a_j+\sqrt{\kappa_{\rm w}^j}b^{j}_{\rm in}(t),\label{lineardeltab}
\end{align}
with $H_{\rm pa}$ the parametric Hamiltonian given in Eq.~(\ref{effH}) and ${\cal I}_{jl}$ the matrix in Eq.~(\ref{Imatrix}). 

Under the conditions of perfect impedance matching discussed above, $\Delta_{\rm w}=0$ and $\kappa_{\rm w}^j=2J_{\rm w}(\delta_{j1}+\delta_{jN})$, the waveguide reaches its steady-state in a short time-scale and we can adiabatically eliminate the waveguide fluctuations $\delta b_j$. To do so, we set $\delta \dot{b}_j\approx 0$ in Eq.~(\ref{lineardeltab}) and using (\ref{ExactSol}) we formally solve for $\dot{b}_j$, obtaining,
\begin{align}
    \delta b_j(t)\approx{}&\!-i\sum_{l=1}^N \frac{J_{\rm w}'{}^l}{2J_{\rm w}}e^{-i\bar{\varphi}|j-l|}\delta a_l+\sum_{l=1,N}\frac{e^{-i\bar{\varphi}|j-l|}}{\sqrt{2J_{\rm w}}}b^{l}_{\rm in}(t).
\end{align}
Replacing this into Eq.(\ref{lineardeltaa}), we obtain
\begin{align}
    \delta\dot{a}_j={}&i[H_{\rm pa},\delta a_j]-\frac{\kappa}{2}\delta a_j+\sqrt{\kappa} a_{\rm in}^j(t)\label{lineardeltaa2}\\
    -{}&\sum_{l=1}^N \sqrt{\gamma_j\gamma_l}e^{-i\bar{\varphi}|j-l|}\delta a_l-i\sum_{l=1,N}\sqrt{\gamma_j}e^{-i\bar{\varphi}|j-l|}b_{\rm in}^{l}(t).\nonumber
\end{align}
The last line in Eq.~(\ref{lineardeltaa2}) corresponds to the standard terms for collective superradiant decay mediated by a waveguide. The decay rate $\gamma_j=(J_{\rm w}{}^j)^2/(2J_{\rm w})$ describes emission of excitations from the JJA at site $j$ to traveling waves with phase difference $\bar{\varphi}=\pi/2$ in the auxiliary array. These traveling waves can propagate to the right or the left, leaving the auxiliary array at sites $j=1$ or $j=N$, respectively. These two types of decay processes are described by noise input operators $b_{\rm in}^1(t)$ and $b_{\rm in}^N(t)$, respectively. These decay processes can be neglected provided $\gamma_j$ is much smaller than all other system quantities as done for the classical displacements in Eq.~(\ref{alphaEq2}). In this case, Eq.~(\ref{lineardeltaa2}) reduces to Eq.~(\ref{QLE2}) as there was no waveguide.

\subsection{Simulation of mean-field dynamics including non-counter-rotating terms}\label{SimNonRWA}

In the main text and in the adiabatic elimination Sec.~\ref{app:adiabatic_elimination}, we have applied the rotating waveguide approximation (RWA) to describe the dynamics of the JJA weakly coupled the waveguide, provided couplings, detunings, and pump strengths are much smaller than free frequencies, i.e.~$J_{\rm w}, \bar{J}_{\rm w}',\bar{\Delta}_b, \Omega_{\rm pw},  \ll \bar{\omega}_b,\omega_{\rm w},\omega_p$. Therefore, when having a large pump $\Omega_{\rm pw}\gtrsim \omega_{\rm w}$, this approximation may be not justified. However, in this subsection, we numerically solve for the fully coupled mean-field dynamics of the JJA-waveguide system and show that including the non-counter-rotating terms does not change the dynamics qualitatively, and when slightly adjusting the pump strength $\Omega_{\rm pw}$, one can achieve the same quasi-homogeneous steady-state configuration predicted by Eqs.~(\ref{classicalTime_coupled})-(\ref{betaEq}) (with RWA).

To solve for the full mean-field dynamics without RWA, we describe JJA-waveguide coupling and waveguide pump via the full Hamiltonians in Eqs.~(\ref{Hb_app}) and (\ref{HpRWA2}) instead of the approximated ones in Eqs.~(\ref{Hb}) and (\ref{HpRWA2_main}), respectively. This modifies Eqs.~(\ref{classicalTime_coupled})-(\ref{betaEq}) by adding fast rotating terms of the form $\sim \Omega_{\rm pw} e^{2i\omega_pt}$, making the dynamics explicitly time-dependent despite being in the rotating frame with the pump frequency $\omega_p$. 

\begin{figure}[t]
\centering
\includegraphics[width=\columnwidth]{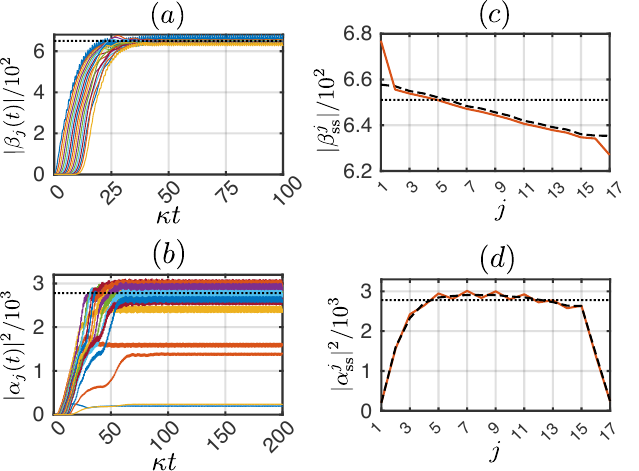}
\caption{Non-linear mean-field dynamics of JJA coupled to waveguide including non-counter-rotating terms. (a)-(b) Time evolution of (a) waveguide mean displacements $|\beta_j(t)|$ and (b) JJA mean occupations $|\alpha_j(t)|^2$, for the same parameters as in Figs.~\ref{Fig_inhomogeneities_auxiliary}(a)-(b), except for not doing RWA and using a smaller waveguide pump $\Omega_{\rm pw}=1.0078\Omega_{\rm pw}^{\rm theo}$. The dotted black lines indicate the steady-state estimations in the central region. (c)-(d) Spatial steady-state profiles for (c) waveguide $|\beta_{\rm ss}^j|$ and (d) JJA $|\alpha_{\rm ss}^j|^2$, obtained from the dynamics in (a)-(b) (red curves). For comparison, the dashed black line indicates the profiles obtained in Fig.~\ref{Fig_inhomogeneities_auxiliary}(e)-(f) for $\Omega_{\rm pw}=1.01\Omega_{\rm pw}^{\rm theo}$ and assuming RWA.}
\label{Fig_nonRWA}
\end{figure}

Figs.~\ref{Fig_nonRWA}(a)-(b) display the time evolution of the mean-field displacements for waveguide $|\beta_j(t)|$ and JJA array $|\alpha_j(t)|^2$ at all sites $j=1,\dots,N$, for the same parameters as in Figs.~\ref{Fig_inhomogeneities_auxiliary}(a)-(b), but including non-RWA terms and applying a smaller pump $\Omega_{\rm pw}=1.0078\Omega_{\rm pw}^{\rm theo}$. We observe that even in the presence of non-RWA terms, waveguide and JJA reach the desired quasi-homogeneous steady-state configuration at the central region (marked by dotted lines). The main difference is that there appears a small but fast oscillation of the occupations that persists in steady-state causing a slight increase of the inhomogeneities around the predicted values. To see this more precisely, Figs.~\ref{Fig_nonRWA}(c)-(d) compares the spatial dependence of the steady state profiles obtained without RWA (red) and with RWA (dashed black), for the same parameters as in Figs.~\ref{Fig_inhomogeneities_auxiliary}(e)-(f). We confirm the good agreement up to negligible inhomogeneities, provided the waveguide pump is properly calibrated. In this case: $\Omega_{\rm pw}=1.0078\Omega_{\rm pw}^{\rm theo}$ without RWA and $\Omega_{\rm pw}=1.01\Omega_{\rm pw}^{\rm theo}$ with RWA.

To avoid complicating the model unnecessarily, we thus use RWA approximation everywhere in the main text, but bearing in mind that small differences due to non-counter rotating terms or small decay $\gamma_j\ll \kappa$ induced by the waveguide-JJA interaction can be compensated by calibrating $\Omega_{\rm pw}$ at the end.

\section{Amplifier's properties from Green's functions}\label{app:amplifier_properties}

In this Appendix, we derive the expressions for the amplifier's properties such as gain, reverse gain, noise, and saturation, in terms of the Green's function matrix. We follow a procedure similar to Ref.~\cite{ramos_topological_2021}.

First, we look for the steady-state solution of the JJA fluctuation dynamics in Eq.~(\ref{FullQLangevin}). Since these equations are linear, they can be solved exactly in Fourier space.  Defining the Fourier transform of the fluctuation and input operators as $\delta a_j(\tilde{\omega})=(2\pi)^{-1/2}\int dt e^{i\tilde{\omega} t}\delta a_j(t)$ and $a^j_{\rm in}(\tilde{\omega})=(2\pi)^{-1/2}\int d\tilde{\omega} e^{i\tilde{\omega} t}\delta a^j_{\rm in}(t)$, we obtain
\begin{align}
\delta \vec{a}(\tilde{\omega}) = i\sqrt{\kappa} G(\tilde{\omega})\vec{a}_{\rm in}(\tilde{\omega}).\label{fourierVector}
\end{align}
Here, $G(\tilde{\omega})= (\tilde{\omega}-H_{\rm nh})^{-1}$ is the $2N\times 2N$ Green's function matrix, whereas $\delta \vec{a}(\tilde{\omega})=[\delta a_j(\tilde{\omega}),\delta a_j^\dag(-\tilde{\omega})]^T$ and $\vec{a}_{\rm in}(\tilde{\omega})=[a^j_{\rm in}(\tilde{\omega}),a^j_{\rm in}{}^\dag (-\tilde{\omega})]^T$ are the Fourier transformed Nambu vectors. Notice that the frequency variable $\tilde{\omega}$ is actually a detuning with respect to the pump's frequency, namely
\begin{align}
\tilde{\omega}=\omega-\omega_p.\label{omega_tilde}
\end{align}
In this Appendix, we express all Fourier quantities in terms of $\tilde{\omega}$ to simplify notation. When re-writing them in terms of the absolute frequency $\omega$, we obtain the expressions with the conventions of the main text. 

From Eq.~(\ref{fourierVector}), we can explicitly solve for frequency-resolved fluctuation operator $\delta a_j(\tilde{\omega})$ in terms of the Green's function components:
\begin{align}
    \delta a_j(\tilde{\omega}) ={}&i\sqrt{\kappa}\sum_l[G_{jl}(\tilde{\omega})a^l_{\rm in}(\tilde{\omega})+G_{j,N+l}(\tilde{\omega}) a^l_{\rm in}{}^{\dag}(-\tilde{\omega})].\label{formalSolution}
\end{align}
Notice that the signal contribution $\sim a^l_{\rm in}(\tilde{\omega})$ depends on the direct components, $G_{jl}(\tilde{\omega})$, while the idler contribution $a^l_{\rm in}{}^{\dag}(-\tilde{\omega})$ on the anomalous ones, $G_{j,N+l}(\tilde{\omega})$. In addition, recall that the Fourier input operators satisfy standard commutation relations, $[a^j_{\rm in}(\tilde{\omega}),a^{l}_{\rm in}{}^{\dag}(\tilde{\omega}')]= \delta_{jl}\delta (\tilde{\omega}-\tilde{\omega}')$.

The output operator $a_{\rm out}^j(t)$ describes the amplified field leaving the array at output port $j$. Using the input-output equation (\ref{dispInOut1}), the decomposition (\ref{displacement}), and absorbing the phases $a_{\rm out/in}^j(t)\rightarrow a_{\rm out/in}^j(t) e^{i\theta_{\rm ss}^j}$ as in the main text, the output field operator takes the form,
\begin{align}
a^j_{\rm out}(t) = a^j_{\rm in}(t)-\sqrt{\kappa}|\alpha_{\rm ss}^j|-\sqrt{\kappa}\delta a_j(t).
\end{align}
To obtain the frequency-resolved output operator $a^j_{\rm out}(\tilde{\omega})$ we apply Fourier transform on the above equation,
\begin{align}
a^j_{\rm out}(\tilde{\omega}) = a^j_{\rm in}(\tilde{\omega})-\sqrt{2\pi\kappa}|\alpha_{\rm ss}^j|\delta(\tilde{\omega})-\sqrt{\kappa}\delta a_j(\tilde{\omega}).\label{output_freq_resolved}
\end{align}
Replacing Eq.~(\ref{formalSolution}) in the above equation, we obtain Eq.~(\ref{eq:aout}) of the main text which gives the output fields as a result of the scattered input field, idler field, and the pump.

To characterize the amplifying properties at any output $j$ of the array, we evaluate the output field $a^j_{\rm out}(t)$ in the special case of a coherent input signal (\ref{inputsignal}) of frequency $\tilde{\omega}_s=\omega_s-\omega_p$ entering at site $j=I$. The Fourier input operator associated with this coherent signal reads, 
\begin{align}
a^j_{\rm in}(\tilde{\omega})=\sqrt{2\pi}\alpha_{\rm sig} \delta_{jI}\delta(\tilde{\omega}-\tilde{\omega}_s)+\delta a^j_{\rm in}(\tilde{\omega}),\label{inputnoise_Fourier}
\end{align}
where $\alpha_{\rm sig}$ is the coherent amplitude of the signal and $\delta a^j_{\rm in}(\tilde{\omega})$ the vacuum noise fluctuations at frequency $\tilde{\omega}$ and input channel $j$, satisfying $\braket{\delta a^j_{\rm in}(\tilde{\omega})\delta a^{l}_{\rm in}{}^{\dag}(\tilde{\omega}')}=\delta_{jl}\delta(\tilde{\omega}-\tilde{\omega}')$. Using the monochromatic signal input (\ref{inputnoise_Fourier}) in Eq.~(\ref{output_freq_resolved}) we obtain the corresponding output field, which can be conveniently decomposed into coherent and incoherent parts, $a^j_{\rm out}(\tilde{\omega})=\braket{a^j_{\rm out}(\tilde{\omega})}+\delta a^j_{\rm out}(\tilde{\omega})$. The coherent part has a monochromatic contribution at the signal, idler, and pump frequencies,  
\begin{align}
    \langle a^j_{\rm out}(\tilde{\omega}) \rangle =&{} [\delta_{jI}-i\kappa G_{jI}(\tilde{\omega}_s)]\sqrt{2\pi}\alpha_{\rm sig} \delta(\tilde{\omega}-\tilde{\omega}_s) \nonumber\\
    {}&-i\kappa G_{j,N+I}(-\tilde{\omega}_s)\sqrt{2\pi}\alpha_{\rm sig}^\ast \delta(\tilde{\omega}+\tilde{\omega}_s)\nonumber\\
    {}&+\sqrt{2\pi\kappa}|\alpha_{\rm ss}^j|\delta(\tilde{\omega}).\label{OutputcoherentFreq}
\end{align}
The incoherent part, $\delta a_{\rm out}^j(\tilde{\omega})= a_{\rm out}^j(\tilde{\omega})-\langle a^j_{\rm out}(\tilde{\omega}) \rangle$, describes noise added by the amplifier at frequency $\tilde{\omega}$ and has the same form of Eq.~(\ref{formalSolution}), replacing $a^j_{\rm in}(\tilde{\omega})\rightarrow\delta a^j_{\rm in}(\tilde{\omega})$.

Using an inverse Fourier transform, we obtain the desired output field $a^j_{\rm out}(t)=\braket{a^j_{\rm out}(t)}+\delta a^j_{\rm out}(t)$, which also decomposes into coherent and incoherent contributions. The total output photon flux $F_{\rm out}^j=\langle a_{\rm out}^j{}^\dag(t) a_{\rm out}^j(t)\rangle$ at channel $j$ then reads \cite{ramos_topological_2021}:
\begin{align}
    F_{\rm out}^j= {\cal G}_j(\tilde{\omega}_s) |\alpha_{\rm sig}|^2 + N^j_{\rm out}.\label{output_flux_gain}
\end{align}
Here, ${\cal G}_j(\tilde{\omega}_s)=|\delta_{jI}-i\kappa G_{jI}(\tilde{\omega}_s)|^2$ is the amplifier's gain, which reduces to Eq.~(\ref{gainjI}) of the main text when $j\neq I$. In addition, $N^j_{\rm out}=\langle \delta a_{\rm out}^j{}^\dag(t)\delta a_{\rm out}^j(t)\rangle$ is the noise flux generated by the amplifier, which can be computed as
\begin{align}
    N^j_{\rm out} ={}&\frac{1}{2\pi}\iint d\tilde{\omega} d\tilde{\omega}' e^{i(\tilde{\omega}-\tilde{\omega}')t} \braket{\delta a_j^{\rm out}{}^\dag(\tilde{\omega})\delta a_j^{\rm out}(\tilde{\omega}')}\nonumber\\
    ={}&\frac{1}{2\pi}\int d\tilde{\omega} n^j_{\rm amp}(\tilde{\omega}),
\end{align}
with $n^j_{\rm amp}(\tilde{\omega}) = \kappa^2 \sum_{l=1}^N |G_{j,N+l}(\tilde{\omega})|^2$ the number of incoherent noise photons generated by the amplifier per unit frequency $\tilde{\omega}$. The added noise $n_j^{\rm add}(\tilde{\omega})=n_{\rm amp}^j(\tilde{\omega})/{\cal G}_j(\tilde{\omega})$ is defined as the noise normalized by the gain, and it is lower bounded by $1$ as discussed in the main text.

The reverse gain ${\cal G}_j^{(R)}(\tilde{\omega}_s)$ is obtained as in Eq.~(\ref{output_flux_gain}) but when sending the signal in the opposite direction, i.e.~entering at site $j>I$ and retrieving it at site $j=I$. The result is given in Eq.(\ref{reversegainjI}) of the main text, for $j\neq I$. Analogously, the gain and reverse gain of the idler field is obtained when measuring the output field at frequency $\tilde{\omega}=-\tilde{\omega}_s$, obtaining 
\begin{align}
   {\cal G}^{(I)}_j = {}&\kappa^2 |G_{j,N+I}(-\tilde{\omega}_s)|^2,\\
   {\cal G}^{(I,R)}_j ={}& \kappa^2 |G_{I,N+j}(-\tilde{\omega}_s)|^2.\label{GainIdler}
\end{align}

Besides characterizing quantities associated with the output field, we can also use the Green's function matrix to calculate quantities of the JJA such as the average occupation of the quantum fluctuations $\langle \delta a_j^\dag(t) \delta a_j(t)\rangle$, as well as correlations of flux $\phi_j(t)$ or charge $q_j(t)$ operators. Important to estimate the saturation of the amplifier, we compute
\begin{align}
    \langle \delta a_j^\dag(t) \delta a_j(t)\rangle ={}&\frac{1}{2\pi}\iint d\tilde{\omega} d\tilde{\omega}' e^{i(\tilde{\omega}-\tilde{\omega}')t}\langle \delta a_j^\dag(\tilde{\omega}) \delta a_j(\tilde{\omega}') \rangle\nonumber\\
    ={}&\frac{1}{2\pi}\int d\tilde{\omega} n_{\rm amp}^j + n_{\rm sig}^j(\tilde{\omega}_s),\label{Occupation_Fluctuation_j}
\end{align}
which should satisfy $\langle \delta a_j^\dag(t)\delta a_j(t)\rangle \ll |\alpha_{\rm ss}^j|^2$ to validate the operation of the JJA as a linear amplifier array. Note that the total occupation at site $j$ depends on the integrated amplifier noise $n_{\rm amp}^j$ at that site, as well as on a coherent contribution $n_{\rm sig}^j(\tilde{\omega}_s)$ that grows with the signal amplitude $\alpha_{\rm sig}$ as
\begin{align}
    n_{\rm sig}^j(\tilde{\omega})={}&\kappa|\alpha_{\rm sig}|^2(|G_{jI}(\tilde{\omega}_s)|^2 + |G_{j,N+I}(-\tilde{\omega}_s)|^2 )\\
    +{}&\kappa(\alpha_{\rm sig}^2 e^{2i\theta_{\rm ss}^I}G_{jI}(\tilde{\omega}_s)G_{j,N+I}^{\ast}(-\tilde{\omega}_s)e^{-2i\tilde{\omega}_s t}+{\rm h.c.})\nonumber.
\end{align}

\begin{figure*}
    \centering
    \includegraphics[width=\textwidth]{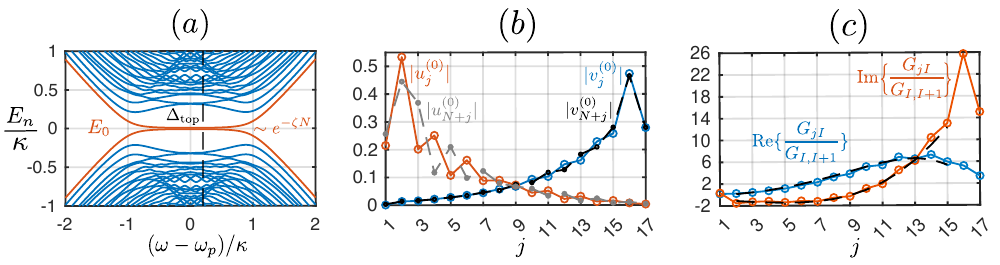}
    \caption{Topological Interpretation: Connection between edge states of $\cal{H}(\omega)$ and exponential dependence of Green's functions. (a) Eigen-spectrum $E_n$ of the extended Hamiltonian ${\cal H}(\omega)$ in Eq.~(\ref{Hext}) as function of frequency $\omega$. In a topologically non-trivial steady state, $E_n(\omega)$ manifests pairs of zero-energy modes $E_0(\omega) \sim \pm e^{-N\zeta(\omega)}$ (in red) within a topological band-gap. \tr{The vertical dashed line corresponds to the frequency $\omega=\tr{\omega_p+0.2\kappa}$ at which panels (b)-(c) are evaluated.} (b) Absolute value of edge states $|u_j^{(0)}|$ and $|u_{N+j}^{(0)}|$ (red) and $|v_j^{(0)}|$ and $|v_{N+j}^{(0)}|$ (blue) as function of array site $j$ and for given frequency $\omega=\tr{\omega_p+0.2\kappa}$. (c) Real part \tr{(blue)} and imaginary part \tr{(red)} of the normalized Green`s function \tr{$G_{jI}/G_{I+1,I}$} as a function of site index $j$. Dashed lines correspond to the exponential fit \tr{$G_{jI}/G_{I,I+1}=e^{\zeta(j-I)}$} with \tr{an inverse} localization length $\zeta\approx \tr{0.19+0.20i}$. \tr{Other parameters are the same as in Fig.~\ref{Fig_directional_amplification}(a)-(d)} and row $P_3$ in Tables \ref{tab:circuit_param}-\ref{tab:four_wave_mixing}.}
    \label{Fig_topology}
\end{figure*}

Interesting circuit quantities to compute are the flux averages $\langle \phi_j \rangle = 2\phi_{\rm zpf}^j{\rm Re}(\langle a_j(t) \rangle)$, and $\langle \phi_j^2 \rangle = 2(\phi_{\rm zpf}^j)^2(\langle a_j^\dag(t) a_j(t)\rangle + {\rm Re}\langle a_j^2(t)\rangle +1/2)$, which need to satisfy $\langle \phi_j^2 \rangle / \Phi_0^2\ll 1$ for the Kerr non-linear physics to be valid. We obtain these quantities by computing the required averages of JJA operators as,
\begin{align}
    \langle a_j(t) \rangle ={}&|\alpha_{\rm ss}^j|+\langle \delta a_j(t)\rangle,\\
    \langle a_j^\dag(t)a_j(t)\rangle ={}& |\alpha_{\rm ss}^j|^2 + 2 |\alpha_{\rm ss}^j| {\rm Re}\langle \delta a_j(t)\rangle + \langle \delta a_j^\dag(t) \delta a_j(t)\rangle,\nonumber\\
    \langle a_j^2(t)\rangle ={}& |\alpha_{\rm ss}^j|^2 + 2 |\alpha_{\rm ss}^j|\langle \delta a_j(t)\rangle + \langle \delta a_j^2(t)\rangle,\nonumber
\end{align}
where $\langle \delta a_j^\dag(t) \delta a_j(t)\rangle$ is given in Eq.~(\ref{Occupation_Fluctuation_j}) and the other averages read, 
\begin{align}
    \langle \delta a_j(t)\rangle ={}& i \sqrt{\kappa}\alpha_{\rm sig} e^{i\theta_{\rm ss}^I}G_{jI}(\tilde{\omega}_s)e^{-i\tilde{\omega}_s t}+{\rm h.c.},\\
    \langle \delta a_j^2(t)\rangle ={}&(1/2\pi)\int d\tilde{\omega} \kappa G_{jl}(\tilde{\omega})G_{j,N+l}(-\tilde{\omega})\\
    +{}& 2\kappa|\alpha_{\rm sig}|^2 G_{jI}(\tilde{\omega}_s) G_{j,N+I}(-\tilde{\omega}_s)\nonumber\\
    +{}&\kappa\left\lbrace (\alpha_{\rm sig}^\ast)^2 e^{-2i\theta_{\rm ss}^I} [G_{j,N+I}(-\tilde{\omega}_s)]^2e^{2i\tilde{\omega}_s t}\right.\nonumber\\
    {}&\hspace{0.3cm}+\left. (\alpha_{\rm sig})^2 e^{2i\theta_{\rm ss}^I} [G_{jI}(\tilde{\omega}_s)]^2e^{-2i\tilde{\omega}_s t}\right\rbrace\nonumber. 
\end{align}

\section{Directional amplification from edge states of extended Hamiltonian}\label{TopRelationExtHam}

In this Appendix, we give details on the fundamental relation between the edge states of the extended Hamiltonian ${\cal H}(\omega)$ in Eq.~(\ref{Hext}) and the topological amplification properties of the system, such as the exponential dependence of the Green's functions. A more extended description of this theoretical framework can be found in our other works \cite{ramos_topological_2021,GomezLeon2022}.

The starting point of our formalism is the expression of the Green's function matrix \tr{$G(\omega)$ in Eq.~(\ref{eq:GreensFunctionMatrix})} by means of the following singular value decomposition (SVD):
\begin{equation}
\tr{G^{-1}(\omega)} = \omega-\omega_p-H_{\rm nh} = U(\omega) S(\omega) V^\dag(\omega).
\label{SVD}
\end{equation}
Here, $U(\omega)$, $V(\omega)$, are unitary matrices and 
$S(\omega)$ is a semi-positive diagonal matrix, 
$S(\omega)_{n m} = \delta_{nm} E_n(\omega)$. 
Using this decomposition we can write,
\begin{align}
    G(\omega) &=
    V(\omega) S^{-1}(\omega) U^\dagger(\omega) \nonumber\\
    {}&=\sum_{n} E_n^{-1}(\omega) \vec{v}^{(n)}(\omega) \vec{u}^{(n)}{}^\dag(\omega),
\label{Gw}
\end{align}
where $E_n(\omega)\geq 0$ are singular values, whereas $\vec{u}^{(n)}(\omega)$ and $\vec{v}^{(n)}(\omega)$ are the associated singular vectors given by the columns of 
$U(\omega)$ and $V(\omega)$, respectively. 
As originally demonstrated in Ref.~\cite{porras_topological_2019}, the quantities $E_n(\omega)$, $\vec{v}^{(n)}(\omega)$, and $\vec{u}^{(n)}(\omega)$ can also be interpreted as the positive eigenenergies and the associated eigenvectors of the extended Hamiltonian ${\cal H}(\omega)$ defined in Eq.~(\ref{Hext}). Therefore, the decomposition (\ref{Gw}) can also be obtained by solving the eigenvalue problem
\begin{align}
    {\cal H}(\omega)
    \begin{pmatrix}
    \vec{u}^{(n)} \\
    \vec{v}^{(n)}
    \end{pmatrix}
    =\pm E_n \begin{pmatrix}
    \vec{u}^{(n)} \\
    \pm \vec{v}^{(n)}
    \end{pmatrix}.\label{eig}
\end{align}
The extended Hamiltonian ${\cal H}(\omega)$ has an intrinsic chiral symmetry which may induce the emergence of zero-energy modes in the spectrum $E_n$.
A crucial observation is that those zero-energy modes enhance Green's functions since $E_n$ appears inverted in Eq.~\eqref{Gw}.

Following the reasoning above, the characterization of the spectral amplifying properties of the system can be obtained by exploiting the relation between Green's function and the eigenstates of the effective Hamiltonian
In particular, we can identify topological amplification regimes in which the extended Hamiltonian ${\cal H}(\omega)$ is in a topologically non-trivial phase according to the ten-fold way \cite{ryu_topological_2010}. 
In that case, the eigenspectrum $E_n$ presents at least one pair of zero-energy modes $\pm E_0$ with exponentially small energy, $E_0\propto e^{-N \zeta}$.
Also, the associated zero-energy eigenvectors behave as edge states localized either on the right boundary, $v^{(0)}_j\sim v^{(0)}_{N+j}\sim e^{j\zeta}$, or on the left boundary, $u^{(0)}_j\sim u^{(0)}_{N+j}\sim e^{-j\zeta}$, with $\zeta$ the inverse localization length. 
In Fig.~\ref{Fig_topology}\tr{(a)}, we numerically calculate the eigen-spectrum $E_n/\kappa$ associated to ${\cal H}(\omega)$ for the same parameters used in \tr{Figs.~\ref{Fig_directional_amplification}(a)-(d) of the main text}. We indeed observe that zero-energy modes appear within a topological band gap and a topological bandwidth because the topological region is frequency-dependent. For a finite system of size $N$, this region is defined as $E_0(\omega)\leq 1/N$, where $E_0(\omega)$ is the energy of the zero-mode. In addition, Fig.~\ref{Fig_topology}(b), shows the spatial dependence of the eigenvectors $u^{(0)}_j$ and $v^{(0)}_j$ associated with $E_0(\omega)$, confirming the exponential localization at both boundaries of the array.

The exponential suppression with system size of the zero-energy eigenvectors implies that the first term of the SVD decomposition, $G_{jl}^{(0)}(\omega) = E_0^{-1}(\omega) v^{(0)}_j(\omega) u^{(0)}_l{}^\ast(\omega)$, is the dominant contribution to the Green's function components,
\begin{align}
    G_{jl}(\omega)={}&G_{jl}^{(0)}(\omega)+\sum_{n>0} E_n^{-1}(\omega) v^{(n)}_j(\omega) u^{(n)}_l{}^\ast(\omega),\label{decomp2}
\end{align}
for $j>l$. Therefore, Green's function can be well-approximated by an exponential function $G_{jl}(\omega)\approx G_{jl}^{(0)}(\omega)\sim e^{\zeta (j-l)}$, for $j>l$. This occurs analogously for the anomalous components $G_{j,N+l}(\omega)$. 

Inhomogeneities and finite size effects slightly change this dependence on the boundaries, but in the central region of the array, $G_{jI}(\omega)$ can be well approximated as stated in Eq.~(\ref{exponentialdep}) of the main text. In Fig.~\ref{Fig_topology}(c), we numerically confirm this approximation by fitting it to the real and imaginary parts of the exact normalized Green's function components \tr{$G_{jI}/G_{I,I+1}$}, obtained numerically from $G_{jl}=(\omega-\omega_p-H_{\rm nh})^{-1}_{jl}$. We see that the fit \tr{$G_{jI}/G_{I,I+1}=e^{\zeta(j-I-1)}$} with $\zeta\approx \tr{0.19-0.20i}$ agrees very well with the numerical data, up to finite size effects near the boundary $j\sim N$ [cf.~dashed lines in Fig.~\tr{\ref{Fig_topology}(c)}]. Using Eq.~(\ref{exponentialdep}) with the fitted inverse localization length $\zeta\approx 0.19-0.20i$, we approximate the gain in Eq.~(\ref{gainjI}) as,
\begin{align}
    {\cal G}_j\approx e^{2{\rm Re}(\zeta)(j-[I+1])}\kappa^2|G_{I+1,I}(\omega_s)|^2, \quad j\geq I+1.
\end{align}
This theoretical prediction is plotted by the dashed lines in Fig.~\ref{Fig_topology}(a) and agrees very well with the full numerical results. Note that using the decimation technique, the inverse localization length and the exponential dependence of Green's functions can be analytically demonstrated for homogeneous systems without resorting to approximate fits \cite{gomez-leon_decimation_2022,GomezLeon2022}.

For the theoretical estimation of the reverse gain in Fig.~\ref{Fig_directional_amplification}(b, dashed line), we use the approximation
\begin{align}
    {\cal G}_j^{(R)}\approx e^{-2{\rm Re}(\zeta)(j-[I+1])}\kappa^2|G_{I,I+1}^{(0)}(\omega_s)|^2,\label{exp_estimation_reverse}
\end{align}
for $j\geq I+1$. Notice that here we use the same fitted $\zeta$ as above, but we multiply the exponential factor by the modulus squared of $G_{I,I+1}^{(0)}(\omega_s)=E_0^{-1}(\omega) v^{(0)}_j(\omega) u^{(0)}_l{}^\ast(\omega)$ instead of the full $G_{I,I+1}(\omega_s)$. This is due to decomposition (\ref{decomp2}). For $j<l$, the first term $\sim E_0^{-1}$ shows indeed the exponential dependence $\sim e^{-\zeta|j-l|}$, but this exponentially suppressed term does not dominate the behavior of $G_{jl}$, as it is on the same order as the sum of all the others. Therefore, using $|G_{I,I+1}^{(0)}(\omega_s)|^2$ in Eq.~(\ref{exp_estimation_reverse}) we estimate more precisely the scaling of the reverse gain. Still, this approximation can over- or under-estimate the exact expression of ${\cal G}_j^{(R)}$. Remarkably, we have found numerically that for a large distance, $|j-l|\gg 1$, contributions from singular vectors with $n \neq 0$ tend to interfere destructively with each other, resulting in a reverse gain that is even more suppressed than one can expect from the exponential dependence of the edge states alone. Therefore, increasing the array size $N$ is doubly beneficial since it enhances exponentially both directional amplification and suppression of backward emission. An example of this behavior is shown in Fig.~\ref{Fig_directional_amplification}(b), where for $j\sim N$ the actual reverse gain of the amplifier is well below the estimation of the dashed line.

\begin{figure}[t]
\centering
\includegraphics[width=\columnwidth]{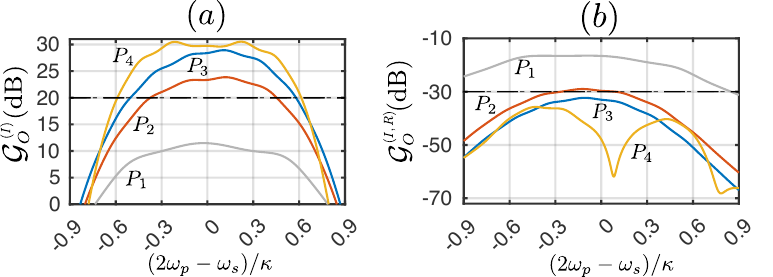}
\caption{Frequency dependence of gain and reverse gain for the idler field leaving at output site $O=N-1$, when a signal of frequency $\omega_s$ is sent via input $I=2$. Parameters sets $P_1-P_4$ are the same as Figs.~\ref{Fig_performance}(a)-(b).}
\label{Fig_Performance_idler}
\end{figure}

Finally, we comment on the behavior of the gain ${\cal G}^{(I)}_O$ and reverse gain ${\cal G}^{(I,R)}_O$ for the idler field at frequency $\omega=2\omega_p-\omega_s$. As shown in Fig.~\ref{Fig_Performance_idler}(a)-(b), the frequency-dependent response of the idler field is qualitatively equal and quantitatively very similar to the amplification of the signal field at frequency $\omega=\omega_s$ shown in Figs.~\ref{Fig_performance}(a)-(b) of the main text for the same parameter sets $P_1-P_4$.

\section{Increase bandwidth via Quarton nonlinearities}\label{app:quarton}

This Appendix discusses the possibility of using Quarton non-linearities \cite{ye_engineering_2021} to induce larger Kerr couplings $\bar{K}_{c,s}$ in the JJA and thereby increase the amplification bandwidth of the topological JPA array.

In Sec.~\ref{Peformance_ExpParameters}, we discussed four feasible operation points to realize the topological JPA array with state-of-the-art superconducting circuit parameters. We have shown a high directional amplification performance with gains above 20 dB over a bandwidth of $\omega_{\rm 20dB}\sim \kappa\sim 2\pi\cdot 300$ MHz. Reaching a larger amplifier's bandwidth on the order of GHz, for instance, requires scaling up all effective parameters of the device $\bar{J}$, $\bar{g}_c$, $\bar{g}_c$, and $\kappa$, while keeping fixed the ratios $\bar{J}/\kappa$, $\bar{g}_c/\kappa$, and $\bar{g}_s/\kappa$ of the operation points $P_1-P_4$. Designing $\kappa$ up to a GHz is possible, but the biggest limitation is having larger squeezing terms $\bar{g}_{c,s}$. Evaluating Eqs.~(\ref{gchom})-(\ref{gsHom}) at the homogeneous central region, we find a simple expression for the squeezing terms in the relevant case $\bar{\varphi}=\pi/2$:
\begin{align}
    \bar{g}_c=\frac{\bar{g}_s}{\lambda}=\bar{K}_c|\bar{\alpha}_{\rm ss}|^2= \frac{\bar{E}_C}{\hbar} \frac{|\bar{\alpha}_{\rm ss}|^2}{1+\lambda}\ll \frac{1}{1+\lambda}\frac{\bar{\omega}_b}{16},\label{bandwith_lim}
\end{align} 
with the factor $\lambda=L_J/(2L_J')$ controlled by the ratio between intra- and inter-site Josephson inductances and $\bar{E}_C=e^2/(2\bar{C}_{\rm eq})$ the charging energy at the central region. The last inequality in Eq.~(\ref{bandwith_lim}) comes from the low phase drop condition, which restricts the maximum occupation as $|\bar{\alpha}_{\rm ss}|^2\ll \hbar \bar{\omega}_b/(16\bar{E}_C)$. Since 
a topological amplifying phase requires $\lambda = \bar{g}_c/\bar{g}_s\gtrsim 1$, Eq.~(\ref{bandwith_lim}) implies that the squeezing terms $\bar{g}_{c,s}$ are limited in the JJA by $\bar{g}_c=\bar{g}_s/\lambda \ll \bar{\omega}_b/16$. Importantly, notice that having sub-arrays of $M$ JJs does not help for increasing $\bar{g}_{c,s}$ since $|\bar{\alpha}_{\rm ss}|^2$ increases in equal proportion as $\bar{K}_{c,s}$ reduce. Also notice that the limitation in $\bar{g}_{c,s}$ originates ultimately from the design of the JJA, which fixes the equivalent inductance per site as $L_{\rm eq}=L_J'/(2[1+\lambda])$ and thereby the Kerr non-linearities as $\bar{K}_s =\bar{E}_C/\hbar$, and $\bar{K}_c =(\bar{E}_C/\hbar)(1+\lambda)^{-1}=\bar{T} = \bar{T}'$. 

\begin{figure}[t!]
    \includegraphics[width=1\columnwidth]{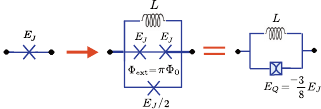}
    \caption{Josephson junctions in the JJA are replaced by Quarton nonlinearities in parallel to a linear inductor $L$. This can achieve larger Kerr-nonlinearities $\bar{K}_c$ and $\bar{K}_s$ and thereby increase the bandwidth of the topological amplifier array.}
    \label{fig:quarton}
\end{figure}

We now explore the possibility of replacing all JJs of the setup with Quarton non-linearities in parallel with a linear inductor $L$, as this can relax the condition $L_{\rm eq}=L_J'/(2[1+\lambda])$ and thereby increase the Kerr non-linearities without reducing $|\bar{\alpha}_{\rm ss}|^2$. As shown in Fig.~\ref{fig:quarton}, a Quarton non-linearity is formed by a closed loop of three JJs, two identical with Josephson energy $E_J$ on one arm, and another with half energy $E_J/2$ on the other arm. To this Quarton, we couple a linear inductor $L$ on parallel, and we threaten an external flux of half flux quantum, $\Phi_{\rm ext}=\pi\Phi_0$ around the Quarton loop. If we replace all non-linearities of the JJA by these Quartons, the Lagrangian of the JJA in Eq.~(\ref{LagJJ}) is modified to:
\begin{align}
     {\cal L}_{\rm JJA}^{(Q)}= {}&\sum_{j=1}^N \frac{C_j}{2}(\dot{\phi}_j)^2 +\sum_{j=0}^{N}\frac{C'_j}{2}(\dot{\phi}_{j+1}-\dot{\phi}_j)^2\label{Lag_Quarton}\\
    +{}&\sum_{j=1}^N \left[2 E_J \cos\left(\frac{\phi_j}{2\Phi_0}\right)-\frac{E_J}{2} \cos\left(\frac{\phi_j}{\Phi_0}\right)-\frac{1}{2L}\phi_j^2\right]\nonumber\\
    +{}&\sum_{j=0}^N \left[2 E_J' \cos\left(\frac{\phi_{j+1}-\phi_j}{2\Phi_0}\right)\!-\!\frac{E_J'}{2}\cos\left(\frac{\phi_{j+1}-\phi_j}{\Phi_0}\right)\right.\nonumber\\
    {}&\hspace{0.8cm}\left.-\frac{1}{2L'}(\phi_{j+1}-\phi_j)^2\right]\nonumber,\qquad \phi_0=\phi_{N+1}=0,
\end{align}
where $E_J$, $E_J'$ are the Josephson energies of JJs for intra- and inter-site Quartons. Similarly, $L$ and $L'$ are the intra- and inter-site linear inductances coupled in parallel to each Quarton. Expanding this Lagrangian for small phase drop $\langle \phi_j^2 \rangle/(\Phi_0)^2\ll 1$, and applying the same quantization procedure as done for the standard JJA in Appendix~\ref{TotalH}, we obtain the associated quantum Hamiltonian:
\begin{align}
    H_{\rm JJA}^{(Q)}={}&\sum_{j=1}^N\frac{1}{2C^j_{\rm eq}}q_j^2+\sum_{j=1}^N\frac{1}{2L}\phi_j^2\label{HexpandQ}\\
    +{}&\sum_{j=1}^{N-1}\frac{C'_j}{C^j_{\rm eq}C^{j+1}_{\rm eq}}q_jq_{j+1}-\sum_{j=1}^{N-1}\frac{1}{L'}\phi_j\phi_{j+1}\nonumber\\
    -{}&\sum_{j=1}^N \frac{E_Q}{24(\Phi_0)^4}\phi_j^4-\sum_{j=0}^{N}\frac{E'_{Q}}{24(\Phi_0)^4}(\phi_{j+1}-\phi_j)^4+{\cal O}^6\nonumber.
\end{align}
Here, the effective intra- and inter-site Quarton energies are related to the corresponding Josephson energies as
\begin{align}
E_Q = -(3/8)E_J,\qquad E_Q' = -(3/8)E_J'.
\end{align}
Note that these Quarton energies do not add a quadratic term, and therefore the linear inductances only depend on $L$ and $L'$, which can be tuned independently of $E_Q$ and $E_Q'$. This will be key to increasing the bandwidth in our setup. Notice also that the non-linear inductances effectively change signs leading to negative Kerr non-linearities.

The above Hamiltonian (\ref{HexpandQ}) has the same form as the JJA Hamiltonian in Eq.~(\ref{Hexpand}), and therefore all formulas used in the main text can used for Quarton non-linearities with the replacements: $L_{\rm eq}\rightarrow L$, $L_J'\rightarrow L'$, $E_J\rightarrow E_Q$, and $E_J'\rightarrow E_Q'$. In this idealized model of identical Quartons, it is simple to estimate the consequences of having pure Quarton nonlinearities instead of JJs. The linear properties of the JJA remain of the same form: $\bar{\omega}_b =1/\sqrt{C_{\rm eq}L}$ and $\bar{J}_b = (\bar{\omega}_b/2)(\bar{C}/\bar{C}_{\rm eq}-L/L')$, but the Kerr nonlinearities are modified to:
\begin{align}
    \bar{K}_c= \frac{\bar{K}_s}{1+\lambda}=\frac{\hbar \bar{E}'_Q}{4(\Phi_0)^4}\frac{L}{\bar{C}_{\rm eq}}=-\frac{3}{4}\frac{\bar{E}_C}{\hbar}\frac{L}{L'_J},\label{quarton_Kerr}
\end{align}
with $L_J'=(\Phi_0)^2/E_J'$ the Josephson inductance associated with the Josephson energy $E_J'$ of the large JJs in the inter-site Quartons.

The key to using a Quarton non-linearity is that it allows for the increase of $\bar{K}_{c,s}$ independently of $\bar{E}_C$ via $L/L'_J$ so this restriction in inequality (\ref{bandwith_lim}) is lifted. The restriction $|\bar{\alpha}_{\rm ss}|^2\ll \hbar\bar{\omega}_b/(16\bar{E}_C)$ on the maximum mean occupation remains leading to the condition: $|\bar{g}_c|\ll (3/64)(L/L'_J)\bar{\omega}_b$. However, it is now possible to approach $|\bar{g}_c|\lesssim \bar{\omega}_b$, by having $L'_J\ll L=1/(\bar{\omega}_b^2 \bar{C}_{\rm eq})$. The only practical limitation of having minimal values of Josephson inductances $L_J/M=L_J'/(2\lambda M)$ above $\sim 40$ pH, so that stray inductances do not affect \cite{planat_understanding_2019,eichler_quantum-limited_2014}. 

In practice, Eq.~(\ref{quarton_Kerr}) implies that for given total capacitance $\bar{C}_{\rm eq}$, bare frequency $\tilde{\omega}_b$ of a JJA, the implementation of Quarton non-linearities can scale the Kerr-non-linearities as
\begin{align}
    \bar{K}_c, \bar{K}_s \rightarrow -\chi (\bar{K}_c, \bar{K}_s), 
\end{align}
with a factor $\chi$ larger with respect to the case of simple JJs, given by
\begin{align}
\chi=(3/4)(1+\lambda)(L/L_J').
\end{align}
Since sub-arrays of $M$ JJs in series can also be implemented with Quartons, the JJA can effectively sustain the same $|\bar{\alpha}_{\rm ss}|^2$, and thus the scaled Kerr non-linearities imply larger squeezing terms in Eqs.~(\ref{gchom})-(\ref{gsHom}) by the same factor as $(\bar{g}_c, \bar{g}_s) \rightarrow -\chi (\bar{g}_c, \bar{g}_s)$. In addition, note that because Kerr non-linearities become negative, which has various consequences. First, squeezing terms change the sign, but this has no physical implication in the effective dynamics given by $H_{\rm pa}$ as this constant phase can be absorbed in the operators $a_j\rightarrow a_j e^{i\pi/2}$. Second, the frequency shift in the effective detuning $\bar{\Delta}$ (\ref{DeltaHom}) also changes the sign, which can be compensated by scaling and reversing the sign of the bare detuning as $\bar{\Delta}_b\rightarrow -\chi \bar{\Delta}_b$. Third, we need to reverse the sign also of the bare hopping as $\bar{J}_b\rightarrow -\chi \bar{J}_b$, so that the effective hopping in Eq.~(\ref{JeffHom}) scales and reverses sign accordingly $\bar{J}\rightarrow -\chi \bar{J}$. This can be done by selecting $L'$ appropriately without modifying capacitances. Fourth, change the phase difference induced by the external pump as $\bar{\varphi}\rightarrow \bar{\varphi}-\pi$ to compensate for the change of sign of $\bar{J}$. In addition, we also need to scale up the JJA decay rate $\kappa\rightarrow \chi \kappa$, which we do by increasing the impedance of the transmission $Z_{\rm TL}$ so that we avoid modifying capacitances. Finally, we adjust the pump power and the quantities of the auxiliary array to scale the effective pump as $\Omega_p\rightarrow \chi \Omega_p$. The pump frequency may also be slightly re-adjusted.

In summary, with Quarton non-linearities we may implement the same operation points $P_1-P_4$ in Table~\ref{tab:four_wave_mixing}, but with all effective quantities rescaled by $\chi$, up to signs. The practical limitations will be on the superconducting circuit technology to implement Quartons, for instance, the restriction that each JJ forming the Quartons must have Josephson inductances $L_J/M$ above 40 pH and that the JJA bare frequency is in the range $\bar{\omega}_b\sim 4-8$ GHz. Taking into account these restrictions, we demostrate that the operation point $P_2$ can be indeed implemented with state-of-the-art parameters.

In particular, we keep the same size $N=15$, number of JJs $M=16$, ratio $\lambda=L_J/2L_J'=1$, capacitances $\bar{C}, \bar{C}', \bar{C}_{\rm TL}, \bar{C}_{\rm eq}$, and tapering the as in row $P_2$ of Tables~\ref{tab:circuit_param} and \ref{tab:tapering_f}. Regarding the modifications to implement Quartons, we take Josephson inductances as $L_J/M=L_J'/(2M)\sim 41 $pH, and linear inductances as $L=1.93$ nH and $L'= 6.89$ nH, such that $L/L_J'\approx 3$, and Kerr non-linearities and hopping rates scale by factor $\chi\approx 4.4$ as discussed above. To compensate for the negative shift in detuning (\ref{DeltaHom}), we now require a much lower frequency of each site: $\bar{\omega}_b/2\pi \approx 5.7 {\rm GHz}$ and also a slightly smaller pump frequency: $\bar{\omega}_p\approx 7.43$ GHz. Finally, to scale up the decay rate and the pump strength by the same factor $\chi\approx 4.4$ and without changing capacitances, we increase the impedance of the lines and the pump power to $Z_{\rm TL} = 450 \Omega$, and $P_p\rightarrow -78.4$ dBm, respectively. This scales all effective parameters of row $P_2$ of Table~\ref{tab:four_wave_mixing} by $\chi$, and in particular, the decay rate as $\kappa\rightarrow \chi \kappa \approx 2\pi \cdot 1.2$ GHz. Under these considerations, we realize the operation point $P_2$, and thus we estimate that the topological JPA with Quarton non-linearities can have the same performance in gain, reverse gain, and added noise as shown in Table~\ref{Fig_performance}(red). The bandwidth above 20 dB in this case is $\Delta \omega_{\rm 20 dB}\approx 0.83 \kappa$, and with the scaled $\kappa$ the devie can reach $\Delta \omega_{\rm 20 dB}/2\pi\approx 1$ GHz.

In the case of implementing the auxiliary waveguide, we estimate that the same above dyanamics can be obtained when setting the parameters: $L_{\rm w} = 0.43$ nH, $C_{\rm w} = 567$ fF, $C_{\rm w}' = 244$ fF, $C_{\rm pw}=723$ fF, $\bar{C}''=17$ fF, $\bar{R} = 324.7 \Omega$, $P_{\rm pw}= -46.5$ dBm, and the impedance of the line is kept as $Z_{\rm pw} = 10 \Omega$. This leads to $\omega_{\rm w}/2\pi\approx 7.43$ GHz, $J_{\rm w}/2\pi \sim 846$ MHz, $J_{\rm w}'/2\pi \sim 85$ MHz, and $\kappa_{\rm w}^{1,N}/2\pi \sim 1.69$ GHz.

\begin{widetext}

\section{Summary of parameters}\label{app:Table_Parameters}

In this Appendix, we provide all superconducting circuit parameters considered in the simulations of this work for each of the four operation points $P_1$-$P_4$, which lead to the effective parameters in Table~\ref{tab:four_wave_mixing} of the main text.

The required parameters for the JJA and the auxiliary waveguide are summarized in Table~\ref{tab:circuit_param} and Table~\ref{tab:auxiliar_waveguide}, respectively. The specific values for the tarering functions used for operation points $P_1-P_4$ are summarized in Table~\ref{tab:tapering_f}.

\begin{table*}[!ht]
\center
\begin{tabular}{|c||c|c|c||c|c|c|c|c|c|c||c|c|c|c||c|c|c|c|}
\hline
    & $N$ & $M$ & $N_{JJ}$ & $\bar{C}$ &  $\bar{C}'$ & $\bar{C}_{\rm TL}$ & $\bar{C}_{\rm eq}$ & $L_J/M$ & $L'_J/M$ & $L_{\rm eq}$ & $\bar{J}_{b}/2\pi$ & $\bar{K}_c/2\pi M^2$ & $\bar{K}_s/2\pi M^2$ & $\kappa/2\pi$ & $\bar{P}_p$ & $\bar{\Omega}_p/2\pi$ & $\bar{\Delta}_b/2\pi$ & $|\bar{\alpha}_{\rm ss}|^2$  \\
    & & & &(fF) &(fF) & (fF) & (fF) & (pH) & (pH) & (pH) & (MHz) & (kHz)  & (kHz) &(MHz) & (dBm)  & (GHz)  & (MHz)  &  \\
\hline
$P_1$ & 8 & 1 & 17 & 1513 & 1014 & 459 & 4000 & 188 & 376 & 94 & 17 & 2413 & 4827 & 279 & -97.4 & 1.22 & -385 & 27 \\
\hline
$P_2$ & 15 & 16 & 496 & 51.5 & 101.6 & 145.2 & 400 & 117.7 & 235.4 & 941.7 & 17 & 94& 188 & 279 & -83.3 & 6.19 & -384 & 680 \\
\hline
$P_3$ & 17 & 25 & 875 & 141.6 & 170.2 & 188.0 & 670 & 45.0 & 90.0 & 562.3 & 17 & 23 & 46 & 279 & -77.2 & 12.5 & -384 & 2800 \\
\hline
$P_4$ & 11 & 30 & 690 & 159.6 & 76.0 & 138.4 & 450 & 41.9 & 167.4 & 837.1 & 9 & 16 & 48 & 225 & -77.7 & 10.6 & -342 & 2700 \\
\hline
\end{tabular}
 \caption{{Circuit parameters for various operation points of the \tr{topological JPA array}.} The first three columns describe the size $N$ of the array, the number $M$ of JJs per sub-array, and the total number of JJs in the array $N_{\rm JJ}=M(2N+1)$ for each opearation point $P_1-P_4$, The next seven columns show the capacitances and inductances used in the circuit for the JJA, such as on-site capacitance \tr{$\bar{C}$}, inter-site capacitance \tr{$\bar{C}'$}, \tr{capacitive coupling to the transmission lines $\bar{C}_{\rm TL}$}, equivalent (total) capacitance per site $\bar{C}_{\rm eq}$, \tr{on-site Josephson inductance $L_J/M$}, \tr{inter-site Josephson inductance $L_J'/M$}, and equivalent on-site inductance $L_{\rm eq}$. The next four columns show couplings characterizing the JJA as a Kerr resonator array: bare hopping \tr{$\bar{J}_b$, cross- and self-Kerr nonlinearities $\bar{K}_{c,s}/M^2$, and local decay rate $\kappa$. Notice that for all parameter sets the bare frequency of the JJA is $\bar{\omega}_b/2\pi=8.2$ GHz. The last \tr{four} columns show the power \tr{$\bar{P}_p$} of the external pumps at the central region, the \tr{corresponding effective pump strength $\bar{\Omega}_p$, the detuning $\bar{\Delta}_b$} between pump and JJA frequency, and the mean number of photons \tr{$|\bar{\alpha}_{\rm ss}|^2$} induced at the central region of the JJA. The impedance of all transmission lines coupled to the JJA is $Z_{\rm TL}=50\Omega$.}}\label{tab:circuit_param}
\end{table*}

\begin{table*}[!ht]
\center
\begin{tabular}{|c||c||c|c|c|c|c||c|c|c||c|c||c|c|c|c|c||c|c|}
\hline
    & $N$ & $\bar{C}_{\rm w}$ & $\bar{C}'_{\rm w}$ & $\bar{C}''$ & $\bar{C}_{\rm eq}^{\rm w}$ & $L_{\rm w}$ & $C_{\rm pw}$ & $C_{\rm TL}^{I}$ & $C_{\rm TL}^{O}$ & $\bar{R}$ & $\bar{C}$ & $\omega_{\rm w}/2\pi$ & $J_{\rm w}/2\pi$ & $\bar{J}'_{\rm w}/2\pi$ & $\kappa_{\rm w}^{1,N}/2\pi$ & $\bar{\gamma}_{\rm w}/2\pi$ & $P_{\rm pw}$ & $|\bar{\beta}_{\rm ss}|$ \\
    & & (fF) & (fF) & (fF) & (fF) & (pH) & (fF) & (fF) & (fF) & ($\Omega$) & (fF) & (GHz) & (MHz) & (MHz) & (MHz) & (MHz) & (dBm) & \\
\hline
$P_1$ & 8 & 908.3 & 50.1 & 9.7 & 1018 & 407.3 & 319.5 & 472.2 & 473.7 & 142.7 & 1962 & 7.82 & 192 & 19 & 384 & 0.96 & -66.8 & 64 \\
\hline
$P_2$ & 15 & 915.0 & 50.1 & 3.1 & 1018 & 407.3 & 319.3 & 149.0 & 150.7 & 1429 & 193.7 & 7.82 & 192 & 19 & 384 & 0.96 &  -52.7 & 322 \\
\hline
$P_3$ & 17 & 914.1 & 50.1 & 4.0 & 1018 & 407.3 & 319.3 & 193.0 & 194.5 & 852.8 & 325.6 & 7.82 & 192 & 19 & 384 & 0.96 & -46.6 & 651 \\
\hline
$P_4$ & 11 & 921.7 & 44.0 & 2.9 &  1013 & 405.1 & 298.6 & 141.4 & 139.8 & 1573 & 295.2 & 7.86 &  171 & 17 & 342 & 0.85 & -47.6 & 621\\
\hline
\end{tabular}
\caption{Parameters describing the auxiliary waveguide to distribute the pump over the JJA for each operation point $P_1-P_4$. The first column indicates the size of the auxiliary waveguide which is the same size as the JJA. The next five columns indicate the capacitances and inductances used in the circuit for the auxiliary waveguide, such as on-site capacitance $\bar{C}_{\rm w}$, inter-site capacitance $\bar{C}'_{\rm w}$, capacitive coupling $\bar{C}''$ to the JJA, total equivalent capacitance $\bar{C}_{\rm eq}^{\rm w}$, and local inductance $L_{\rm w}$. The next three columns indicate the capacitance to the pump transmission line $C_{\rm pw}$ on sites $j=1,N$ and to the signal transmission lines $C_{\rm TL}^{I,O}$ at sites $j=I,O$. The next two columns indicate the resistance $R_j$ and the modified local JJA capacitance needed at sites $j\neq I,O$ in case of inducing local decay $\kappa$ via on-chip attenuators. The next five columns show the couplings characterizing the auxiliary waveguide as a coupled cavity array: bare frequency $\omega_{\rm w}$, intra-array hopping $\bar{J}_{\rm w}$, hopping to JJA $\bar{J}_{\rm w}'$, waveguide decay at borders $\kappa_{\rm w}^{1,N}$, and collective decay rate $\bar{\gamma}_{\rm w}$ induced by the auxiliary on the JJA. The last two columns show the power $\bar{P}_{\rm pw}$ of the waveguide pump at site $j=1$ and the mean waveguide displacement $|\bar{\beta}_{\rm ss}|$ in steady state. Finally, the impedance of the transmission lines coupled to waveguides is $Z_{\rm pw}=10 \Omega$ and the impedance of the auxiliary waveguide is $Z_{\rm w}=20 \Omega$ for all parameter sets.}\label{tab:auxiliar_waveguide}
\end{table*}

\begin{table*}[!ht]
\center
\begin{tabular}{|c||c|c|c||c|c|c|c|c||c|c|c|c|c||c|c|c|c|c|}
\hline
    & $N$ & $N_L$ & $N_R$ & $f_1$ & $f_2$ & $f_{N-2}$ & $f_{N-1}$ & $f_{N}$ & $q_1$ & $q_2$ & $q_{N-2}$ & $q_{N-1}$ & $q_{N}$ & $h_1$ & $h_2$ & $h_{N-2}$ & $h_{N-1}$ & $h_{N}$\\
    & & & & & & & & & & & & & & & & & & \\
\hline
$P_1$ & 8 & 2 & 3 & 0.25 & 0.75 & 0.85 & 0.50 & 0.15 & 1.11 & 1.06 & 1.01 & 1.06 & 1.11 & 1.08 & 1.03 & 1.00 & 1.03 & 1.08 \\
\hline
$P_2$ & 15 & 2 & 3 & 0.25 & 0.75 & 0.85 & 0.50 & 0.15 & 1.14 &  1.05 & 1.01 & 1.06 & 1.10 & 1.10 & 1.03 & 1.01 & 1.04 & 1.08 \\
\hline
$P_3$ & 17 & 2 & 3 & 0.25 & 0.75 & 0.85 & 0.50 & 0.15 & 1.16 & 1.05 & 1.01 & 1.06 & 1.11 & 1.11 & 1.03 & 1.01 & 1.04 & 1.08 \\
\hline
$P_4$ & 11 & 2 & 2 & 0.25 & 0.75 & 1 & 0.75 & 0.25 & 1.10 & 1.04 & 1 & 1.02 & 1.10 & 1.07 & 1.02 & 1 & 1.01 & 1.06 \\
\hline
\end{tabular}
\caption{Specific values of tapering functions $f_j$, $q_j$, and $h_j$ used in the left and right buffer regions of the JJA for parameter sets $P_1-P_4$. The function $f_j$ is given in Eq.~(\ref{tapering}), whereas $q_j$ and $h_j=(q_j+q_{j-1})/2$ are obtained from optimization as detailed in Sec.~\ref{Sec:Stabilization_inhomogeneities}.}\label{tab:tapering_f}
\end{table*}

\end{widetext}

\input{bib.bbl}

\end{document}

%% file: bib.bbl
%

%% file: main.bbl
\begin{thebibliography}{82}%
\makeatletter
\providecommand \@ifxundefined [1]{%
 \@ifx{#1\undefined}
}%
\providecommand \@ifnum [1]{%
 \ifnum #1\expandafter \@firstoftwo
 \else \expandafter \@secondoftwo
 \fi
}%
\providecommand \@ifx [1]{%
 \ifx #1\expandafter \@firstoftwo
 \else \expandafter \@secondoftwo
 \fi
}%
\providecommand \natexlab [1]{#1}%
\providecommand \enquote  [1]{``#1''}%
\providecommand \bibnamefont  [1]{#1}%
\providecommand \bibfnamefont [1]{#1}%
\providecommand \citenamefont [1]{#1}%
\providecommand \href@noop [0]{\@secondoftwo}%
\providecommand \href [0]{\begingroup \@sanitize@url \@href}%
\providecommand \@href[1]{\@@startlink{#1}\@@href}%
\providecommand \@@href[1]{\endgroup#1\@@endlink}%
\providecommand \@sanitize@url [0]{\catcode `\\12\catcode `\$12\catcode `\&12\catcode `\#12\catcode `\^12\catcode `\_12\catcode `\%12\relax}%
\providecommand \@@startlink[1]{}%
\providecommand \@@endlink[0]{}%
\providecommand \url  [0]{\begingroup\@sanitize@url \@url }%
\providecommand \@url [1]{\endgroup\@href {#1}{\urlprefix }}%
\providecommand \urlprefix  [0]{URL }%
\providecommand \Eprint [0]{\href }%
\providecommand \doibase [0]{https://doi.org/}%
\providecommand \selectlanguage [0]{\@gobble}%
\providecommand \bibinfo  [0]{\@secondoftwo}%
\providecommand \bibfield  [0]{\@secondoftwo}%
\providecommand \translation [1]{[#1]}%
\providecommand \BibitemOpen [0]{}%
\providecommand \bibitemStop [0]{}%
\providecommand \bibitemNoStop [0]{.\EOS\space}%
\providecommand \EOS [0]{\spacefactor3000\relax}%
\providecommand \BibitemShut  [1]{\csname bibitem#1\endcsname}%
\let\auto@bib@innerbib\@empty
\bibitem [{\citenamefont {Smith}\ \emph {et~al.}(2013)\citenamefont {Smith}, \citenamefont {Bakker}, \citenamefont {Witvers}, \citenamefont {Woestenburg},\ and\ \citenamefont {Palmer}}]{smith_low_2013}%
  \BibitemOpen
  \bibfield  {author} {\bibinfo {author} {\bibfnamefont {D.~M.~P.}\ \bibnamefont {Smith}}, \bibinfo {author} {\bibfnamefont {L.}~\bibnamefont {Bakker}}, \bibinfo {author} {\bibfnamefont {R.~H.}\ \bibnamefont {Witvers}}, \bibinfo {author} {\bibfnamefont {B.~E.~M.}\ \bibnamefont {Woestenburg}},\ and\ \bibinfo {author} {\bibfnamefont {K.~D.}\ \bibnamefont {Palmer}},\ }\bibfield  {title} {\bibinfo {title} {Low noise amplifier for radio astronomy},\ }\href {https://doi.org/10.1017/S1759078712000840} {\bibfield  {journal} {\bibinfo  {journal} {International Journal of Microwave and Wireless Technologies}\ }\textbf {\bibinfo {volume} {5}},\ \bibinfo {pages} {453} (\bibinfo {year} {2013})}\BibitemShut {NoStop}%
\bibitem [{\citenamefont {Cleland}\ \emph {et~al.}(2002)\citenamefont {Cleland}, \citenamefont {Aldridge}, \citenamefont {Driscoll},\ and\ \citenamefont {Gossard}}]{cleland_nanomechanical_2002}%
  \BibitemOpen
  \bibfield  {author} {\bibinfo {author} {\bibfnamefont {A.~N.}\ \bibnamefont {Cleland}}, \bibinfo {author} {\bibfnamefont {J.~S.}\ \bibnamefont {Aldridge}}, \bibinfo {author} {\bibfnamefont {D.~C.}\ \bibnamefont {Driscoll}},\ and\ \bibinfo {author} {\bibfnamefont {A.~C.}\ \bibnamefont {Gossard}},\ }\bibfield  {title} {\bibinfo {title} {Nanomechanical displacement sensing using a quantum point contact},\ }\href {https://doi.org/10.1063/1.1497436} {\bibfield  {journal} {\bibinfo  {journal} {Appl. Phys. Lett.}\ }\textbf {\bibinfo {volume} {81}},\ \bibinfo {pages} {1699} (\bibinfo {year} {2002})}\BibitemShut {NoStop}%
\bibitem [{\citenamefont {Blais}\ \emph {et~al.}(2021)\citenamefont {Blais}, \citenamefont {Grimsmo}, \citenamefont {Girvin},\ and\ \citenamefont {Wallraff}}]{blais_circuit_2021}%
  \BibitemOpen
  \bibfield  {author} {\bibinfo {author} {\bibfnamefont {A.}~\bibnamefont {Blais}}, \bibinfo {author} {\bibfnamefont {A.~L.}\ \bibnamefont {Grimsmo}}, \bibinfo {author} {\bibfnamefont {S.}~\bibnamefont {Girvin}},\ and\ \bibinfo {author} {\bibfnamefont {A.}~\bibnamefont {Wallraff}},\ }\bibfield  {title} {\bibinfo {title} {Circuit quantum electrodynamics},\ }\href {https://doi.org/10.1103/RevModPhys.93.025005} {\bibfield  {journal} {\bibinfo  {journal} {Rev. Mod. Phys.}\ }\textbf {\bibinfo {volume} {93}},\ \bibinfo {pages} {025005} (\bibinfo {year} {2021})}\BibitemShut {NoStop}%
\bibitem [{\citenamefont {Garc\'{\i}a-Ripoll}(2022)}]{Ripoll2022}%
  \BibitemOpen
  \bibfield  {author} {\bibinfo {author} {\bibfnamefont {J.~J.}\ \bibnamefont {Garc\'{\i}a-Ripoll}},\ }\href@noop {} {\emph {\bibinfo {title} {Quantum Information and Quantum Optics with Superconducting Circuits}}}\ (\bibinfo  {publisher} {Cambridge University Press},\ \bibinfo {address} {Cambridge},\ \bibinfo {year} {2022})\BibitemShut {NoStop}%
\bibitem [{\citenamefont {Vijay}\ \emph {et~al.}(2011)\citenamefont {Vijay}, \citenamefont {Slichter},\ and\ \citenamefont {Siddiqi}}]{vijay_observation_2011}%
  \BibitemOpen
  \bibfield  {author} {\bibinfo {author} {\bibfnamefont {R.}~\bibnamefont {Vijay}}, \bibinfo {author} {\bibfnamefont {D.~H.}\ \bibnamefont {Slichter}},\ and\ \bibinfo {author} {\bibfnamefont {I.}~\bibnamefont {Siddiqi}},\ }\bibfield  {title} {\bibinfo {title} {Observation of {Quantum} {Jumps} in a {Superconducting} {Artificial} {Atom}},\ }\href {https://doi.org/10.1103/PhysRevLett.106.110502} {\bibfield  {journal} {\bibinfo  {journal} {Phys. Rev. Lett.}\ }\textbf {\bibinfo {volume} {106}},\ \bibinfo {pages} {110502} (\bibinfo {year} {2011})}\BibitemShut {NoStop}%
\bibitem [{\citenamefont {Walter}\ \emph {et~al.}(2017)\citenamefont {Walter}, \citenamefont {Kurpiers}, \citenamefont {Gasparinetti}, \citenamefont {Magnard}, \citenamefont {Potočnik}, \citenamefont {Salathé}, \citenamefont {Pechal}, \citenamefont {Mondal}, \citenamefont {Oppliger}, \citenamefont {Eichler},\ and\ \citenamefont {Wallraff}}]{walter_rapid_2017}%
  \BibitemOpen
  \bibfield  {author} {\bibinfo {author} {\bibfnamefont {T.}~\bibnamefont {Walter}}, \bibinfo {author} {\bibfnamefont {P.}~\bibnamefont {Kurpiers}}, \bibinfo {author} {\bibfnamefont {S.}~\bibnamefont {Gasparinetti}}, \bibinfo {author} {\bibfnamefont {P.}~\bibnamefont {Magnard}}, \bibinfo {author} {\bibfnamefont {A.}~\bibnamefont {Potočnik}}, \bibinfo {author} {\bibfnamefont {Y.}~\bibnamefont {Salathé}}, \bibinfo {author} {\bibfnamefont {M.}~\bibnamefont {Pechal}}, \bibinfo {author} {\bibfnamefont {M.}~\bibnamefont {Mondal}}, \bibinfo {author} {\bibfnamefont {M.}~\bibnamefont {Oppliger}}, \bibinfo {author} {\bibfnamefont {C.}~\bibnamefont {Eichler}},\ and\ \bibinfo {author} {\bibfnamefont {A.}~\bibnamefont {Wallraff}},\ }\bibfield  {title} {\bibinfo {title} {Rapid {High}-{Fidelity} {Single}-{Shot} {Dispersive} {Readout} of {Superconducting} {Qubits}},\ }\href {https://doi.org/10.1103/PhysRevApplied.7.054020} {\bibfield  {journal} {\bibinfo  {journal} {Phys. Rev. Applied}\ }\textbf {\bibinfo {volume} {7}},\
  \bibinfo {pages} {054020} (\bibinfo {year} {2017})}\BibitemShut {NoStop}%
\bibitem [{\citenamefont {Dassonneville}\ \emph {et~al.}(2020)\citenamefont {Dassonneville}, \citenamefont {Ramos}, \citenamefont {Milchakov}, \citenamefont {Planat}, \citenamefont {Dumur}, \citenamefont {Foroughi}, \citenamefont {Puertas}, \citenamefont {Leger}, \citenamefont {Bharadwaj}, \citenamefont {Delaforce}, \citenamefont {Naud}, \citenamefont {Hasch-Guichard}, \citenamefont {García-Ripoll}, \citenamefont {Roch},\ and\ \citenamefont {Buisson}}]{dassonneville_fast_2020}%
  \BibitemOpen
  \bibfield  {author} {\bibinfo {author} {\bibfnamefont {R.}~\bibnamefont {Dassonneville}}, \bibinfo {author} {\bibfnamefont {T.}~\bibnamefont {Ramos}}, \bibinfo {author} {\bibfnamefont {V.}~\bibnamefont {Milchakov}}, \bibinfo {author} {\bibfnamefont {L.}~\bibnamefont {Planat}}, \bibinfo {author} {\bibfnamefont {E.}~\bibnamefont {Dumur}}, \bibinfo {author} {\bibfnamefont {F.}~\bibnamefont {Foroughi}}, \bibinfo {author} {\bibfnamefont {J.}~\bibnamefont {Puertas}}, \bibinfo {author} {\bibfnamefont {S.}~\bibnamefont {Leger}}, \bibinfo {author} {\bibfnamefont {K.}~\bibnamefont {Bharadwaj}}, \bibinfo {author} {\bibfnamefont {J.}~\bibnamefont {Delaforce}}, \bibinfo {author} {\bibfnamefont {C.}~\bibnamefont {Naud}}, \bibinfo {author} {\bibfnamefont {W.}~\bibnamefont {Hasch-Guichard}}, \bibinfo {author} {\bibfnamefont {J.}~\bibnamefont {García-Ripoll}}, \bibinfo {author} {\bibfnamefont {N.}~\bibnamefont {Roch}},\ and\ \bibinfo {author} {\bibfnamefont {O.}~\bibnamefont {Buisson}},\ }\bibfield  {title} {\bibinfo
  {title} {Fast {High}-{Fidelity} {Quantum} {Nondemolition} {Qubit} {Readout} via a {Nonperturbative} {Cross}-{Kerr} {Coupling}},\ }\href {https://doi.org/10.1103/PhysRevX.10.011045} {\bibfield  {journal} {\bibinfo  {journal} {Phys. Rev. X}\ }\textbf {\bibinfo {volume} {10}},\ \bibinfo {pages} {011045} (\bibinfo {year} {2020})}\BibitemShut {NoStop}%
\bibitem [{\citenamefont {Pereira}\ \emph {et~al.}(2023)\citenamefont {Pereira}, \citenamefont {García-Ripoll},\ and\ \citenamefont {Ramos}}]{pereira_parallel_2022}%
  \BibitemOpen
  \bibfield  {author} {\bibinfo {author} {\bibfnamefont {L.}~\bibnamefont {Pereira}}, \bibinfo {author} {\bibfnamefont {J.~J.}\ \bibnamefont {García-Ripoll}},\ and\ \bibinfo {author} {\bibfnamefont {T.}~\bibnamefont {Ramos}},\ }\bibfield  {title} {\bibinfo {title} {Parallel tomography of quantum non-demolition measurements in multi-qubit devices},\ }\href {https://doi.org/10.1038/s41534-023-00688-7} {\bibfield  {journal} {\bibinfo  {journal} {npj Quantum Inf}\ }\textbf {\bibinfo {volume} {9}},\ \bibinfo {pages} {1} (\bibinfo {year} {2023})}\BibitemShut {NoStop}%
\bibitem [{\citenamefont {Aumentado}(2020)}]{aumentado_superconducting_2020}%
  \BibitemOpen
  \bibfield  {author} {\bibinfo {author} {\bibfnamefont {J.}~\bibnamefont {Aumentado}},\ }\bibfield  {title} {\bibinfo {title} {Superconducting {Parametric} {Amplifiers}: {The} {State} of the {Art} in {Josephson} {Parametric} {Amplifiers}},\ }\href {https://doi.org/10.1109/MMM.2020.2993476} {\bibfield  {journal} {\bibinfo  {journal} {IEEE Microwave Magazine}\ }\textbf {\bibinfo {volume} {21}},\ \bibinfo {pages} {45} (\bibinfo {year} {2020})}\BibitemShut {NoStop}%
\bibitem [{\citenamefont {Castellanos-Beltran}\ \emph {et~al.}(2008)\citenamefont {Castellanos-Beltran}, \citenamefont {Irwin}, \citenamefont {Hilton}, \citenamefont {Vale},\ and\ \citenamefont {Lehnert}}]{castellanos-beltran_amplification_2008}%
  \BibitemOpen
  \bibfield  {author} {\bibinfo {author} {\bibfnamefont {M.~A.}\ \bibnamefont {Castellanos-Beltran}}, \bibinfo {author} {\bibfnamefont {K.~D.}\ \bibnamefont {Irwin}}, \bibinfo {author} {\bibfnamefont {G.~C.}\ \bibnamefont {Hilton}}, \bibinfo {author} {\bibfnamefont {L.~R.}\ \bibnamefont {Vale}},\ and\ \bibinfo {author} {\bibfnamefont {K.~W.}\ \bibnamefont {Lehnert}},\ }\bibfield  {title} {\bibinfo {title} {Amplification and squeezing of quantum noise with a tunable {Josephson} metamaterial},\ }\href {https://doi.org/10.1038/nphys1090} {\bibfield  {journal} {\bibinfo  {journal} {Nature Phys}\ }\textbf {\bibinfo {volume} {4}},\ \bibinfo {pages} {929} (\bibinfo {year} {2008})}\BibitemShut {NoStop}%
\bibitem [{\citenamefont {Ho~Eom}\ \emph {et~al.}(2012)\citenamefont {Ho~Eom}, \citenamefont {Day}, \citenamefont {LeDuc},\ and\ \citenamefont {Zmuidzinas}}]{ho_eom_wideband_2012}%
  \BibitemOpen
  \bibfield  {author} {\bibinfo {author} {\bibfnamefont {B.}~\bibnamefont {Ho~Eom}}, \bibinfo {author} {\bibfnamefont {P.~K.}\ \bibnamefont {Day}}, \bibinfo {author} {\bibfnamefont {H.~G.}\ \bibnamefont {LeDuc}},\ and\ \bibinfo {author} {\bibfnamefont {J.}~\bibnamefont {Zmuidzinas}},\ }\bibfield  {title} {\bibinfo {title} {A wideband, low-noise superconducting amplifier with high dynamic range},\ }\href {https://doi.org/10.1038/nphys2356} {\bibfield  {journal} {\bibinfo  {journal} {Nature Phys}\ }\textbf {\bibinfo {volume} {8}},\ \bibinfo {pages} {623} (\bibinfo {year} {2012})}\BibitemShut {NoStop}%
\bibitem [{\citenamefont {Parker}\ \emph {et~al.}(2022)\citenamefont {Parker}, \citenamefont {Savytskyi}, \citenamefont {Vine}, \citenamefont {Laucht}, \citenamefont {Duty}, \citenamefont {Morello}, \citenamefont {Grimsmo},\ and\ \citenamefont {Pla}}]{parker_degenerate_2022}%
  \BibitemOpen
  \bibfield  {author} {\bibinfo {author} {\bibfnamefont {D.~J.}\ \bibnamefont {Parker}}, \bibinfo {author} {\bibfnamefont {M.}~\bibnamefont {Savytskyi}}, \bibinfo {author} {\bibfnamefont {W.}~\bibnamefont {Vine}}, \bibinfo {author} {\bibfnamefont {A.}~\bibnamefont {Laucht}}, \bibinfo {author} {\bibfnamefont {T.}~\bibnamefont {Duty}}, \bibinfo {author} {\bibfnamefont {A.}~\bibnamefont {Morello}}, \bibinfo {author} {\bibfnamefont {A.~L.}\ \bibnamefont {Grimsmo}},\ and\ \bibinfo {author} {\bibfnamefont {J.~J.}\ \bibnamefont {Pla}},\ }\bibfield  {title} {\bibinfo {title} {Degenerate {Parametric} {Amplification} via {Three}-{Wave} {Mixing} {Using} {Kinetic} {Inductance}},\ }\href {https://doi.org/10.1103/PhysRevApplied.17.034064} {\bibfield  {journal} {\bibinfo  {journal} {Phys. Rev. Applied}\ }\textbf {\bibinfo {volume} {17}},\ \bibinfo {pages} {034064} (\bibinfo {year} {2022})}\BibitemShut {NoStop}%
\bibitem [{\citenamefont {Frasca}\ \emph {et~al.}(2024)\citenamefont {Frasca}, \citenamefont {Roy}, \citenamefont {Beaulieu},\ and\ \citenamefont {Scarlino}}]{Frasca2024}%
  \BibitemOpen
  \bibfield  {author} {\bibinfo {author} {\bibfnamefont {S.}~\bibnamefont {Frasca}}, \bibinfo {author} {\bibfnamefont {C.}~\bibnamefont {Roy}}, \bibinfo {author} {\bibfnamefont {G.}~\bibnamefont {Beaulieu}},\ and\ \bibinfo {author} {\bibfnamefont {P.}~\bibnamefont {Scarlino}},\ }\bibfield  {title} {\bibinfo {title} {Three-wave-mixing quantum-limited kinetic inductance parametric amplifier operating at 6 t near 1 k},\ }\href {https://doi.org/10.1103/PhysRevApplied.21.024011} {\bibfield  {journal} {\bibinfo  {journal} {Phys. Rev. Appl.}\ }\textbf {\bibinfo {volume} {21}},\ \bibinfo {pages} {024011} (\bibinfo {year} {2024})}\BibitemShut {NoStop}%
\bibitem [{\citenamefont {Eichler}\ and\ \citenamefont {Wallraff}(2014)}]{eichler_controlling_2014}%
  \BibitemOpen
  \bibfield  {author} {\bibinfo {author} {\bibfnamefont {C.}~\bibnamefont {Eichler}}\ and\ \bibinfo {author} {\bibfnamefont {A.}~\bibnamefont {Wallraff}},\ }\bibfield  {title} {\bibinfo {title} {Controlling the dynamic range of a {Josephson} parametric amplifier},\ }\href {https://doi.org/10.1140/epjqt2} {\bibfield  {journal} {\bibinfo  {journal} {EPJ Quantum Technol.}\ }\textbf {\bibinfo {volume} {1}},\ \bibinfo {pages} {2} (\bibinfo {year} {2014})}\BibitemShut {NoStop}%
\bibitem [{\citenamefont {Roy}\ and\ \citenamefont {Devoret}(2016)}]{roy_introduction_2016}%
  \BibitemOpen
  \bibfield  {author} {\bibinfo {author} {\bibfnamefont {A.}~\bibnamefont {Roy}}\ and\ \bibinfo {author} {\bibfnamefont {M.}~\bibnamefont {Devoret}},\ }\bibfield  {title} {\bibinfo {title} {Introduction to parametric amplification of quantum signals with {Josephson} circuits},\ }\href {https://doi.org/10.1016/j.crhy.2016.07.012} {\bibfield  {journal} {\bibinfo  {journal} {Comptes Rendus Physique}\ }\bibinfo {series} {Quantum microwaves / {Micro}-ondes quantiques},\ \textbf {\bibinfo {volume} {17}},\ \bibinfo {pages} {740} (\bibinfo {year} {2016})}\BibitemShut {NoStop}%
\bibitem [{\citenamefont {Macklin}\ \emph {et~al.}(2015)\citenamefont {Macklin}, \citenamefont {O’Brien}, \citenamefont {Hover}, \citenamefont {Schwartz}, \citenamefont {Bolkhovsky}, \citenamefont {Zhang}, \citenamefont {Oliver},\ and\ \citenamefont {Siddiqi}}]{macklin_nearquantum-limited_2015}%
  \BibitemOpen
  \bibfield  {author} {\bibinfo {author} {\bibfnamefont {C.}~\bibnamefont {Macklin}}, \bibinfo {author} {\bibfnamefont {K.}~\bibnamefont {O’Brien}}, \bibinfo {author} {\bibfnamefont {D.}~\bibnamefont {Hover}}, \bibinfo {author} {\bibfnamefont {M.~E.}\ \bibnamefont {Schwartz}}, \bibinfo {author} {\bibfnamefont {V.}~\bibnamefont {Bolkhovsky}}, \bibinfo {author} {\bibfnamefont {X.}~\bibnamefont {Zhang}}, \bibinfo {author} {\bibfnamefont {W.~D.}\ \bibnamefont {Oliver}},\ and\ \bibinfo {author} {\bibfnamefont {I.}~\bibnamefont {Siddiqi}},\ }\bibfield  {title} {\bibinfo {title} {A near–quantum-limited {Josephson} traveling-wave parametric amplifier},\ }\href {https://doi.org/10.1126/science.aaa8525} {\bibfield  {journal} {\bibinfo  {journal} {Science}\ }\textbf {\bibinfo {volume} {350}},\ \bibinfo {pages} {307} (\bibinfo {year} {2015})}\BibitemShut {NoStop}%
\bibitem [{\citenamefont {White}\ \emph {et~al.}(2015)\citenamefont {White}, \citenamefont {Mutus}, \citenamefont {Hoi}, \citenamefont {Barends}, \citenamefont {Campbell}, \citenamefont {Chen}, \citenamefont {Chen}, \citenamefont {Chiaro}, \citenamefont {Dunsworth}, \citenamefont {Jeffrey}, \citenamefont {Kelly}, \citenamefont {Megrant}, \citenamefont {Neill}, \citenamefont {O'Malley}, \citenamefont {Roushan}, \citenamefont {Sank}, \citenamefont {Vainsencher}, \citenamefont {Wenner}, \citenamefont {Chaudhuri}, \citenamefont {Gao},\ and\ \citenamefont {Martinis}}]{white_traveling_2015}%
  \BibitemOpen
  \bibfield  {author} {\bibinfo {author} {\bibfnamefont {T.~C.}\ \bibnamefont {White}}, \bibinfo {author} {\bibfnamefont {J.~Y.}\ \bibnamefont {Mutus}}, \bibinfo {author} {\bibfnamefont {I.-C.}\ \bibnamefont {Hoi}}, \bibinfo {author} {\bibfnamefont {R.}~\bibnamefont {Barends}}, \bibinfo {author} {\bibfnamefont {B.}~\bibnamefont {Campbell}}, \bibinfo {author} {\bibfnamefont {Y.}~\bibnamefont {Chen}}, \bibinfo {author} {\bibfnamefont {Z.}~\bibnamefont {Chen}}, \bibinfo {author} {\bibfnamefont {B.}~\bibnamefont {Chiaro}}, \bibinfo {author} {\bibfnamefont {A.}~\bibnamefont {Dunsworth}}, \bibinfo {author} {\bibfnamefont {E.}~\bibnamefont {Jeffrey}}, \bibinfo {author} {\bibfnamefont {J.}~\bibnamefont {Kelly}}, \bibinfo {author} {\bibfnamefont {A.}~\bibnamefont {Megrant}}, \bibinfo {author} {\bibfnamefont {C.}~\bibnamefont {Neill}}, \bibinfo {author} {\bibfnamefont {P.~J.~J.}\ \bibnamefont {O'Malley}}, \bibinfo {author} {\bibfnamefont {P.}~\bibnamefont {Roushan}}, \bibinfo {author} {\bibfnamefont {D.}~\bibnamefont
  {Sank}}, \bibinfo {author} {\bibfnamefont {A.}~\bibnamefont {Vainsencher}}, \bibinfo {author} {\bibfnamefont {J.}~\bibnamefont {Wenner}}, \bibinfo {author} {\bibfnamefont {S.}~\bibnamefont {Chaudhuri}}, \bibinfo {author} {\bibfnamefont {J.}~\bibnamefont {Gao}},\ and\ \bibinfo {author} {\bibfnamefont {J.~M.}\ \bibnamefont {Martinis}},\ }\bibfield  {title} {\bibinfo {title} {Traveling wave parametric amplifier with {Josephson} junctions using minimal resonator phase matching},\ }\href {https://doi.org/10.1063/1.4922348} {\bibfield  {journal} {\bibinfo  {journal} {Appl. Phys. Lett.}\ }\textbf {\bibinfo {volume} {106}},\ \bibinfo {pages} {242601} (\bibinfo {year} {2015})}\BibitemShut {NoStop}%
\bibitem [{\citenamefont {Winkel}\ \emph {et~al.}(2020)\citenamefont {Winkel}, \citenamefont {Takmakov}, \citenamefont {Rieger}, \citenamefont {Planat}, \citenamefont {Hasch-Guichard}, \citenamefont {Grünhaupt}, \citenamefont {Maleeva}, \citenamefont {Foroughi}, \citenamefont {Henriques}, \citenamefont {Borisov}, \citenamefont {Ferrero}, \citenamefont {Ustinov}, \citenamefont {Wernsdorfer}, \citenamefont {Roch},\ and\ \citenamefont {Pop}}]{winkel_nondegenerate_2020}%
  \BibitemOpen
  \bibfield  {author} {\bibinfo {author} {\bibfnamefont {P.}~\bibnamefont {Winkel}}, \bibinfo {author} {\bibfnamefont {I.}~\bibnamefont {Takmakov}}, \bibinfo {author} {\bibfnamefont {D.}~\bibnamefont {Rieger}}, \bibinfo {author} {\bibfnamefont {L.}~\bibnamefont {Planat}}, \bibinfo {author} {\bibfnamefont {W.}~\bibnamefont {Hasch-Guichard}}, \bibinfo {author} {\bibfnamefont {L.}~\bibnamefont {Grünhaupt}}, \bibinfo {author} {\bibfnamefont {N.}~\bibnamefont {Maleeva}}, \bibinfo {author} {\bibfnamefont {F.}~\bibnamefont {Foroughi}}, \bibinfo {author} {\bibfnamefont {F.}~\bibnamefont {Henriques}}, \bibinfo {author} {\bibfnamefont {K.}~\bibnamefont {Borisov}}, \bibinfo {author} {\bibfnamefont {J.}~\bibnamefont {Ferrero}}, \bibinfo {author} {\bibfnamefont {A.~V.}\ \bibnamefont {Ustinov}}, \bibinfo {author} {\bibfnamefont {W.}~\bibnamefont {Wernsdorfer}}, \bibinfo {author} {\bibfnamefont {N.}~\bibnamefont {Roch}},\ and\ \bibinfo {author} {\bibfnamefont {I.~M.}\ \bibnamefont {Pop}},\ }\bibfield  {title} {\bibinfo
  {title} {Nondegenerate {Parametric} {Amplifiers} {Based} on {Dispersion}-{Engineered} {Josephson}-{Junction} {Arrays}},\ }\href {https://doi.org/10.1103/PhysRevApplied.13.024015} {\bibfield  {journal} {\bibinfo  {journal} {Phys. Rev. Applied}\ }\textbf {\bibinfo {volume} {13}},\ \bibinfo {pages} {024015} (\bibinfo {year} {2020})}\BibitemShut {NoStop}%
\bibitem [{\citenamefont {Planat}\ \emph {et~al.}(2020)\citenamefont {Planat}, \citenamefont {Ranadive}, \citenamefont {Dassonneville}, \citenamefont {Puertas~Martínez}, \citenamefont {Léger}, \citenamefont {Naud}, \citenamefont {Buisson}, \citenamefont {Hasch-Guichard}, \citenamefont {Basko},\ and\ \citenamefont {Roch}}]{planat_photonic-crystal_2020}%
  \BibitemOpen
  \bibfield  {author} {\bibinfo {author} {\bibfnamefont {L.}~\bibnamefont {Planat}}, \bibinfo {author} {\bibfnamefont {A.}~\bibnamefont {Ranadive}}, \bibinfo {author} {\bibfnamefont {R.}~\bibnamefont {Dassonneville}}, \bibinfo {author} {\bibfnamefont {J.}~\bibnamefont {Puertas~Martínez}}, \bibinfo {author} {\bibfnamefont {S.}~\bibnamefont {Léger}}, \bibinfo {author} {\bibfnamefont {C.}~\bibnamefont {Naud}}, \bibinfo {author} {\bibfnamefont {O.}~\bibnamefont {Buisson}}, \bibinfo {author} {\bibfnamefont {W.}~\bibnamefont {Hasch-Guichard}}, \bibinfo {author} {\bibfnamefont {D.~M.}\ \bibnamefont {Basko}},\ and\ \bibinfo {author} {\bibfnamefont {N.}~\bibnamefont {Roch}},\ }\bibfield  {title} {\bibinfo {title} {Photonic-{Crystal} {Josephson} {Traveling}-{Wave} {Parametric} {Amplifier}},\ }\href {https://doi.org/10.1103/PhysRevX.10.021021} {\bibfield  {journal} {\bibinfo  {journal} {Phys. Rev. X}\ }\textbf {\bibinfo {volume} {10}},\ \bibinfo {pages} {021021} (\bibinfo {year} {2020})}\BibitemShut {NoStop}%
\bibitem [{\citenamefont {Ranzani}\ \emph {et~al.}(2018)\citenamefont {Ranzani}, \citenamefont {Bal}, \citenamefont {Fong}, \citenamefont {Ribeill}, \citenamefont {Wu}, \citenamefont {Long}, \citenamefont {Ku}, \citenamefont {Erickson}, \citenamefont {Pappas},\ and\ \citenamefont {Ohki}}]{ranzani_kinetic_2018}%
  \BibitemOpen
  \bibfield  {author} {\bibinfo {author} {\bibfnamefont {L.}~\bibnamefont {Ranzani}}, \bibinfo {author} {\bibfnamefont {M.}~\bibnamefont {Bal}}, \bibinfo {author} {\bibfnamefont {K.~C.}\ \bibnamefont {Fong}}, \bibinfo {author} {\bibfnamefont {G.}~\bibnamefont {Ribeill}}, \bibinfo {author} {\bibfnamefont {X.}~\bibnamefont {Wu}}, \bibinfo {author} {\bibfnamefont {J.}~\bibnamefont {Long}}, \bibinfo {author} {\bibfnamefont {H.-S.}\ \bibnamefont {Ku}}, \bibinfo {author} {\bibfnamefont {R.~P.}\ \bibnamefont {Erickson}}, \bibinfo {author} {\bibfnamefont {D.}~\bibnamefont {Pappas}},\ and\ \bibinfo {author} {\bibfnamefont {T.~A.}\ \bibnamefont {Ohki}},\ }\bibfield  {title} {\bibinfo {title} {Kinetic inductance traveling-wave amplifiers for multiplexed qubit readout},\ }\href {https://doi.org/10.1063/1.5063252} {\bibfield  {journal} {\bibinfo  {journal} {Applied Physics Letters}\ }\textbf {\bibinfo {volume} {113}},\ \bibinfo {pages} {242602} (\bibinfo {year} {2018})}\BibitemShut {NoStop}%
\bibitem [{\citenamefont {Malnou}\ \emph {et~al.}(2021)\citenamefont {Malnou}, \citenamefont {Vissers}, \citenamefont {Wheeler}, \citenamefont {Aumentado}, \citenamefont {Hubmayr}, \citenamefont {Ullom},\ and\ \citenamefont {Gao}}]{malnou_three-wave_2021}%
  \BibitemOpen
  \bibfield  {author} {\bibinfo {author} {\bibfnamefont {M.}~\bibnamefont {Malnou}}, \bibinfo {author} {\bibfnamefont {M.}~\bibnamefont {Vissers}}, \bibinfo {author} {\bibfnamefont {J.}~\bibnamefont {Wheeler}}, \bibinfo {author} {\bibfnamefont {J.}~\bibnamefont {Aumentado}}, \bibinfo {author} {\bibfnamefont {J.}~\bibnamefont {Hubmayr}}, \bibinfo {author} {\bibfnamefont {J.}~\bibnamefont {Ullom}},\ and\ \bibinfo {author} {\bibfnamefont {J.}~\bibnamefont {Gao}},\ }\bibfield  {title} {\bibinfo {title} {Three-{Wave} {Mixing} {Kinetic} {Inductance} {Traveling}-{Wave} {Amplifier} with {Near}-{Quantum}-{Limited} {Noise} {Performance}},\ }\href {https://doi.org/10.1103/PRXQuantum.2.010302} {\bibfield  {journal} {\bibinfo  {journal} {PRX Quantum}\ }\textbf {\bibinfo {volume} {2}},\ \bibinfo {pages} {010302} (\bibinfo {year} {2021})}\BibitemShut {NoStop}%
\bibitem [{\citenamefont {Renberg~Nilsson}\ \emph {et~al.}(2023)\citenamefont {Renberg~Nilsson}, \citenamefont {Fadavi~Roudsari}, \citenamefont {Shiri}, \citenamefont {Delsing},\ and\ \citenamefont {Shumeiko}}]{renberg_nilsson_high-gain_2023}%
  \BibitemOpen
  \bibfield  {author} {\bibinfo {author} {\bibfnamefont {H.}~\bibnamefont {Renberg~Nilsson}}, \bibinfo {author} {\bibfnamefont {A.}~\bibnamefont {Fadavi~Roudsari}}, \bibinfo {author} {\bibfnamefont {D.}~\bibnamefont {Shiri}}, \bibinfo {author} {\bibfnamefont {P.}~\bibnamefont {Delsing}},\ and\ \bibinfo {author} {\bibfnamefont {V.}~\bibnamefont {Shumeiko}},\ }\bibfield  {title} {\bibinfo {title} {High-{Gain} {Traveling}-{Wave} {Parametric} {Amplifier} {Based} on {Three}-{Wave} {Mixing}},\ }\href {https://doi.org/10.1103/PhysRevApplied.19.044056} {\bibfield  {journal} {\bibinfo  {journal} {Phys. Rev. Appl.}\ }\textbf {\bibinfo {volume} {19}},\ \bibinfo {pages} {044056} (\bibinfo {year} {2023})}\BibitemShut {NoStop}%
\bibitem [{\citenamefont {Kuznetsov}\ \emph {et~al.}(2024)\citenamefont {Kuznetsov}, \citenamefont {Nardi}, \citenamefont {Riemensberger}, \citenamefont {Davydova}, \citenamefont {Churaev}, \citenamefont {Seidler},\ and\ \citenamefont {Kippenberg}}]{kuznetsov_ultra-broadband_2024}%
  \BibitemOpen
  \bibfield  {author} {\bibinfo {author} {\bibfnamefont {N.}~\bibnamefont {Kuznetsov}}, \bibinfo {author} {\bibfnamefont {A.}~\bibnamefont {Nardi}}, \bibinfo {author} {\bibfnamefont {J.}~\bibnamefont {Riemensberger}}, \bibinfo {author} {\bibfnamefont {A.}~\bibnamefont {Davydova}}, \bibinfo {author} {\bibfnamefont {M.}~\bibnamefont {Churaev}}, \bibinfo {author} {\bibfnamefont {P.}~\bibnamefont {Seidler}},\ and\ \bibinfo {author} {\bibfnamefont {T.~J.}\ \bibnamefont {Kippenberg}},\ }\bibfield  {title} {\bibinfo {title} {An ultra-broadband photonic-chip-based traveling-wave parametric amplifier},\ }\href {http://arxiv.org/abs/2404.08609} {\bibfield  {journal} {\bibinfo  {journal} {arXiv:2404.08609}\ } (\bibinfo {year} {2024})}\BibitemShut {NoStop}%
\bibitem [{\citenamefont {Esposito}\ \emph {et~al.}(2021)\citenamefont {Esposito}, \citenamefont {Ranadive}, \citenamefont {Planat},\ and\ \citenamefont {Roch}}]{esposito_perspective_2021}%
  \BibitemOpen
  \bibfield  {author} {\bibinfo {author} {\bibfnamefont {M.}~\bibnamefont {Esposito}}, \bibinfo {author} {\bibfnamefont {A.}~\bibnamefont {Ranadive}}, \bibinfo {author} {\bibfnamefont {L.}~\bibnamefont {Planat}},\ and\ \bibinfo {author} {\bibfnamefont {N.}~\bibnamefont {Roch}},\ }\bibfield  {title} {\bibinfo {title} {Perspective on traveling wave microwave parametric amplifiers},\ }\href {https://doi.org/10.1063/5.0064892} {\bibfield  {journal} {\bibinfo  {journal} {Appl. Phys. Lett.}\ }\textbf {\bibinfo {volume} {119}},\ \bibinfo {pages} {120501} (\bibinfo {year} {2021})}\BibitemShut {NoStop}%
\bibitem [{\citenamefont {Bell}\ and\ \citenamefont {Samolov}(2015)}]{bell_traveling-wave_2015}%
  \BibitemOpen
  \bibfield  {author} {\bibinfo {author} {\bibfnamefont {M.}~\bibnamefont {Bell}}\ and\ \bibinfo {author} {\bibfnamefont {A.}~\bibnamefont {Samolov}},\ }\bibfield  {title} {\bibinfo {title} {Traveling-{Wave} {Parametric} {Amplifier} {Based} on a {Chain} of {Coupled} {Asymmetric} {SQUIDs}},\ }\href {https://doi.org/10.1103/PhysRevApplied.4.024014} {\bibfield  {journal} {\bibinfo  {journal} {Phys. Rev. Appl.}\ }\textbf {\bibinfo {volume} {4}},\ \bibinfo {pages} {024014} (\bibinfo {year} {2015})}\BibitemShut {NoStop}%
\bibitem [{\citenamefont {Kow}\ \emph {et~al.}(2022)\citenamefont {Kow}, \citenamefont {Podolskiy},\ and\ \citenamefont {Kamal}}]{kow_self_2024}%
  \BibitemOpen
  \bibfield  {author} {\bibinfo {author} {\bibfnamefont {C.}~\bibnamefont {Kow}}, \bibinfo {author} {\bibfnamefont {V.}~\bibnamefont {Podolskiy}},\ and\ \bibinfo {author} {\bibfnamefont {A.}~\bibnamefont {Kamal}},\ }\bibfield  {title} {\bibinfo {title} {Self phase-matched broadband amplification with a left-handed {Josephson} transmission line},\ }\href {http://arxiv.org/abs/2201.04660} {\bibfield  {journal} {\bibinfo  {journal} {arXiv:2201.04660}\ } (\bibinfo {year} {2022})}\BibitemShut {NoStop}%
\bibitem [{\citenamefont {Abdo}\ \emph {et~al.}(2013)\citenamefont {Abdo}, \citenamefont {Sliwa}, \citenamefont {Frunzio},\ and\ \citenamefont {Devoret}}]{abdo_directional_2013}%
  \BibitemOpen
  \bibfield  {author} {\bibinfo {author} {\bibfnamefont {B.}~\bibnamefont {Abdo}}, \bibinfo {author} {\bibfnamefont {K.}~\bibnamefont {Sliwa}}, \bibinfo {author} {\bibfnamefont {L.}~\bibnamefont {Frunzio}},\ and\ \bibinfo {author} {\bibfnamefont {M.}~\bibnamefont {Devoret}},\ }\bibfield  {title} {\bibinfo {title} {Directional {Amplification} with a {Josephson} {Circuit}},\ }\href {https://doi.org/10.1103/PhysRevX.3.031001} {\bibfield  {journal} {\bibinfo  {journal} {Phys. Rev. X}\ }\textbf {\bibinfo {volume} {3}},\ \bibinfo {pages} {031001} (\bibinfo {year} {2013})}\BibitemShut {NoStop}%
\bibitem [{\citenamefont {Sliwa}\ \emph {et~al.}(2015)\citenamefont {Sliwa}, \citenamefont {Hatridge}, \citenamefont {Narla}, \citenamefont {Shankar}, \citenamefont {Frunzio}, \citenamefont {Schoelkopf},\ and\ \citenamefont {Devoret}}]{sliwa_reconfigurable_2015}%
  \BibitemOpen
  \bibfield  {author} {\bibinfo {author} {\bibfnamefont {K.}~\bibnamefont {Sliwa}}, \bibinfo {author} {\bibfnamefont {M.}~\bibnamefont {Hatridge}}, \bibinfo {author} {\bibfnamefont {A.}~\bibnamefont {Narla}}, \bibinfo {author} {\bibfnamefont {S.}~\bibnamefont {Shankar}}, \bibinfo {author} {\bibfnamefont {L.}~\bibnamefont {Frunzio}}, \bibinfo {author} {\bibfnamefont {R.}~\bibnamefont {Schoelkopf}},\ and\ \bibinfo {author} {\bibfnamefont {M.}~\bibnamefont {Devoret}},\ }\bibfield  {title} {\bibinfo {title} {Reconfigurable {Josephson} {Circulator}/{Directional} {Amplifier}},\ }\href {https://doi.org/10.1103/PhysRevX.5.041020} {\bibfield  {journal} {\bibinfo  {journal} {Phys. Rev. X}\ }\textbf {\bibinfo {volume} {5}},\ \bibinfo {pages} {041020} (\bibinfo {year} {2015})}\BibitemShut {NoStop}%
\bibitem [{\citenamefont {Lecocq}\ \emph {et~al.}(2017)\citenamefont {Lecocq}, \citenamefont {Ranzani}, \citenamefont {Peterson}, \citenamefont {Cicak}, \citenamefont {Simmonds}, \citenamefont {Teufel},\ and\ \citenamefont {Aumentado}}]{lecocq_nonreciprocal_2017}%
  \BibitemOpen
  \bibfield  {author} {\bibinfo {author} {\bibfnamefont {F.}~\bibnamefont {Lecocq}}, \bibinfo {author} {\bibfnamefont {L.}~\bibnamefont {Ranzani}}, \bibinfo {author} {\bibfnamefont {G.}~\bibnamefont {Peterson}}, \bibinfo {author} {\bibfnamefont {K.}~\bibnamefont {Cicak}}, \bibinfo {author} {\bibfnamefont {R.}~\bibnamefont {Simmonds}}, \bibinfo {author} {\bibfnamefont {J.}~\bibnamefont {Teufel}},\ and\ \bibinfo {author} {\bibfnamefont {J.}~\bibnamefont {Aumentado}},\ }\bibfield  {title} {\bibinfo {title} {Nonreciprocal {Microwave} {Signal} {Processing} with a {Field}-{Programmable} {Josephson} {Amplifier}},\ }\href {https://doi.org/10.1103/PhysRevApplied.7.024028} {\bibfield  {journal} {\bibinfo  {journal} {Phys. Rev. Applied}\ }\textbf {\bibinfo {volume} {7}},\ \bibinfo {pages} {024028} (\bibinfo {year} {2017})}\BibitemShut {NoStop}%
\bibitem [{\citenamefont {Metelmann}\ and\ \citenamefont {Clerk}(2015)}]{metelmann_nonreciprocal_2015}%
  \BibitemOpen
  \bibfield  {author} {\bibinfo {author} {\bibfnamefont {A.}~\bibnamefont {Metelmann}}\ and\ \bibinfo {author} {\bibfnamefont {A.}~\bibnamefont {Clerk}},\ }\bibfield  {title} {\bibinfo {title} {Nonreciprocal {Photon} {Transmission} and {Amplification} via {Reservoir} {Engineering}},\ }\href {https://doi.org/10.1103/PhysRevX.5.021025} {\bibfield  {journal} {\bibinfo  {journal} {Phys. Rev. X}\ }\textbf {\bibinfo {volume} {5}},\ \bibinfo {pages} {021025} (\bibinfo {year} {2015})}\BibitemShut {NoStop}%
\bibitem [{\citenamefont {Fang}\ \emph {et~al.}(2017)\citenamefont {Fang}, \citenamefont {Luo}, \citenamefont {Metelmann}, \citenamefont {Matheny}, \citenamefont {Marquardt}, \citenamefont {Clerk},\ and\ \citenamefont {Painter}}]{fang_generalized_2017}%
  \BibitemOpen
  \bibfield  {author} {\bibinfo {author} {\bibfnamefont {K.}~\bibnamefont {Fang}}, \bibinfo {author} {\bibfnamefont {J.}~\bibnamefont {Luo}}, \bibinfo {author} {\bibfnamefont {A.}~\bibnamefont {Metelmann}}, \bibinfo {author} {\bibfnamefont {M.~H.}\ \bibnamefont {Matheny}}, \bibinfo {author} {\bibfnamefont {F.}~\bibnamefont {Marquardt}}, \bibinfo {author} {\bibfnamefont {A.~A.}\ \bibnamefont {Clerk}},\ and\ \bibinfo {author} {\bibfnamefont {O.}~\bibnamefont {Painter}},\ }\bibfield  {title} {\bibinfo {title} {Generalized non-reciprocity in an optomechanical circuit via synthetic magnetism and reservoir engineering},\ }\href {https://doi.org/10.1038/nphys4009} {\bibfield  {journal} {\bibinfo  {journal} {Nature Phys}\ }\textbf {\bibinfo {volume} {13}},\ \bibinfo {pages} {465} (\bibinfo {year} {2017})}\BibitemShut {NoStop}%
\bibitem [{\citenamefont {Pucher}\ \emph {et~al.}(2022)\citenamefont {Pucher}, \citenamefont {Liedl}, \citenamefont {Jin}, \citenamefont {Rauschenbeutel},\ and\ \citenamefont {Schneeweiss}}]{pucher_atomic_2022}%
  \BibitemOpen
  \bibfield  {author} {\bibinfo {author} {\bibfnamefont {S.}~\bibnamefont {Pucher}}, \bibinfo {author} {\bibfnamefont {C.}~\bibnamefont {Liedl}}, \bibinfo {author} {\bibfnamefont {S.}~\bibnamefont {Jin}}, \bibinfo {author} {\bibfnamefont {A.}~\bibnamefont {Rauschenbeutel}},\ and\ \bibinfo {author} {\bibfnamefont {P.}~\bibnamefont {Schneeweiss}},\ }\bibfield  {title} {\bibinfo {title} {Atomic spin-controlled non-reciprocal {Raman} amplification of fibre-guided light},\ }\href {https://doi.org/10.1038/s41566-022-00987-z} {\bibfield  {journal} {\bibinfo  {journal} {Nat. Photon.}\ }\textbf {\bibinfo {volume} {16}},\ \bibinfo {pages} {380} (\bibinfo {year} {2022})}\BibitemShut {NoStop}%
\bibitem [{\citenamefont {Heinsoo}\ \emph {et~al.}(2018)\citenamefont {Heinsoo}, \citenamefont {Andersen}, \citenamefont {Remm}, \citenamefont {Krinner}, \citenamefont {Walter}, \citenamefont {Salathé}, \citenamefont {Gasparinetti}, \citenamefont {Besse}, \citenamefont {Potočnik}, \citenamefont {Wallraff},\ and\ \citenamefont {Eichler}}]{heinsoo_rapid_2018}%
  \BibitemOpen
  \bibfield  {author} {\bibinfo {author} {\bibfnamefont {J.}~\bibnamefont {Heinsoo}}, \bibinfo {author} {\bibfnamefont {C.~K.}\ \bibnamefont {Andersen}}, \bibinfo {author} {\bibfnamefont {A.}~\bibnamefont {Remm}}, \bibinfo {author} {\bibfnamefont {S.}~\bibnamefont {Krinner}}, \bibinfo {author} {\bibfnamefont {T.}~\bibnamefont {Walter}}, \bibinfo {author} {\bibfnamefont {Y.}~\bibnamefont {Salathé}}, \bibinfo {author} {\bibfnamefont {S.}~\bibnamefont {Gasparinetti}}, \bibinfo {author} {\bibfnamefont {J.-C.}\ \bibnamefont {Besse}}, \bibinfo {author} {\bibfnamefont {A.}~\bibnamefont {Potočnik}}, \bibinfo {author} {\bibfnamefont {A.}~\bibnamefont {Wallraff}},\ and\ \bibinfo {author} {\bibfnamefont {C.}~\bibnamefont {Eichler}},\ }\bibfield  {title} {\bibinfo {title} {Rapid {High}-fidelity {Multiplexed} {Readout} of {Superconducting} {Qubits}},\ }\href {https://doi.org/10.1103/PhysRevApplied.10.034040} {\bibfield  {journal} {\bibinfo  {journal} {Phys. Rev. Applied}\ }\textbf {\bibinfo {volume} {10}},\ \bibinfo
  {pages} {034040} (\bibinfo {year} {2018})}\BibitemShut {NoStop}%
\bibitem [{\citenamefont {Peano}\ \emph {et~al.}(2016)\citenamefont {Peano}, \citenamefont {Houde}, \citenamefont {Marquardt},\ and\ \citenamefont {Clerk}}]{peano_topological_2016}%
  \BibitemOpen
  \bibfield  {author} {\bibinfo {author} {\bibfnamefont {V.}~\bibnamefont {Peano}}, \bibinfo {author} {\bibfnamefont {M.}~\bibnamefont {Houde}}, \bibinfo {author} {\bibfnamefont {F.}~\bibnamefont {Marquardt}},\ and\ \bibinfo {author} {\bibfnamefont {A.~A.}\ \bibnamefont {Clerk}},\ }\bibfield  {title} {\bibinfo {title} {Topological {Quantum} {Fluctuations} and {Traveling} {Wave} {Amplifiers}},\ }\href {https://doi.org/10.1103/PhysRevX.6.041026} {\bibfield  {journal} {\bibinfo  {journal} {Phys. Rev. X}\ }\textbf {\bibinfo {volume} {6}},\ \bibinfo {pages} {041026} (\bibinfo {year} {2016})}\BibitemShut {NoStop}%
\bibitem [{\citenamefont {Porras}\ and\ \citenamefont {Fernández-Lorenzo}(2019)}]{porras_topological_2019}%
  \BibitemOpen
  \bibfield  {author} {\bibinfo {author} {\bibfnamefont {D.}~\bibnamefont {Porras}}\ and\ \bibinfo {author} {\bibfnamefont {S.}~\bibnamefont {Fernández-Lorenzo}},\ }\bibfield  {title} {\bibinfo {title} {Topological {Amplification} in {Photonic} {Lattices}},\ }\href {https://doi.org/10.1103/PhysRevLett.122.143901} {\bibfield  {journal} {\bibinfo  {journal} {Phys. Rev. Lett.}\ }\textbf {\bibinfo {volume} {122}},\ \bibinfo {pages} {143901} (\bibinfo {year} {2019})}\BibitemShut {NoStop}%
\bibitem [{\citenamefont {Wanjura}\ \emph {et~al.}(2020)\citenamefont {Wanjura}, \citenamefont {Brunelli},\ and\ \citenamefont {Nunnenkamp}}]{wanjura_topological_2020}%
  \BibitemOpen
  \bibfield  {author} {\bibinfo {author} {\bibfnamefont {C.~C.}\ \bibnamefont {Wanjura}}, \bibinfo {author} {\bibfnamefont {M.}~\bibnamefont {Brunelli}},\ and\ \bibinfo {author} {\bibfnamefont {A.}~\bibnamefont {Nunnenkamp}},\ }\bibfield  {title} {\bibinfo {title} {Topological framework for directional amplification in driven-dissipative cavity arrays},\ }\href {https://doi.org/10.1038/s41467-020-16863-9} {\bibfield  {journal} {\bibinfo  {journal} {Nature Communications}\ }\textbf {\bibinfo {volume} {11}},\ \bibinfo {pages} {3149} (\bibinfo {year} {2020})}\BibitemShut {NoStop}%
\bibitem [{\citenamefont {Ramos}\ \emph {et~al.}(2021)\citenamefont {Ramos}, \citenamefont {García-Ripoll},\ and\ \citenamefont {Porras}}]{ramos_topological_2021}%
  \BibitemOpen
  \bibfield  {author} {\bibinfo {author} {\bibfnamefont {T.}~\bibnamefont {Ramos}}, \bibinfo {author} {\bibfnamefont {J.~J.}\ \bibnamefont {García-Ripoll}},\ and\ \bibinfo {author} {\bibfnamefont {D.}~\bibnamefont {Porras}},\ }\bibfield  {title} {\bibinfo {title} {Topological input-output theory for directional amplification},\ }\href {https://doi.org/10.1103/PhysRevA.103.033513} {\bibfield  {journal} {\bibinfo  {journal} {Phys. Rev. A}\ }\textbf {\bibinfo {volume} {103}},\ \bibinfo {pages} {033513} (\bibinfo {year} {2021})}\BibitemShut {NoStop}%
\bibitem [{\citenamefont {Ozawa}\ \emph {et~al.}(2019)\citenamefont {Ozawa}, \citenamefont {Price}, \citenamefont {Amo}, \citenamefont {Goldman}, \citenamefont {Hafezi}, \citenamefont {Lu}, \citenamefont {Rechtsman}, \citenamefont {Schuster}, \citenamefont {Simon}, \citenamefont {Zilberberg},\ and\ \citenamefont {Carusotto}}]{ozawa_topological_2019}%
  \BibitemOpen
  \bibfield  {author} {\bibinfo {author} {\bibfnamefont {T.}~\bibnamefont {Ozawa}}, \bibinfo {author} {\bibfnamefont {H.~M.}\ \bibnamefont {Price}}, \bibinfo {author} {\bibfnamefont {A.}~\bibnamefont {Amo}}, \bibinfo {author} {\bibfnamefont {N.}~\bibnamefont {Goldman}}, \bibinfo {author} {\bibfnamefont {M.}~\bibnamefont {Hafezi}}, \bibinfo {author} {\bibfnamefont {L.}~\bibnamefont {Lu}}, \bibinfo {author} {\bibfnamefont {M.~C.}\ \bibnamefont {Rechtsman}}, \bibinfo {author} {\bibfnamefont {D.}~\bibnamefont {Schuster}}, \bibinfo {author} {\bibfnamefont {J.}~\bibnamefont {Simon}}, \bibinfo {author} {\bibfnamefont {O.}~\bibnamefont {Zilberberg}},\ and\ \bibinfo {author} {\bibfnamefont {I.}~\bibnamefont {Carusotto}},\ }\bibfield  {title} {\bibinfo {title} {Topological photonics},\ }\href {https://doi.org/10.1103/RevModPhys.91.015006} {\bibfield  {journal} {\bibinfo  {journal} {Rev. Mod. Phys.}\ }\textbf {\bibinfo {volume} {91}},\ \bibinfo {pages} {015006} (\bibinfo {year} {2019})}\BibitemShut {NoStop}%
\bibitem [{\citenamefont {Kreikebaum}\ \emph {et~al.}(2020)\citenamefont {Kreikebaum}, \citenamefont {O’Brien}, \citenamefont {Morvan},\ and\ \citenamefont {Siddiqi}}]{kreikebaum_improving_2020}%
  \BibitemOpen
  \bibfield  {author} {\bibinfo {author} {\bibfnamefont {J.~M.}\ \bibnamefont {Kreikebaum}}, \bibinfo {author} {\bibfnamefont {K.~P.}\ \bibnamefont {O’Brien}}, \bibinfo {author} {\bibfnamefont {A.}~\bibnamefont {Morvan}},\ and\ \bibinfo {author} {\bibfnamefont {I.}~\bibnamefont {Siddiqi}},\ }\bibfield  {title} {\bibinfo {title} {Improving wafer-scale {Josephson} junction resistance variation in superconducting quantum coherent circuits},\ }\href {https://doi.org/10.1088/1361-6668/ab8617} {\bibfield  {journal} {\bibinfo  {journal} {Supercond. Sci. Technol.}\ }\textbf {\bibinfo {volume} {33}},\ \bibinfo {pages} {06LT02} (\bibinfo {year} {2020})}\BibitemShut {NoStop}%
\bibitem [{\citenamefont {Eichler}\ \emph {et~al.}(2014)\citenamefont {Eichler}, \citenamefont {Salathe}, \citenamefont {Mlynek}, \citenamefont {Schmidt},\ and\ \citenamefont {Wallraff}}]{eichler_quantum-limited_2014}%
  \BibitemOpen
  \bibfield  {author} {\bibinfo {author} {\bibfnamefont {C.}~\bibnamefont {Eichler}}, \bibinfo {author} {\bibfnamefont {Y.}~\bibnamefont {Salathe}}, \bibinfo {author} {\bibfnamefont {J.}~\bibnamefont {Mlynek}}, \bibinfo {author} {\bibfnamefont {S.}~\bibnamefont {Schmidt}},\ and\ \bibinfo {author} {\bibfnamefont {A.}~\bibnamefont {Wallraff}},\ }\bibfield  {title} {\bibinfo {title} {Quantum-{Limited} {Amplification} and {Entanglement} in {Coupled} {Nonlinear} {Resonators}},\ }\href {https://doi.org/10.1103/PhysRevLett.113.110502} {\bibfield  {journal} {\bibinfo  {journal} {Phys. Rev. Lett.}\ }\textbf {\bibinfo {volume} {113}},\ \bibinfo {pages} {110502} (\bibinfo {year} {2014})}\BibitemShut {NoStop}%
\bibitem [{\citenamefont {Mutus}\ \emph {et~al.}(2014)\citenamefont {Mutus}, \citenamefont {White}, \citenamefont {Barends}, \citenamefont {Chen}, \citenamefont {Chen}, \citenamefont {Chiaro}, \citenamefont {Dunsworth}, \citenamefont {Jeffrey}, \citenamefont {Kelly}, \citenamefont {Megrant}, \citenamefont {Neill}, \citenamefont {O'Malley}, \citenamefont {Roushan}, \citenamefont {Sank}, \citenamefont {Vainsencher}, \citenamefont {Wenner}, \citenamefont {Sundqvist}, \citenamefont {Cleland},\ and\ \citenamefont {Martinis}}]{mutus_strong_2014}%
  \BibitemOpen
  \bibfield  {author} {\bibinfo {author} {\bibfnamefont {J.~Y.}\ \bibnamefont {Mutus}}, \bibinfo {author} {\bibfnamefont {T.~C.}\ \bibnamefont {White}}, \bibinfo {author} {\bibfnamefont {R.}~\bibnamefont {Barends}}, \bibinfo {author} {\bibfnamefont {Y.}~\bibnamefont {Chen}}, \bibinfo {author} {\bibfnamefont {Z.}~\bibnamefont {Chen}}, \bibinfo {author} {\bibfnamefont {B.}~\bibnamefont {Chiaro}}, \bibinfo {author} {\bibfnamefont {A.}~\bibnamefont {Dunsworth}}, \bibinfo {author} {\bibfnamefont {E.}~\bibnamefont {Jeffrey}}, \bibinfo {author} {\bibfnamefont {J.}~\bibnamefont {Kelly}}, \bibinfo {author} {\bibfnamefont {A.}~\bibnamefont {Megrant}}, \bibinfo {author} {\bibfnamefont {C.}~\bibnamefont {Neill}}, \bibinfo {author} {\bibfnamefont {P.~J.~J.}\ \bibnamefont {O'Malley}}, \bibinfo {author} {\bibfnamefont {P.}~\bibnamefont {Roushan}}, \bibinfo {author} {\bibfnamefont {D.}~\bibnamefont {Sank}}, \bibinfo {author} {\bibfnamefont {A.}~\bibnamefont {Vainsencher}}, \bibinfo {author} {\bibfnamefont {J.}~\bibnamefont
  {Wenner}}, \bibinfo {author} {\bibfnamefont {K.~M.}\ \bibnamefont {Sundqvist}}, \bibinfo {author} {\bibfnamefont {A.~N.}\ \bibnamefont {Cleland}},\ and\ \bibinfo {author} {\bibfnamefont {J.~M.}\ \bibnamefont {Martinis}},\ }\bibfield  {title} {\bibinfo {title} {Strong environmental coupling in a {Josephson} parametric amplifier},\ }\href {https://doi.org/10.1063/1.4886408} {\bibfield  {journal} {\bibinfo  {journal} {Appl. Phys. Lett.}\ }\textbf {\bibinfo {volume} {104}},\ \bibinfo {pages} {263513} (\bibinfo {year} {2014})}\BibitemShut {NoStop}%
\bibitem [{\citenamefont {Kounalakis}\ \emph {et~al.}(2018)\citenamefont {Kounalakis}, \citenamefont {Dickel}, \citenamefont {Bruno}, \citenamefont {Langford},\ and\ \citenamefont {Steele}}]{kounalakis_tuneable_2018}%
  \BibitemOpen
  \bibfield  {author} {\bibinfo {author} {\bibfnamefont {M.}~\bibnamefont {Kounalakis}}, \bibinfo {author} {\bibfnamefont {C.}~\bibnamefont {Dickel}}, \bibinfo {author} {\bibfnamefont {A.}~\bibnamefont {Bruno}}, \bibinfo {author} {\bibfnamefont {N.~K.}\ \bibnamefont {Langford}},\ and\ \bibinfo {author} {\bibfnamefont {G.~A.}\ \bibnamefont {Steele}},\ }\bibfield  {title} {\bibinfo {title} {Tuneable hopping and nonlinear cross-{Kerr} interactions in a high-coherence superconducting circuit},\ }\href {https://doi.org/10.1038/s41534-018-0088-9} {\bibfield  {journal} {\bibinfo  {journal} {npj Quantum Inf}\ }\textbf {\bibinfo {volume} {4}},\ \bibinfo {pages} {38} (\bibinfo {year} {2018})}\BibitemShut {NoStop}%
\bibitem [{\citenamefont {Ranadive}\ \emph {et~al.}(2024)\citenamefont {Ranadive}, \citenamefont {Fazliji}, \citenamefont {Gal}, \citenamefont {Cappelli}, \citenamefont {Butseraen}, \citenamefont {Bonet}, \citenamefont {Eyraud}, \citenamefont {Böhling}, \citenamefont {Planat}, \citenamefont {Metelmann},\ and\ \citenamefont {Roch}}]{ranadive_traveling_2024}%
  \BibitemOpen
  \bibfield  {author} {\bibinfo {author} {\bibfnamefont {A.}~\bibnamefont {Ranadive}}, \bibinfo {author} {\bibfnamefont {B.}~\bibnamefont {Fazliji}}, \bibinfo {author} {\bibfnamefont {G.~L.}\ \bibnamefont {Gal}}, \bibinfo {author} {\bibfnamefont {G.}~\bibnamefont {Cappelli}}, \bibinfo {author} {\bibfnamefont {G.}~\bibnamefont {Butseraen}}, \bibinfo {author} {\bibfnamefont {E.}~\bibnamefont {Bonet}}, \bibinfo {author} {\bibfnamefont {E.}~\bibnamefont {Eyraud}}, \bibinfo {author} {\bibfnamefont {S.}~\bibnamefont {Böhling}}, \bibinfo {author} {\bibfnamefont {L.}~\bibnamefont {Planat}}, \bibinfo {author} {\bibfnamefont {A.}~\bibnamefont {Metelmann}},\ and\ \bibinfo {author} {\bibfnamefont {N.}~\bibnamefont {Roch}},\ }\bibfield  {title} {\bibinfo {title} {A {Traveling} {Wave} {Parametric} {Amplifier} {Isolator}},\ }\href {http://arxiv.org/abs/2406.19752} {\bibfield  {journal} {\bibinfo  {journal} {arXiv:2406.19752}\ } (\bibinfo {year} {2024})}\BibitemShut {NoStop}%
\bibitem [{\citenamefont {Malnou}\ \emph {et~al.}(2024)\citenamefont {Malnou}, \citenamefont {Miller}, \citenamefont {Estrada}, \citenamefont {Genter}, \citenamefont {Cicak}, \citenamefont {Teufel}, \citenamefont {Aumentado},\ and\ \citenamefont {Lecocq}}]{malnou_traveling-wave_2024}%
  \BibitemOpen
  \bibfield  {author} {\bibinfo {author} {\bibfnamefont {M.}~\bibnamefont {Malnou}}, \bibinfo {author} {\bibfnamefont {B.~T.}\ \bibnamefont {Miller}}, \bibinfo {author} {\bibfnamefont {J.~A.}\ \bibnamefont {Estrada}}, \bibinfo {author} {\bibfnamefont {K.}~\bibnamefont {Genter}}, \bibinfo {author} {\bibfnamefont {K.}~\bibnamefont {Cicak}}, \bibinfo {author} {\bibfnamefont {J.~D.}\ \bibnamefont {Teufel}}, \bibinfo {author} {\bibfnamefont {J.}~\bibnamefont {Aumentado}},\ and\ \bibinfo {author} {\bibfnamefont {F.}~\bibnamefont {Lecocq}},\ }\bibfield  {title} {\bibinfo {title} {A {Traveling}-{Wave} {Parametric} {Amplifier} and {Converter}},\ }\href {http://arxiv.org/abs/2406.19476} {\bibfield  {journal} {\bibinfo  {journal} {arXiv:2406.19476}\ } (\bibinfo {year} {2024})}\BibitemShut {NoStop}%
\bibitem [{\citenamefont {Jouanny}\ \emph {et~al.}(2024)\citenamefont {Jouanny}, \citenamefont {Frasca}, \citenamefont {Weibel}, \citenamefont {Peyruchat}, \citenamefont {Scigliuzzo}, \citenamefont {Oppliger}, \citenamefont {Palma}, \citenamefont {Sbroggio}, \citenamefont {Beaulieu}, \citenamefont {Zilberberg},\ and\ \citenamefont {Scarlino}}]{jouanny_band_2024}%
  \BibitemOpen
  \bibfield  {author} {\bibinfo {author} {\bibfnamefont {V.}~\bibnamefont {Jouanny}}, \bibinfo {author} {\bibfnamefont {S.}~\bibnamefont {Frasca}}, \bibinfo {author} {\bibfnamefont {V.~J.}\ \bibnamefont {Weibel}}, \bibinfo {author} {\bibfnamefont {L.}~\bibnamefont {Peyruchat}}, \bibinfo {author} {\bibfnamefont {M.}~\bibnamefont {Scigliuzzo}}, \bibinfo {author} {\bibfnamefont {F.}~\bibnamefont {Oppliger}}, \bibinfo {author} {\bibfnamefont {F.~D.}\ \bibnamefont {Palma}}, \bibinfo {author} {\bibfnamefont {D.}~\bibnamefont {Sbroggio}}, \bibinfo {author} {\bibfnamefont {G.}~\bibnamefont {Beaulieu}}, \bibinfo {author} {\bibfnamefont {O.}~\bibnamefont {Zilberberg}},\ and\ \bibinfo {author} {\bibfnamefont {P.}~\bibnamefont {Scarlino}},\ }\bibfield  {title} {\bibinfo {title} {Band engineering and study of disorder using topology in compact high kinetic inductance cavity arrays},\ }\href {http://arxiv.org/abs/2403.18150} {\bibfield  {journal} {\bibinfo  {journal} {arXiv:2403.18150}\ } (\bibinfo {year}
  {2024})}\BibitemShut {NoStop}%
\bibitem [{\citenamefont {Roushan}\ \emph {et~al.}(2017)\citenamefont {Roushan}, \citenamefont {Neill}, \citenamefont {Megrant}, \citenamefont {Chen}, \citenamefont {Babbush}, \citenamefont {Barends}, \citenamefont {Campbell}, \citenamefont {Chen}, \citenamefont {Chiaro}, \citenamefont {Dunsworth}, \citenamefont {Fowler}, \citenamefont {Jeffrey}, \citenamefont {Kelly}, \citenamefont {Lucero}, \citenamefont {Mutus}, \citenamefont {O’Malley}, \citenamefont {Neeley}, \citenamefont {Quintana}, \citenamefont {Sank}, \citenamefont {Vainsencher}, \citenamefont {Wenner}, \citenamefont {White}, \citenamefont {Kapit}, \citenamefont {Neven},\ and\ \citenamefont {Martinis}}]{roushan_chiral_2017}%
  \BibitemOpen
  \bibfield  {author} {\bibinfo {author} {\bibfnamefont {P.}~\bibnamefont {Roushan}}, \bibinfo {author} {\bibfnamefont {C.}~\bibnamefont {Neill}}, \bibinfo {author} {\bibfnamefont {A.}~\bibnamefont {Megrant}}, \bibinfo {author} {\bibfnamefont {Y.}~\bibnamefont {Chen}}, \bibinfo {author} {\bibfnamefont {R.}~\bibnamefont {Babbush}}, \bibinfo {author} {\bibfnamefont {R.}~\bibnamefont {Barends}}, \bibinfo {author} {\bibfnamefont {B.}~\bibnamefont {Campbell}}, \bibinfo {author} {\bibfnamefont {Z.}~\bibnamefont {Chen}}, \bibinfo {author} {\bibfnamefont {B.}~\bibnamefont {Chiaro}}, \bibinfo {author} {\bibfnamefont {A.}~\bibnamefont {Dunsworth}}, \bibinfo {author} {\bibfnamefont {A.}~\bibnamefont {Fowler}}, \bibinfo {author} {\bibfnamefont {E.}~\bibnamefont {Jeffrey}}, \bibinfo {author} {\bibfnamefont {J.}~\bibnamefont {Kelly}}, \bibinfo {author} {\bibfnamefont {E.}~\bibnamefont {Lucero}}, \bibinfo {author} {\bibfnamefont {J.}~\bibnamefont {Mutus}}, \bibinfo {author} {\bibfnamefont {P.~J.~J.}\ \bibnamefont
  {O’Malley}}, \bibinfo {author} {\bibfnamefont {M.}~\bibnamefont {Neeley}}, \bibinfo {author} {\bibfnamefont {C.}~\bibnamefont {Quintana}}, \bibinfo {author} {\bibfnamefont {D.}~\bibnamefont {Sank}}, \bibinfo {author} {\bibfnamefont {A.}~\bibnamefont {Vainsencher}}, \bibinfo {author} {\bibfnamefont {J.}~\bibnamefont {Wenner}}, \bibinfo {author} {\bibfnamefont {T.}~\bibnamefont {White}}, \bibinfo {author} {\bibfnamefont {E.}~\bibnamefont {Kapit}}, \bibinfo {author} {\bibfnamefont {H.}~\bibnamefont {Neven}},\ and\ \bibinfo {author} {\bibfnamefont {J.}~\bibnamefont {Martinis}},\ }\bibfield  {title} {\bibinfo {title} {Chiral ground-state currents of interacting photons in a synthetic magnetic field},\ }\href {https://doi.org/10.1038/nphys3930} {\bibfield  {journal} {\bibinfo  {journal} {Nature Physics}\ }\textbf {\bibinfo {volume} {13}},\ \bibinfo {pages} {146} (\bibinfo {year} {2017})}\BibitemShut {NoStop}%
\bibitem [{\citenamefont {Peng}\ \emph {et~al.}(2022)\citenamefont {Peng}, \citenamefont {Naghiloo}, \citenamefont {Wang}, \citenamefont {Cunningham}, \citenamefont {Ye},\ and\ \citenamefont {O'Brien}}]{Peng2022}%
  \BibitemOpen
  \bibfield  {author} {\bibinfo {author} {\bibfnamefont {K.}~\bibnamefont {Peng}}, \bibinfo {author} {\bibfnamefont {M.}~\bibnamefont {Naghiloo}}, \bibinfo {author} {\bibfnamefont {J.}~\bibnamefont {Wang}}, \bibinfo {author} {\bibfnamefont {G.~D.}\ \bibnamefont {Cunningham}}, \bibinfo {author} {\bibfnamefont {Y.}~\bibnamefont {Ye}},\ and\ \bibinfo {author} {\bibfnamefont {K.~P.}\ \bibnamefont {O'Brien}},\ }\bibfield  {title} {\bibinfo {title} {Floquet-mode traveling-wave parametric amplifiers},\ }\href {https://doi.org/10.1103/PRXQuantum.3.020306} {\bibfield  {journal} {\bibinfo  {journal} {PRX Quantum}\ }\textbf {\bibinfo {volume} {3}},\ \bibinfo {pages} {020306} (\bibinfo {year} {2022})}\BibitemShut {NoStop}%
\bibitem [{\citenamefont {Carrasco}\ \emph {et~al.}(2022)\citenamefont {Carrasco}, \citenamefont {Valenzuela}, \citenamefont {Falcón}, \citenamefont {Finger},\ and\ \citenamefont {Mena}}]{carrasco_effect_2022}%
  \BibitemOpen
  \bibfield  {author} {\bibinfo {author} {\bibfnamefont {J.}~\bibnamefont {Carrasco}}, \bibinfo {author} {\bibfnamefont {D.}~\bibnamefont {Valenzuela}}, \bibinfo {author} {\bibfnamefont {C.}~\bibnamefont {Falcón}}, \bibinfo {author} {\bibfnamefont {R.}~\bibnamefont {Finger}},\ and\ \bibinfo {author} {\bibfnamefont {F.~P.}\ \bibnamefont {Mena}},\ }\bibfield  {title} {\bibinfo {title} {The effect of complex dispersion and characteristic impedance on the gain of superconducting traveling-wave kinetic inductance parametric amplifiers},\ }\href {http://arxiv.org/abs/2210.00626} {\bibfield  {journal} {\bibinfo  {journal} {arXiv:2210.00626}\ } (\bibinfo {year} {2022})}\BibitemShut {NoStop}%
\bibitem [{\citenamefont {Yurke}\ and\ \citenamefont {Denker}(1984)}]{yurke_quantum_1984}%
  \BibitemOpen
  \bibfield  {author} {\bibinfo {author} {\bibfnamefont {B.}~\bibnamefont {Yurke}}\ and\ \bibinfo {author} {\bibfnamefont {J.~S.}\ \bibnamefont {Denker}},\ }\bibfield  {title} {\bibinfo {title} {Quantum network theory},\ }\href {https://doi.org/10.1103/PhysRevA.29.1419} {\bibfield  {journal} {\bibinfo  {journal} {Phys. Rev. A}\ }\textbf {\bibinfo {volume} {29}},\ \bibinfo {pages} {1419} (\bibinfo {year} {1984})}\BibitemShut {NoStop}%
\bibitem [{\citenamefont {Gardiner}\ and\ \citenamefont {Zoller}(2004)}]{QuantumNoise}%
  \BibitemOpen
  \bibfield  {author} {\bibinfo {author} {\bibfnamefont {C.}~\bibnamefont {Gardiner}}\ and\ \bibinfo {author} {\bibfnamefont {P.}~\bibnamefont {Zoller}},\ }\href@noop {} {\emph {\bibinfo {title} {Quantum Noise}}}\ (\bibinfo  {publisher} {SpringerVerlag, Berlin, 3rd Ed.},\ \bibinfo {year} {2004})\BibitemShut {NoStop}%
\bibitem [{\citenamefont {Planat}\ \emph {et~al.}(2019)\citenamefont {Planat}, \citenamefont {Dassonneville}, \citenamefont {Martínez}, \citenamefont {Foroughi}, \citenamefont {Buisson}, \citenamefont {Hasch-Guichard}, \citenamefont {Naud}, \citenamefont {Vijay}, \citenamefont {Murch},\ and\ \citenamefont {Roch}}]{planat_understanding_2019}%
  \BibitemOpen
  \bibfield  {author} {\bibinfo {author} {\bibfnamefont {L.}~\bibnamefont {Planat}}, \bibinfo {author} {\bibfnamefont {R.}~\bibnamefont {Dassonneville}}, \bibinfo {author} {\bibfnamefont {J.~P.}\ \bibnamefont {Martínez}}, \bibinfo {author} {\bibfnamefont {F.}~\bibnamefont {Foroughi}}, \bibinfo {author} {\bibfnamefont {O.}~\bibnamefont {Buisson}}, \bibinfo {author} {\bibfnamefont {W.}~\bibnamefont {Hasch-Guichard}}, \bibinfo {author} {\bibfnamefont {C.}~\bibnamefont {Naud}}, \bibinfo {author} {\bibfnamefont {R.}~\bibnamefont {Vijay}}, \bibinfo {author} {\bibfnamefont {K.}~\bibnamefont {Murch}},\ and\ \bibinfo {author} {\bibfnamefont {N.}~\bibnamefont {Roch}},\ }\bibfield  {title} {\bibinfo {title} {Understanding the {Saturation} {Power} of {Josephson} {Parametric} {Amplifiers} {Made} from {SQUID} {Arrays}},\ }\href {https://doi.org/10.1103/PhysRevApplied.11.034014} {\bibfield  {journal} {\bibinfo  {journal} {Phys. Rev. Applied}\ }\textbf {\bibinfo {volume} {11}},\ \bibinfo {pages} {034014} (\bibinfo {year}
  {2019})}\BibitemShut {NoStop}%
\bibitem [{\citenamefont {Walls}\ and\ \citenamefont {Milburn}(2008)}]{WallsMilburnBook}%
  \BibitemOpen
  \bibfield  {author} {\bibinfo {author} {\bibfnamefont {D.}~\bibnamefont {Walls}}\ and\ \bibinfo {author} {\bibfnamefont {G.}~\bibnamefont {Milburn}},\ }\href@noop {} {\emph {\bibinfo {title} {Quantum Optics}}}\ (\bibinfo  {publisher} {SpringerVerlag, Berlin, 2nd Ed.},\ \bibinfo {year} {2008})\BibitemShut {NoStop}%
\bibitem [{\citenamefont {Koch}\ \emph {et~al.}(2010)\citenamefont {Koch}, \citenamefont {Houck}, \citenamefont {Hur},\ and\ \citenamefont {Girvin}}]{Koch2010}%
  \BibitemOpen
  \bibfield  {author} {\bibinfo {author} {\bibfnamefont {J.}~\bibnamefont {Koch}}, \bibinfo {author} {\bibfnamefont {A.~A.}\ \bibnamefont {Houck}}, \bibinfo {author} {\bibfnamefont {K.~L.}\ \bibnamefont {Hur}},\ and\ \bibinfo {author} {\bibfnamefont {S.~M.}\ \bibnamefont {Girvin}},\ }\bibfield  {title} {\bibinfo {title} {Time-reversal-symmetry breaking in circuit-qed-based photon lattices},\ }\href {https://doi.org/10.1103/PhysRevA.82.043811} {\bibfield  {journal} {\bibinfo  {journal} {Phys. Rev. A}\ }\textbf {\bibinfo {volume} {82}},\ \bibinfo {pages} {043811} (\bibinfo {year} {2010})}\BibitemShut {NoStop}%
\bibitem [{\citenamefont {G\'omez-Leon}\ \emph {et~al.}(2023)\citenamefont {G\'omez-Leon}, \citenamefont {Ramos}, \citenamefont {Gonz\'alez-Tudela},\ and\ \citenamefont {Porras}}]{GomezLeon2022}%
  \BibitemOpen
  \bibfield  {author} {\bibinfo {author} {\bibfnamefont {A.}~\bibnamefont {G\'omez-Leon}}, \bibinfo {author} {\bibfnamefont {T.}~\bibnamefont {Ramos}}, \bibinfo {author} {\bibfnamefont {A.}~\bibnamefont {Gonz\'alez-Tudela}},\ and\ \bibinfo {author} {\bibfnamefont {D.}~\bibnamefont {Porras}},\ }\bibfield  {title} {\bibinfo {title} {Driven-dissipative topological phases in parametric resonator arrays},\ }\href {https://doi.org/10.22331/q-2023-05-23-1016} {\bibfield  {journal} {\bibinfo  {journal} {Quantum}\ }\textbf {\bibinfo {volume} {7}},\ \bibinfo {pages} {1016} (\bibinfo {year} {2023})}\BibitemShut {NoStop}%
\bibitem [{\citenamefont {Yeh}\ \emph {et~al.}(2017)\citenamefont {Yeh}, \citenamefont {LeFebvre}, \citenamefont {Premaratne}, \citenamefont {Wellstood},\ and\ \citenamefont {Palmer}}]{yeh_microwave_2017}%
  \BibitemOpen
  \bibfield  {author} {\bibinfo {author} {\bibfnamefont {J.-H.}\ \bibnamefont {Yeh}}, \bibinfo {author} {\bibfnamefont {J.}~\bibnamefont {LeFebvre}}, \bibinfo {author} {\bibfnamefont {S.}~\bibnamefont {Premaratne}}, \bibinfo {author} {\bibfnamefont {F.~C.}\ \bibnamefont {Wellstood}},\ and\ \bibinfo {author} {\bibfnamefont {B.~S.}\ \bibnamefont {Palmer}},\ }\bibfield  {title} {\bibinfo {title} {Microwave attenuators for use with quantum devices below 100 {mK}},\ }\href {https://doi.org/10.1063/1.4984894} {\bibfield  {journal} {\bibinfo  {journal} {Journal of Applied Physics}\ }\textbf {\bibinfo {volume} {121}},\ \bibinfo {pages} {224501} (\bibinfo {year} {2017})}\BibitemShut {NoStop}%
\bibitem [{\citenamefont {Yeh}\ \emph {et~al.}(2019)\citenamefont {Yeh}, \citenamefont {Huang}, \citenamefont {Zhang}, \citenamefont {Premaratne}, \citenamefont {LeFebvre}, \citenamefont {Wellstood},\ and\ \citenamefont {Palmer}}]{yeh_hot_2019}%
  \BibitemOpen
  \bibfield  {author} {\bibinfo {author} {\bibfnamefont {J.-H.}\ \bibnamefont {Yeh}}, \bibinfo {author} {\bibfnamefont {Y.}~\bibnamefont {Huang}}, \bibinfo {author} {\bibfnamefont {R.}~\bibnamefont {Zhang}}, \bibinfo {author} {\bibfnamefont {S.}~\bibnamefont {Premaratne}}, \bibinfo {author} {\bibfnamefont {J.}~\bibnamefont {LeFebvre}}, \bibinfo {author} {\bibfnamefont {F.~C.}\ \bibnamefont {Wellstood}},\ and\ \bibinfo {author} {\bibfnamefont {B.~S.}\ \bibnamefont {Palmer}},\ }\bibfield  {title} {\bibinfo {title} {Hot electron heatsinks for microwave attenuators below 100{mK}},\ }\href {https://doi.org/10.1063/1.5097369} {\bibfield  {journal} {\bibinfo  {journal} {Appl. Phys. Lett.}\ }\textbf {\bibinfo {volume} {114}},\ \bibinfo {pages} {152602} (\bibinfo {year} {2019})}\BibitemShut {NoStop}%
\bibitem [{\citenamefont {Li}\ \emph {et~al.}(2024)\citenamefont {Li}, \citenamefont {García-Ripoll},\ and\ \citenamefont {Ramos}}]{li_scalable_2024}%
  \BibitemOpen
  \bibfield  {author} {\bibinfo {author} {\bibfnamefont {M.}~\bibnamefont {Li}}, \bibinfo {author} {\bibfnamefont {J.~J.}\ \bibnamefont {García-Ripoll}},\ and\ \bibinfo {author} {\bibfnamefont {T.}~\bibnamefont {Ramos}},\ }\bibfield  {title} {\bibinfo {title} {Scalable multiphoton generation from cavity-synchronized single-photon sources},\ }\href {https://doi.org/10.1103/PhysRevResearch.6.033295} {\bibfield  {journal} {\bibinfo  {journal} {Phys. Rev. Res.}\ }\textbf {\bibinfo {volume} {6}},\ \bibinfo {pages} {033295} (\bibinfo {year} {2024})}\BibitemShut {NoStop}%
\bibitem [{\citenamefont {Caves}\ \emph {et~al.}(2012)\citenamefont {Caves}, \citenamefont {Combes}, \citenamefont {Jiang},\ and\ \citenamefont {Pandey}}]{caves_quantum_2012}%
  \BibitemOpen
  \bibfield  {author} {\bibinfo {author} {\bibfnamefont {C.~M.}\ \bibnamefont {Caves}}, \bibinfo {author} {\bibfnamefont {J.}~\bibnamefont {Combes}}, \bibinfo {author} {\bibfnamefont {Z.}~\bibnamefont {Jiang}},\ and\ \bibinfo {author} {\bibfnamefont {S.}~\bibnamefont {Pandey}},\ }\bibfield  {title} {\bibinfo {title} {Quantum limits on phase-preserving linear amplifiers},\ }\href {https://doi.org/10.1103/PhysRevA.86.063802} {\bibfield  {journal} {\bibinfo  {journal} {Phys. Rev. A}\ }\textbf {\bibinfo {volume} {86}},\ \bibinfo {pages} {063802} (\bibinfo {year} {2012})}\BibitemShut {NoStop}%
\bibitem [{Note1()}]{Note1}%
  \BibitemOpen
  \bibinfo {note} {We use the convention of normally ordered noise moments leading to the fundamental lower bound $n_{\protect \rm add}^j(\omega )\geq 1$ \cite {QuantumNoise}. Alternatively, using the symmetrized convention, the same limit corresponds to $n_{\protect \rm add}^j(\omega )\geq 1/2$ \cite {caves_quantum_1982}}\BibitemShut {NoStop}%
\bibitem [{\citenamefont {G\'{o}mez-Le\'{o}n}\ \emph {et~al.}(2022)\citenamefont {G\'{o}mez-Le\'{o}n}, \citenamefont {Ramos}, \citenamefont {Gonz\'{a}lez-Tudela},\ and\ \citenamefont {Porras}}]{gomezleon_bridging_2021}%
  \BibitemOpen
  \bibfield  {author} {\bibinfo {author} {\bibfnamefont {A.}~\bibnamefont {G\'{o}mez-Le\'{o}n}}, \bibinfo {author} {\bibfnamefont {T.}~\bibnamefont {Ramos}}, \bibinfo {author} {\bibfnamefont {A.}~\bibnamefont {Gonz\'{a}lez-Tudela}},\ and\ \bibinfo {author} {\bibfnamefont {D.}~\bibnamefont {Porras}},\ }\bibfield  {title} {\bibinfo {title} {Bridging the gap between topological non-hermitian physics and open quantum systems},\ }\href {https://doi.org/https://doi.org/10.1103/PhysRevA.106.L011501} {\bibfield  {journal} {\bibinfo  {journal} {Phys. Rev. A}\ }\textbf {\bibinfo {volume} {106}},\ \bibinfo {pages} {L011501} (\bibinfo {year} {2022})}\BibitemShut {NoStop}%
\bibitem [{\citenamefont {Ryu}\ \emph {et~al.}(2010)\citenamefont {Ryu}, \citenamefont {Schnyder}, \citenamefont {Furusaki},\ and\ \citenamefont {Ludwig}}]{ryu_topological_2010}%
  \BibitemOpen
  \bibfield  {author} {\bibinfo {author} {\bibfnamefont {S.}~\bibnamefont {Ryu}}, \bibinfo {author} {\bibfnamefont {A.~P.}\ \bibnamefont {Schnyder}}, \bibinfo {author} {\bibfnamefont {A.}~\bibnamefont {Furusaki}},\ and\ \bibinfo {author} {\bibfnamefont {A.~W.~W.}\ \bibnamefont {Ludwig}},\ }\bibfield  {title} {\bibinfo {title} {Topological insulators and superconductors: tenfold way and dimensional hierarchy},\ }\href {https://doi.org/10.1088/1367-2630/12/6/065010} {\bibfield  {journal} {\bibinfo  {journal} {New J. Phys.}\ }\textbf {\bibinfo {volume} {12}},\ \bibinfo {pages} {065010} (\bibinfo {year} {2010})}\BibitemShut {NoStop}%
\bibitem [{\citenamefont {Ye}\ \emph {et~al.}(2021)\citenamefont {Ye}, \citenamefont {Peng}, \citenamefont {Naghiloo}, \citenamefont {Cunningham},\ and\ \citenamefont {O’Brien}}]{ye_engineering_2021}%
  \BibitemOpen
  \bibfield  {author} {\bibinfo {author} {\bibfnamefont {Y.}~\bibnamefont {Ye}}, \bibinfo {author} {\bibfnamefont {K.}~\bibnamefont {Peng}}, \bibinfo {author} {\bibfnamefont {M.}~\bibnamefont {Naghiloo}}, \bibinfo {author} {\bibfnamefont {G.}~\bibnamefont {Cunningham}},\ and\ \bibinfo {author} {\bibfnamefont {K.~P.}\ \bibnamefont {O’Brien}},\ }\bibfield  {title} {\bibinfo {title} {Engineering {Purely} {Nonlinear} {Coupling} between {Superconducting} {Qubits} {Using} a {Quarton}},\ }\href {https://doi.org/10.1103/PhysRevLett.127.050502} {\bibfield  {journal} {\bibinfo  {journal} {Phys. Rev. Lett.}\ }\textbf {\bibinfo {volume} {127}},\ \bibinfo {pages} {050502} (\bibinfo {year} {2021})}\BibitemShut {NoStop}%
\bibitem [{\citenamefont {Ye}\ \emph {et~al.}(2024{\natexlab{a}})\citenamefont {Ye}, \citenamefont {Kline}, \citenamefont {Yen}, \citenamefont {Cunningham}, \citenamefont {Tan}, \citenamefont {Zang}, \citenamefont {Gingras}, \citenamefont {Niedzielski}, \citenamefont {Stickler}, \citenamefont {Serniak}, \citenamefont {Schwartz},\ and\ \citenamefont {O'Brien}}]{quarton_exp}%
  \BibitemOpen
  \bibfield  {author} {\bibinfo {author} {\bibfnamefont {Y.}~\bibnamefont {Ye}}, \bibinfo {author} {\bibfnamefont {J.~B.}\ \bibnamefont {Kline}}, \bibinfo {author} {\bibfnamefont {A.}~\bibnamefont {Yen}}, \bibinfo {author} {\bibfnamefont {G.}~\bibnamefont {Cunningham}}, \bibinfo {author} {\bibfnamefont {M.}~\bibnamefont {Tan}}, \bibinfo {author} {\bibfnamefont {A.}~\bibnamefont {Zang}}, \bibinfo {author} {\bibfnamefont {M.}~\bibnamefont {Gingras}}, \bibinfo {author} {\bibfnamefont {B.~M.}\ \bibnamefont {Niedzielski}}, \bibinfo {author} {\bibfnamefont {H.}~\bibnamefont {Stickler}}, \bibinfo {author} {\bibfnamefont {K.}~\bibnamefont {Serniak}}, \bibinfo {author} {\bibfnamefont {M.~E.}\ \bibnamefont {Schwartz}},\ and\ \bibinfo {author} {\bibfnamefont {K.~P.}\ \bibnamefont {O'Brien}},\ }\bibfield  {title} {\bibinfo {title} {Near-ultrastrong nonlinear light-matter coupling in superconducting circuits},\ }\href {http://arxiv.org/abs/2404.19199} {\bibfield  {journal} {\bibinfo  {journal} {arXiv:2404.19199}\ }
  (\bibinfo {year} {2024}{\natexlab{a}})}\BibitemShut {NoStop}%
\bibitem [{\citenamefont {Ye}\ \emph {et~al.}(2024{\natexlab{b}})\citenamefont {Ye}, \citenamefont {Kline}, \citenamefont {Chen},\ and\ \citenamefont {O'Brien}}]{quarton_readout}%
  \BibitemOpen
  \bibfield  {author} {\bibinfo {author} {\bibfnamefont {Y.}~\bibnamefont {Ye}}, \bibinfo {author} {\bibfnamefont {J.~B.}\ \bibnamefont {Kline}}, \bibinfo {author} {\bibfnamefont {S.}~\bibnamefont {Chen}},\ and\ \bibinfo {author} {\bibfnamefont {K.~P.}\ \bibnamefont {O'Brien}},\ }\bibfield  {title} {\bibinfo {title} {Ultrafast superconducting qubit readout with the quarton coupler},\ }\href {http://arxiv.org/abs/2402.15664} {\bibfield  {journal} {\bibinfo  {journal} {arXiv:2402.15664}\ } (\bibinfo {year} {2024}{\natexlab{b}})}\BibitemShut {NoStop}%
\bibitem [{\citenamefont {Wanjura}\ \emph {et~al.}(2021)\citenamefont {Wanjura}, \citenamefont {Brunelli},\ and\ \citenamefont {Nunnenkamp}}]{wanjura_correspondence_2021}%
  \BibitemOpen
  \bibfield  {author} {\bibinfo {author} {\bibfnamefont {C.~C.}\ \bibnamefont {Wanjura}}, \bibinfo {author} {\bibfnamefont {M.}~\bibnamefont {Brunelli}},\ and\ \bibinfo {author} {\bibfnamefont {A.}~\bibnamefont {Nunnenkamp}},\ }\bibfield  {title} {\bibinfo {title} {Correspondence between {Non}-{Hermitian} {Topology} and {Directional} {Amplification} in the {Presence} of {Disorder}},\ }\href {https://doi.org/10.1103/PhysRevLett.127.213601} {\bibfield  {journal} {\bibinfo  {journal} {Phys. Rev. Lett.}\ }\textbf {\bibinfo {volume} {127}},\ \bibinfo {pages} {213601} (\bibinfo {year} {2021})}\BibitemShut {NoStop}%
\bibitem [{\citenamefont {Stützer}\ \emph {et~al.}(2018)\citenamefont {Stützer}, \citenamefont {Plotnik}, \citenamefont {Lumer}, \citenamefont {Titum}, \citenamefont {Lindner}, \citenamefont {Segev}, \citenamefont {Rechtsman},\ and\ \citenamefont {Szameit}}]{stutzer_photonic_2018}%
  \BibitemOpen
  \bibfield  {author} {\bibinfo {author} {\bibfnamefont {S.}~\bibnamefont {Stützer}}, \bibinfo {author} {\bibfnamefont {Y.}~\bibnamefont {Plotnik}}, \bibinfo {author} {\bibfnamefont {Y.}~\bibnamefont {Lumer}}, \bibinfo {author} {\bibfnamefont {P.}~\bibnamefont {Titum}}, \bibinfo {author} {\bibfnamefont {N.~H.}\ \bibnamefont {Lindner}}, \bibinfo {author} {\bibfnamefont {M.}~\bibnamefont {Segev}}, \bibinfo {author} {\bibfnamefont {M.~C.}\ \bibnamefont {Rechtsman}},\ and\ \bibinfo {author} {\bibfnamefont {A.}~\bibnamefont {Szameit}},\ }\bibfield  {title} {\bibinfo {title} {Photonic topological {Anderson} insulators},\ }\href {https://doi.org/10.1038/s41586-018-0418-2} {\bibfield  {journal} {\bibinfo  {journal} {Nature}\ }\textbf {\bibinfo {volume} {560}},\ \bibinfo {pages} {461} (\bibinfo {year} {2018})}\BibitemShut {NoStop}%
\bibitem [{Pat(2024)}]{PatentApp}%
  \BibitemOpen
  \href@noop {} {\bibinfo {title} {European patent application ep24382293.9}} (\bibinfo {year} {2024})\BibitemShut {NoStop}%
\bibitem [{\citenamefont {Grimsmo}\ \emph {et~al.}(2021)\citenamefont {Grimsmo}, \citenamefont {Royer}, \citenamefont {Kreikebaum}, \citenamefont {Ye}, \citenamefont {O’Brien}, \citenamefont {Siddiqi},\ and\ \citenamefont {Blais}}]{grimsmo_quantum_2021}%
  \BibitemOpen
  \bibfield  {author} {\bibinfo {author} {\bibfnamefont {A.~L.}\ \bibnamefont {Grimsmo}}, \bibinfo {author} {\bibfnamefont {B.}~\bibnamefont {Royer}}, \bibinfo {author} {\bibfnamefont {J.~M.}\ \bibnamefont {Kreikebaum}}, \bibinfo {author} {\bibfnamefont {Y.}~\bibnamefont {Ye}}, \bibinfo {author} {\bibfnamefont {K.}~\bibnamefont {O’Brien}}, \bibinfo {author} {\bibfnamefont {I.}~\bibnamefont {Siddiqi}},\ and\ \bibinfo {author} {\bibfnamefont {A.}~\bibnamefont {Blais}},\ }\bibfield  {title} {\bibinfo {title} {Quantum {Metamaterial} for {Broadband} {Detection} of {Single} {Microwave} {Photons}},\ }\href {https://doi.org/10.1103/PhysRevApplied.15.034074} {\bibfield  {journal} {\bibinfo  {journal} {Phys. Rev. Applied}\ }\textbf {\bibinfo {volume} {15}},\ \bibinfo {pages} {034074} (\bibinfo {year} {2021})}\BibitemShut {NoStop}%
\bibitem [{\citenamefont {Remm}\ \emph {et~al.}(2023)\citenamefont {Remm}, \citenamefont {Krinner}, \citenamefont {Lacroix}, \citenamefont {Hellings}, \citenamefont {Swiadek}, \citenamefont {Norris}, \citenamefont {Eichler},\ and\ \citenamefont {Wallraff}}]{remm_intermodulation_2022}%
  \BibitemOpen
  \bibfield  {author} {\bibinfo {author} {\bibfnamefont {A.}~\bibnamefont {Remm}}, \bibinfo {author} {\bibfnamefont {S.}~\bibnamefont {Krinner}}, \bibinfo {author} {\bibfnamefont {N.}~\bibnamefont {Lacroix}}, \bibinfo {author} {\bibfnamefont {C.}~\bibnamefont {Hellings}}, \bibinfo {author} {\bibfnamefont {F.~m.~c.}\ \bibnamefont {Swiadek}}, \bibinfo {author} {\bibfnamefont {G.~J.}\ \bibnamefont {Norris}}, \bibinfo {author} {\bibfnamefont {C.}~\bibnamefont {Eichler}},\ and\ \bibinfo {author} {\bibfnamefont {A.}~\bibnamefont {Wallraff}},\ }\bibfield  {title} {\bibinfo {title} {Intermodulation distortion in a josephson traveling-wave parametric amplifier},\ }\href {https://doi.org/10.1103/PhysRevApplied.20.034027} {\bibfield  {journal} {\bibinfo  {journal} {Phys. Rev. Appl.}\ }\textbf {\bibinfo {volume} {20}},\ \bibinfo {pages} {034027} (\bibinfo {year} {2023})}\BibitemShut {NoStop}%
\bibitem [{\citenamefont {Menu}\ and\ \citenamefont {Roscilde}(2023)}]{menu_gaussian-state_2023}%
  \BibitemOpen
  \bibfield  {author} {\bibinfo {author} {\bibfnamefont {R.}~\bibnamefont {Menu}}\ and\ \bibinfo {author} {\bibfnamefont {T.}~\bibnamefont {Roscilde}},\ }\bibfield  {title} {\bibinfo {title} {Gaussian-state {Ansatz} for the non-equilibrium dynamics of quantum spin lattices},\ }\href {https://doi.org/10.21468/SciPostPhys.14.6.151} {\bibfield  {journal} {\bibinfo  {journal} {SciPost Physics}\ }\textbf {\bibinfo {volume} {14}},\ \bibinfo {pages} {151} (\bibinfo {year} {2023})}\BibitemShut {NoStop}%
\bibitem [{\citenamefont {Vicentini}\ \emph {et~al.}(2018)\citenamefont {Vicentini}, \citenamefont {Minganti}, \citenamefont {Rota}, \citenamefont {Orso},\ and\ \citenamefont {Ciuti}}]{Vicentini18}%
  \BibitemOpen
  \bibfield  {author} {\bibinfo {author} {\bibfnamefont {F.}~\bibnamefont {Vicentini}}, \bibinfo {author} {\bibfnamefont {F.}~\bibnamefont {Minganti}}, \bibinfo {author} {\bibfnamefont {R.}~\bibnamefont {Rota}}, \bibinfo {author} {\bibfnamefont {G.}~\bibnamefont {Orso}},\ and\ \bibinfo {author} {\bibfnamefont {C.}~\bibnamefont {Ciuti}},\ }\bibfield  {title} {\bibinfo {title} {Critical slowing down in driven-dissipative bose-hubbard lattices},\ }\href {https://doi.org/10.1103/PhysRevA.97.013853} {\bibfield  {journal} {\bibinfo  {journal} {Phys. Rev. A}\ }\textbf {\bibinfo {volume} {97}},\ \bibinfo {pages} {013853} (\bibinfo {year} {2018})}\BibitemShut {NoStop}%
\bibitem [{\citenamefont {Daley}(2014)}]{daley_quantum_2014}%
  \BibitemOpen
  \bibfield  {author} {\bibinfo {author} {\bibfnamefont {A.~J.}\ \bibnamefont {Daley}},\ }\bibfield  {title} {\bibinfo {title} {Quantum trajectories and open many-body quantum systems},\ }\href {https://doi.org/10.1080/00018732.2014.933502} {\bibfield  {journal} {\bibinfo  {journal} {Advances in Physics}\ }\textbf {\bibinfo {volume} {63}},\ \bibinfo {pages} {77} (\bibinfo {year} {2014})}\BibitemShut {NoStop}%
\bibitem [{\citenamefont {Guimond}\ \emph {et~al.}(2020)\citenamefont {Guimond}, \citenamefont {Vermersch}, \citenamefont {Juan}, \citenamefont {Sharafiev}, \citenamefont {Kirchmair},\ and\ \citenamefont {Zoller}}]{guimond_unidirectional_2020}%
  \BibitemOpen
  \bibfield  {author} {\bibinfo {author} {\bibfnamefont {P.-O.}\ \bibnamefont {Guimond}}, \bibinfo {author} {\bibfnamefont {B.}~\bibnamefont {Vermersch}}, \bibinfo {author} {\bibfnamefont {M.~L.}\ \bibnamefont {Juan}}, \bibinfo {author} {\bibfnamefont {A.}~\bibnamefont {Sharafiev}}, \bibinfo {author} {\bibfnamefont {G.}~\bibnamefont {Kirchmair}},\ and\ \bibinfo {author} {\bibfnamefont {P.}~\bibnamefont {Zoller}},\ }\bibfield  {title} {\bibinfo {title} {A unidirectional on-chip photonic interface for superconducting circuits},\ }\href {https://doi.org/10.1038/s41534-020-0261-9} {\bibfield  {journal} {\bibinfo  {journal} {npj Quantum Information}\ }\textbf {\bibinfo {volume} {6}},\ \bibinfo {pages} {1} (\bibinfo {year} {2020})}\BibitemShut {NoStop}%
\bibitem [{\citenamefont {Flynn}\ \emph {et~al.}(2021)\citenamefont {Flynn}, \citenamefont {Cobanera},\ and\ \citenamefont {Viola}}]{Viola3}%
  \BibitemOpen
  \bibfield  {author} {\bibinfo {author} {\bibfnamefont {V.~P.}\ \bibnamefont {Flynn}}, \bibinfo {author} {\bibfnamefont {E.}~\bibnamefont {Cobanera}},\ and\ \bibinfo {author} {\bibfnamefont {L.}~\bibnamefont {Viola}},\ }\bibfield  {title} {\bibinfo {title} {Topology by dissipation: Majorana bosons in metastable quadratic markovian dynamics},\ }\href {https://doi.org/10.1103/PhysRevLett.127.245701} {\bibfield  {journal} {\bibinfo  {journal} {Phys. Rev. Lett.}\ }\textbf {\bibinfo {volume} {127}},\ \bibinfo {pages} {245701} (\bibinfo {year} {2021})}\BibitemShut {NoStop}%
\bibitem [{\citenamefont {Kawabata}\ \emph {et~al.}(2019)\citenamefont {Kawabata}, \citenamefont {Shiozaki}, \citenamefont {Ueda},\ and\ \citenamefont {Sato}}]{PhysRevX.9.041015}%
  \BibitemOpen
  \bibfield  {author} {\bibinfo {author} {\bibfnamefont {K.}~\bibnamefont {Kawabata}}, \bibinfo {author} {\bibfnamefont {K.}~\bibnamefont {Shiozaki}}, \bibinfo {author} {\bibfnamefont {M.}~\bibnamefont {Ueda}},\ and\ \bibinfo {author} {\bibfnamefont {M.}~\bibnamefont {Sato}},\ }\bibfield  {title} {\bibinfo {title} {Symmetry and topology in non-hermitian physics},\ }\href {https://doi.org/10.1103/PhysRevX.9.041015} {\bibfield  {journal} {\bibinfo  {journal} {Phys. Rev. X}\ }\textbf {\bibinfo {volume} {9}},\ \bibinfo {pages} {041015} (\bibinfo {year} {2019})}\BibitemShut {NoStop}%
\bibitem [{\citenamefont {McDonald}\ \emph {et~al.}(2022)\citenamefont {McDonald}, \citenamefont {Hanai},\ and\ \citenamefont {Clerk}}]{McDonaldPRB2022}%
  \BibitemOpen
  \bibfield  {author} {\bibinfo {author} {\bibfnamefont {A.}~\bibnamefont {McDonald}}, \bibinfo {author} {\bibfnamefont {R.}~\bibnamefont {Hanai}},\ and\ \bibinfo {author} {\bibfnamefont {A.~A.}\ \bibnamefont {Clerk}},\ }\bibfield  {title} {\bibinfo {title} {Nonequilibrium stationary states of quantum non-hermitian lattice models},\ }\href {https://doi.org/10.1103/PhysRevB.105.064302} {\bibfield  {journal} {\bibinfo  {journal} {Phys. Rev. B}\ }\textbf {\bibinfo {volume} {105}},\ \bibinfo {pages} {064302} (\bibinfo {year} {2022})}\BibitemShut {NoStop}%
\bibitem [{\citenamefont {Tangpanitanon}\ \emph {et~al.}(2016)\citenamefont {Tangpanitanon}, \citenamefont {Bastidas}, \citenamefont {Al-Assam}, \citenamefont {Roushan}, \citenamefont {Jaksch},\ and\ \citenamefont {Angelakis}}]{tangpanitanon_topological_2016}%
  \BibitemOpen
  \bibfield  {author} {\bibinfo {author} {\bibfnamefont {J.}~\bibnamefont {Tangpanitanon}}, \bibinfo {author} {\bibfnamefont {V.~M.}\ \bibnamefont {Bastidas}}, \bibinfo {author} {\bibfnamefont {S.}~\bibnamefont {Al-Assam}}, \bibinfo {author} {\bibfnamefont {P.}~\bibnamefont {Roushan}}, \bibinfo {author} {\bibfnamefont {D.}~\bibnamefont {Jaksch}},\ and\ \bibinfo {author} {\bibfnamefont {D.~G.}\ \bibnamefont {Angelakis}},\ }\bibfield  {title} {\bibinfo {title} {Topological {Pumping} of {Photons} in {Nonlinear} {Resonator} {Arrays}},\ }\href {https://doi.org/10.1103/PhysRevLett.117.213603} {\bibfield  {journal} {\bibinfo  {journal} {Phys. Rev. Lett.}\ }\textbf {\bibinfo {volume} {117}},\ \bibinfo {pages} {213603} (\bibinfo {year} {2016})}\BibitemShut {NoStop}%
\bibitem [{\citenamefont {Hafezi}\ \emph {et~al.}(2013)\citenamefont {Hafezi}, \citenamefont {Lukin},\ and\ \citenamefont {Taylor}}]{hafezi_non-equilibrium_2013}%
  \BibitemOpen
  \bibfield  {author} {\bibinfo {author} {\bibfnamefont {M.}~\bibnamefont {Hafezi}}, \bibinfo {author} {\bibfnamefont {M.~D.}\ \bibnamefont {Lukin}},\ and\ \bibinfo {author} {\bibfnamefont {J.~M.}\ \bibnamefont {Taylor}},\ }\bibfield  {title} {\bibinfo {title} {Non-equilibrium fractional quantum {Hall} state of light},\ }\href {https://doi.org/10.1088/1367-2630/15/6/063001} {\bibfield  {journal} {\bibinfo  {journal} {New J. Phys.}\ }\textbf {\bibinfo {volume} {15}},\ \bibinfo {pages} {063001} (\bibinfo {year} {2013})}\BibitemShut {NoStop}%
\bibitem [{\citenamefont {Jin}\ \emph {et~al.}(2013)\citenamefont {Jin}, \citenamefont {Rossini}, \citenamefont {Fazio}, \citenamefont {Leib},\ and\ \citenamefont {Hartmann}}]{jin_photon_2013}%
  \BibitemOpen
  \bibfield  {author} {\bibinfo {author} {\bibfnamefont {J.}~\bibnamefont {Jin}}, \bibinfo {author} {\bibfnamefont {D.}~\bibnamefont {Rossini}}, \bibinfo {author} {\bibfnamefont {R.}~\bibnamefont {Fazio}}, \bibinfo {author} {\bibfnamefont {M.}~\bibnamefont {Leib}},\ and\ \bibinfo {author} {\bibfnamefont {M.~J.}\ \bibnamefont {Hartmann}},\ }\bibfield  {title} {\bibinfo {title} {Photon {Solid} {Phases} in {Driven} {Arrays} of {Nonlinearly} {Coupled} {Cavities}},\ }\href {https://doi.org/10.1103/PhysRevLett.110.163605} {\bibfield  {journal} {\bibinfo  {journal} {Phys. Rev. Lett.}\ }\textbf {\bibinfo {volume} {110}},\ \bibinfo {pages} {163605} (\bibinfo {year} {2013})}\BibitemShut {NoStop}%
\bibitem [{\citenamefont {Ramos}\ \emph {et~al.}(2016)\citenamefont {Ramos}, \citenamefont {Vermersch}, \citenamefont {Hauke}, \citenamefont {Pichler},\ and\ \citenamefont {Zoller}}]{ramos_non-markovian_2016}%
  \BibitemOpen
  \bibfield  {author} {\bibinfo {author} {\bibfnamefont {T.}~\bibnamefont {Ramos}}, \bibinfo {author} {\bibfnamefont {B.}~\bibnamefont {Vermersch}}, \bibinfo {author} {\bibfnamefont {P.}~\bibnamefont {Hauke}}, \bibinfo {author} {\bibfnamefont {H.}~\bibnamefont {Pichler}},\ and\ \bibinfo {author} {\bibfnamefont {P.}~\bibnamefont {Zoller}},\ }\bibfield  {title} {\bibinfo {title} {Non-{Markovian} dynamics in chiral quantum networks with spins and photons},\ }\href {https://doi.org/10.1103/PhysRevA.93.062104} {\bibfield  {journal} {\bibinfo  {journal} {Phys. Rev. A}\ }\textbf {\bibinfo {volume} {93}},\ \bibinfo {pages} {062104} (\bibinfo {year} {2016})}\BibitemShut {NoStop}%
\bibitem [{\citenamefont {G\'omez-Le\'on}\ \emph {et~al.}(2022)\citenamefont {G\'omez-Le\'on}, \citenamefont {Ramos}, \citenamefont {Porras},\ and\ \citenamefont {Gonz\'alez-Tudela}}]{gomez-leon_decimation_2022}%
  \BibitemOpen
  \bibfield  {author} {\bibinfo {author} {\bibfnamefont {A.}~\bibnamefont {G\'omez-Le\'on}}, \bibinfo {author} {\bibfnamefont {T.}~\bibnamefont {Ramos}}, \bibinfo {author} {\bibfnamefont {D.}~\bibnamefont {Porras}},\ and\ \bibinfo {author} {\bibfnamefont {A.}~\bibnamefont {Gonz\'alez-Tudela}},\ }\bibfield  {title} {\bibinfo {title} {Decimation technique for open quantum systems: {A} case study with driven-dissipative bosonic chains},\ }\href {https://doi.org/10.1103/PhysRevA.105.052223} {\bibfield  {journal} {\bibinfo  {journal} {Phys. Rev. A}\ }\textbf {\bibinfo {volume} {105}},\ \bibinfo {pages} {052223} (\bibinfo {year} {2022})}\BibitemShut {NoStop}%
\bibitem [{\citenamefont {Caves}(1982)}]{caves_quantum_1982}%
  \BibitemOpen
  \bibfield  {author} {\bibinfo {author} {\bibfnamefont {C.~M.}\ \bibnamefont {Caves}},\ }\bibfield  {title} {\bibinfo {title} {Quantum limits on noise in linear amplifiers},\ }\href {https://doi.org/10.1103/PhysRevD.26.1817} {\bibfield  {journal} {\bibinfo  {journal} {Phys. Rev. D}\ }\textbf {\bibinfo {volume} {26}},\ \bibinfo {pages} {1817} (\bibinfo {year} {1982})}\BibitemShut {NoStop}%
\end{thebibliography}
